\documentclass[11pt]{article}
\pdfoutput=1
\usepackage{amsmath}
\usepackage{amssymb}
\usepackage{graphicx,bbm,mathrsfs}
\usepackage{nicefrac}
\usepackage{slashed}
\usepackage{bbm}
\usepackage{geometry}
\usepackage{jheppub}

\usepackage{mathtools}

\usepackage{dcolumn}   
\usepackage{bm}        
\usepackage{graphicx,mathrsfs}
\usepackage{nicefrac}
\usepackage{multirow}
\usepackage{color}
\usepackage{mathtools}
\usepackage{makecell}

\usepackage{environ} 
\usepackage{lipsum} 
 \NewEnviron{Smaller11}{
           \scalebox{1.1}{$\BODY$} 
 } 
 \NewEnviron{Smaller08}{
           \scalebox{0.8}{$\BODY$} 
 }

\newcommand{\be}{\begin{equation}}
\newcommand{\ee}{\end{equation}}
\newcommand{\bea}{\begin{eqnarray}}
\newcommand{\eea}{\end{eqnarray}}

\usepackage{wasysym}
\usepackage{hhline,colortbl}

\newcommand{\nn}{\nonumber}

\newcommand{\Gf}{G^{(0)}}
\newcommand{\FPi}{I}
\newcommand{\K}{\Xi}

\usepackage{titlesec}

\titleformat*{\section}{\Large\bfseries}
\titleformat*{\subsection}{\large\bfseries}
\titleformat*{\subsubsection}{\large\bfseries}
\titleformat*{\paragraph}{\large\bfseries}
\titleformat*{\subparagraph}{\large\bfseries}

\makeatletter
\newcommand*{\prodsym}{%
  \DOTSB
  \mathop{
    \mathchoice
      {\rlap{\kern.3em\rotatebox[origin=c]{-90}{}}{\prod}}
      {\vcenter{\rlap{\kern.2em\rotatebox[origin=c]{-90}{}}}{\prod}}
      {\sum}{\sum}
  }\slimits@
}
\makeatother

\makeatletter
\DeclareFontFamily{OMX}{MnSymbolE}{}
\DeclareSymbolFont{MnLargeSymbols}{OMX}{MnSymbolE}{m}{n}
\SetSymbolFont{MnLargeSymbols}{bold}{OMX}{MnSymbolE}{b}{n}
\DeclareFontShape{OMX}{MnSymbolE}{m}{n}{
    <-6>  MnSymbolE5
   <6-7>  MnSymbolE6
   <7-8>  MnSymbolE7
   <8-9>  MnSymbolE8
   <9-10> MnSymbolE9
  <10-12> MnSymbolE10
  <12->   MnSymbolE12
}{}
\DeclareFontShape{OMX}{MnSymbolE}{b}{n}{
    <-6>  MnSymbolE-Bold5
   <6-7>  MnSymbolE-Bold6
   <7-8>  MnSymbolE-Bold7
   <8-9>  MnSymbolE-Bold8
   <9-10> MnSymbolE-Bold9
  <10-12> MnSymbolE-Bold10
  <12->   MnSymbolE-Bold12
}{}

\let\llangle\@undefined
\let\rrangle\@undefined
\DeclareMathDelimiter{\llangle}{\mathopen}%
                     {MnLargeSymbols}{'164}{MnLargeSymbols}{'164}
\DeclareMathDelimiter{\rrangle}{\mathclose}%
                     {MnLargeSymbols}{'171}{MnLargeSymbols}{'171}
\makeatother

\begin{document}

\vspace*{15mm}

\begin{center}

\thispagestyle{empty}
{\huge

 Opacity from Loops in AdS
 }\\[20mm]

\renewcommand{\thefootnote}{\fnsymbol{footnote}}

{\large  
Alexandria~Costantino$^{\,a}$ \footnote{acost007@ucr.edu },\, 
Sylvain~Fichet$^{\,b}$ \footnote{sfichet@caltech.edu }\,
}\\[8mm]

\end{center} 
\noindent
\quad\quad\quad\textit{$^a$ Department of Physics \& Astronomy, 
    	    University of  California, }

\noindent \quad\quad\quad\quad \textit{Riverside, CA
92521} \\

\noindent
\quad\quad\quad \textit{$^b$ ICTP South American Institute for Fundamental Research  \& IFT-UNESP,}

\noindent \quad\quad\quad\quad \textit{R. Dr. Bento Teobaldo Ferraz 271, S\~ao Paulo, Brazil
}

\addtocounter{footnote}{-2}

\vspace*{16mm}

\begin{center}
{  \bf  Abstract }
\end{center}

We investigate how quantum dynamics affects the propagation of a scalar field in Lorentzian AdS.
We work in momentum space, in which the propagator admits
two spectral representations (denoted ``conformal'' and ``momentum'')  in addition to a closed-form one, and all have a simple split structure.
Focusing on scalar bubbles, we compute the imaginary part of the self-energy $ {\rm Im} \Pi$ in the three representations, which involves the evaluation of seemingly very different objects. We explicitly prove their equivalence in any dimension, and derive some elementary and asymptotic properties of  $ {\rm Im} \Pi$. 

Using a WKB-like approach in the timelike region, we evaluate the propagator dressed with the imaginary part of the self-energy. We find that the dressing from loops exponentially dampens the propagator when one of the endpoints is in the IR region, rendering this region {opaque} to propagation.
This suppression may have implications for field-theoretical model-building in AdS.
We argue that in the effective theory (EFT) paradigm, opacity of the IR region induced by higher dimensional operators censors the region of EFT breakdown. This confirms earlier expectations from the literature. Specializing to AdS$_5$, we determine a universal contribution to opacity from gravity.

\newpage
\tableofcontents
\newpage

\section{Introduction}
\label{se:intro}

In Lorentzian flat space, propagators of perturbative quantum field theory are proportional to 
\be  \frac{i}{p^2-m^2+\Pi(p) } \,.
\label{eq:prop_flat}
\ee
$\Pi$ is the self-energy, \textit{i.e.} the bilinear operator arising from  quantum loops. This self-energy dresses the free propagator, yielding a Born series which sums to Eq.\,\eqref{eq:prop_flat}.
Unitarity cuts relate the imaginary part of $\Pi$  to processes ending into asymptotic states.
   ${\rm Im}\Pi$ resolves   singularities occurring in the timelike region $p^2>0$,  making leading loop effects an unavoidable ingredient of QFT in the timelike region.

In Lorentzian Anti-de Sitter (AdS) space---as in any other background---the propagator is also dressed by self-energy insertions, as pictured in Fig.\,\ref{fig:Dressed_propa_gen}.
How does this dressed propagator behave in AdS? Some limited intuition from flat space  might be used (see Sec.\,\ref{se:problem}), but in any case an explicit, quantitative description of the effects of quantum dressing  remains to be obtained. 
In this work we set out to investigate some of these effects.

Unlike in flat space, AdS lacks asymptotic states to define a standard $S$-matrix, and thus a standard optical theorem \cite{Balasubramanian:1999ri,Giddings:1999qu}. Interestingly, it was recently discovered that an AdS unitarity cut in a certain spectral representation plays the role of halving diagrams \cite{Meltzer:2019nbs}. 
In our study we will note the interplay between this AdS cut and the imaginary part of the self-energy. 

A study of the dressed propagator involves evaluating loops in AdS and summing the Born series. But propagators in position space are complicated functions of the AdS geodesic distance, making every step a  challenging calculation. 
Loops in AdS have been an intense topic of study, see
\cite{Heemskerk:2009pn, Penedones:2010ue, Cornalba:2007zb,Fitzpatrick:2011hu,
Alday:2017xua,
Alday:2017vkk,
Alday:2018pdi,
Alday:2018kkw,
Meltzer:2018tnm,
Ponomarev:2019ltz,
Shyani:2019wed,
Alday:2019qrf,
Alday:2019nin,
Meltzer:2019pyl,
Aprile:2017bgs,
Aprile:2017xsp,
Aprile:2017qoy,
Giombi:2017hpr,
Cardona:2017tsw,
Aharony:2016dwx,
Yuan:2017vgp,
Yuan:2018qva,
Bertan:2018afl,
Bertan:2018khc,
Liu:2018jhs,
Carmi:2018qzm,
Aprile:2018efk,
Ghosh:2018bgd,
Mazac:2018ycv,
Beccaria:2019stp,
Chester:2019pvm,
Beccaria:2019dju,
Carmi:2019ocp,
Aprile:2019rep, 
Fichet:2019hkg,
Drummond:2019hel,  Albayrak:2020isk, 
Albayrak:2020bso,
Meltzer:2020qbr}. 
 AdS loops are often evaluated in position space and are given by fairly complex expressions.
A summation of the Born series in the spectral formalism in the $O(N)$ model can be found in \cite{Carmi:2018qzm}. However the expressions involve spectral integrals and are fairly difficult to handle for further analysis.

We choose to investigate the dressed propagator by working in the position-momentum space derived from Poincar\'e coordinates $(p^\mu,z)$. 
Though some symmetry is lost when going in this Poincar\'e position-momentum space, various representations of the propagator become simultaneously available which take a simple form. 
Momentum space is also the natural language to study the loop summation and to connect with flat space QFT knowledge and tools. 
Among the above  references, \cite{Albayrak:2020isk, Albayrak:2020bso} and \cite{Meltzer:2020qbr} 
involve calculations in momentum space---the use of momentum space proves to be instrumental in these works.
Momentum space has also been  used in diverse studies of tree-level amplitudes \cite{Raju:2010by,
Raju:2012zs,
Raju:2012zr,
Raju:2011mp,
Bzowski:2013sza,
Bzowski:2015pba,
Bzowski:2018fql,
Bzowski:2015yxv,
Isono:2019ihz,
Isono:2018rrb,
Isono:2019wex,
Coriano:2018bbe,
Coriano:2019dyc,
Maglio:2019grh,
Gillioz:2018mto,
Coriano:2013jba,
Bzowski:2019kwd,
Gillioz:2019lgs,
Farrow:2018yni,
Nagaraj:2019zmk,
Albayrak:2018tam,Albayrak:2019yve,Albayrak:2019asr} and in the calculation of cosmological observables \cite{Maldacena:2002vr,
Maldacena:2011nz,
Mata:2012bx,
Kundu:2014gxa,
Ghosh:2014kba,
Arkani-Hamed:2015bza,
Kundu:2015xta,
Sleight:2019mgd,
Sleight:2019hfp,
Sleight:2020obc,
Arkani-Hamed:2017fdk,
Arkani-Hamed:2018bjr,
Benincasa:2018ssx,
Benincasa:2019vqr,
Arkani-Hamed:2018kmz,
Baumann:2019oyu,
Baumann:2020dch}.

When considering the effective field theory (EFT) of gravity, the theory necessarily becomes strongly coupled in the IR region of the Poincaré patch. 
This feature was pointed out at a qualitative level in \cite{ArkaniHamed:2000ds}, wherein it was suspected that rapid oscillations of the timelike propagator may render this region inaccessible, hence censoring superPlanckian effects. 
A more accurate analysis involving dressed propagators in the timelike region of AdS has been initiated in \cite{Fichet:2019hkg} and \cite{Costantino:2020msc}.
The present work also serves to reinforce and complete these EFT-oriented analyses.

Why might one study quantum field theory in anti-de Sitter  space in the first place? AdS is a maximally symmetric background, it is a  laboratory to understand QFT in more general curved spacetime. In AdS the negative curvature regulates infrared divergences, which can teach us about flat space QFTs \cite{Callan:1989em}.
The AdS metric is conformally flat, such that boundary conformal theories in flat space can be studied by placing a CFT in the AdS bulk \cite{McAvity:1995zd,Giombi:2020rmc}. 
But, perhaps  more importantly, AdS turns out to be a unique window into   strongly coupled gauge theories and into quantum gravity as a consequence of the AdS/CFT correspondence (for initial works see \cite{Maldacena:1997re,
Gubser:1998bc,
Witten:1998qj,Freedman:1998bj,
Liu:1998ty,
Freedman:1998tz,
DHoker:1999mqo,
DHoker:1999kzh},
for some reviews see
\cite{Aharony:1999ti,Zaffaroni:2000vh,Nastase:2007kj,Kap:lecture}), which identifies a dual conformal field theory (CFT) living on the  boundary of AdS.
QFT in AdS is also interesting for phenomenological reasons.
Models in AdS can solve the electroweak hierarchy problem \cite{Randall:1999ee},  give rise to braneworlds with reasonable modifications to 4d gravity \cite{Randall:1999vf,Brax:2003fv}, and model strongly coupled dark sectors \cite{vonHarling:2012sz,McDonald:2010fe,McDonald:2010iq, Brax:2019koq}.

\subsection{Preliminary Observations}
\label{se:problem}

\begin{figure}
\centering
	\includegraphics[width=0.45\linewidth,trim={5cm 0cm 2cm 1cm},clip]{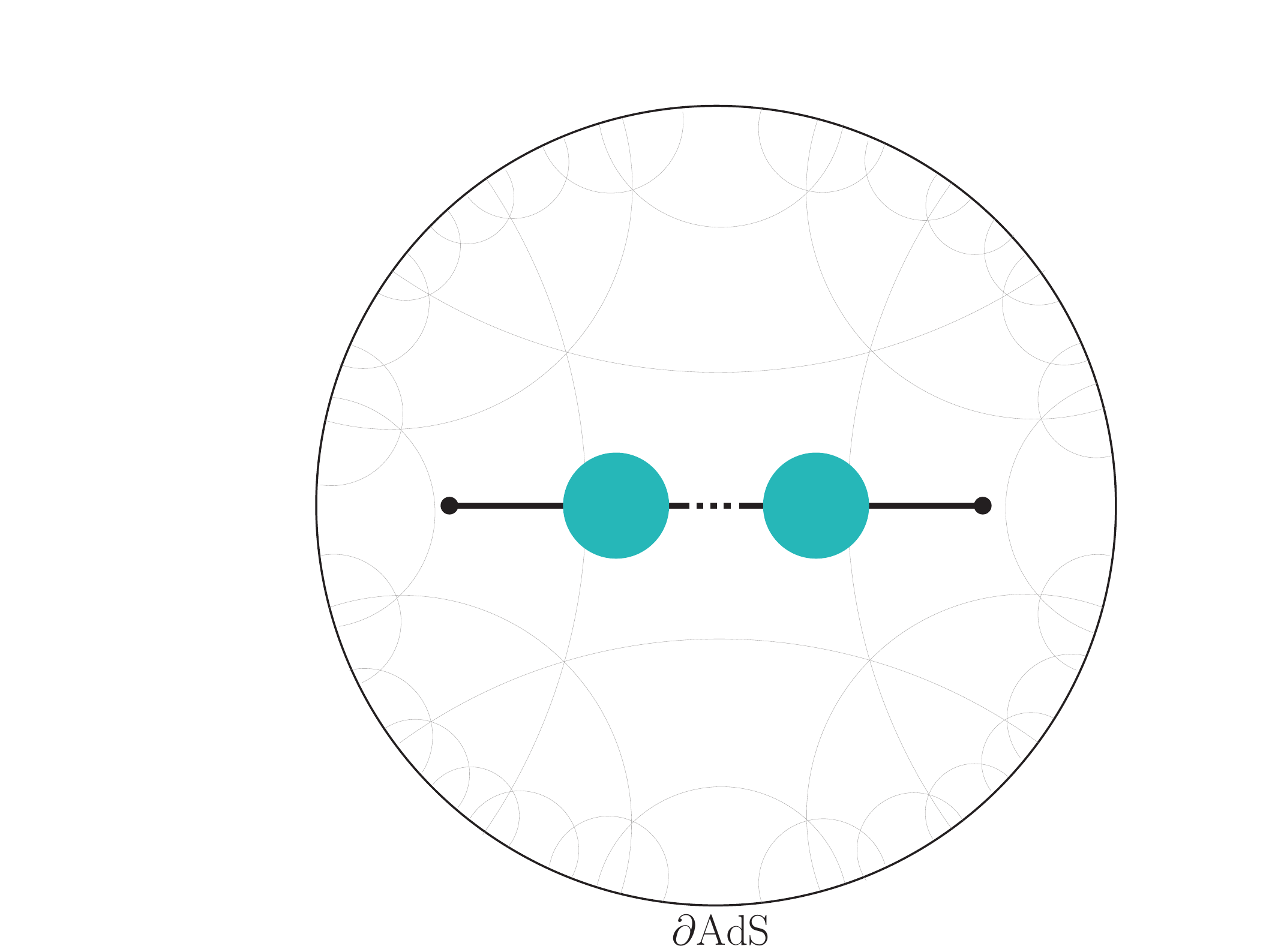}
\caption{
A  propagator  dressed by generic self-energy insertions in the bulk of AdS. 
\label{fig:Dressed_propa_gen}}
\end{figure}

The principal question we want to answer is: How does  quantum dynamics affect the propagation of a field in AdS spacetime?
Here we lay down preliminary observations to set the stage and sharpen the scope of our study. 

Our interest is in interacting field theories. Our notion of field here includes matter fields as well as gravity---treated as an EFT in the sub-Planckian regime.  For the propagating field itself,
we restrict our study to a scalar. We will study quantum effects induced both from matter interactions and from gravity on this propagating scalar.

The presence of interactions results in a  non-trivial self-energy operator $\Pi$ dressing the propagating field. One representation of a dressed propagator is as a Born series---a term of the series is pictured in Fig.\,\ref{fig:Dressed_propa_gen}. 
Another representation of dressing is via the equation of motion following from the quantum effective action, schematically 
\be
\sqrt{|\gamma|}{\cal D}\,G - G * \Pi = -i \delta \,.
\label{eq:EOM_schem}
\ee
Here ${\cal D}$ is the differential operator giving rise to the free equation of motion, $\gamma$ the background metric  on which the field propagates,  $\delta$ a Dirac delta and $G$ is the  propagator.
 A perturbative treatment of the self-energy operator in Eq.\,\eqref{eq:EOM_schem}  generates the  Born series representation.

The qualitative differences in field propagation between the free and interacting theories are expected to occur most clearly in the \textit{timelike} region of spacetime.  
This feature is not specific to AdS.  It is expected simply  because in the timelike regime, the self-energy tends to develop singularities and become complex-valued. 
In particular, while the solutions of the free homogeneous EOM  would tend to be oscillating, in the interacting theory $\Pi$ can develop an  imaginary component  which introduces an exponential behaviour in the solutions. This then translates as an exponential damping in the Feynman propagator described in Eq.\,\eqref{eq:EOM_schem}, where the effect comes from ${\rm Im}\Pi$.
These rough considerations only rely on the analytical structure of $\Pi$, not on the existence of asymptotic states, hence they are valid for QFT in curved spacetime.  

The damping effect can be seen in Lorentzian flat space---though the position space propagator behaves as a power law in the free theory, it exponentially decays in the interacting theory.
Assuming a constant imaginary part ${\rm Im}\Pi\equiv m \Gamma$ in Eq.\,\eqref{eq:prop_flat} gives rise to a  propagator decaying as 
\be
G(X,X')\sim \frac{m^{\frac{d}{2}-1}}{(\Delta X) ^{\frac{d}{2}}}e^{- \frac{\Gamma\, \Delta X}{2}  } \label{eq:Gflat}
\ee
in $d+1-$dimensions, with $\Delta X=\sqrt{\eta_{MN} X^MX'^N}\in \mathbb{R}$ a timelike interval. 
The free theory behaviour amounts to having $\Gamma=0$ in Eq.\,\eqref{eq:Gflat}: there is no exponential decay in this case. This exponential suppression means that particles with $\text{Im}\Pi\neq 0$ decay through time. This interpretation cannot be used in curved spacetimes, where intuitions using asymptotic states are not necessarily valid.

The above points indicate that the imaginary part of the self-energy, ${\rm Im}\Pi$, is likely to play an important role in the field propagation in AdS. Consequently ${\rm Im}\Pi$ will be  a central object in our study.

In flat space, ${\rm Im}\Pi$ implements unitarity cuts. This is not necessarily true in AdS since there are no asymptotic states and hence no standard optical theorem.
However an AdS cut operation has been introduced in \cite{Meltzer:2019nbs}---and a corresponding operation has been identified in  the dual conformal theory \cite{Caron-Huot:2017vep}.
In the scope of the present work, we will note the interplay between the ${\rm Im}()$ and the AdS cut operations.  We do not delve into the CFT side apart from using some elementary results.

At leading order in the loop expansion, ${\rm Im}\Pi$ is expected to be finite in any spacetime dimension. This is because the divergences renormalize local bilinear operators, which contribute to the real part of $\Pi$. The mass term in $\cal D$ may for instance be renormalized. 
We focus on a finite part of $\Pi$ hence renormalization aspects do not require  further discussion and can be ignored.

\subsection{Summary of Results}
\label{se:summary}

Working in Poincar\'e position-momentum space (\textit{i.e.} in   Fourier-transformed Poincar\'e coordinates), three representations of the propagator become simultaneously available and take simple forms: a closed-form ``canonical'' representation, a conformal spectral representation, and a momentum spectral representation.
Each of these representations is useful to illuminate different properties of the AdS loop.

We consider cubic couplings, such that the self-energies considered always have a bubble topology. 
We obtain the following results. 

\begin{figure}
\centering
\includegraphics[angle=0,width=0.48\linewidth,trim={1cm 11cm 2cm 1cm},clip]{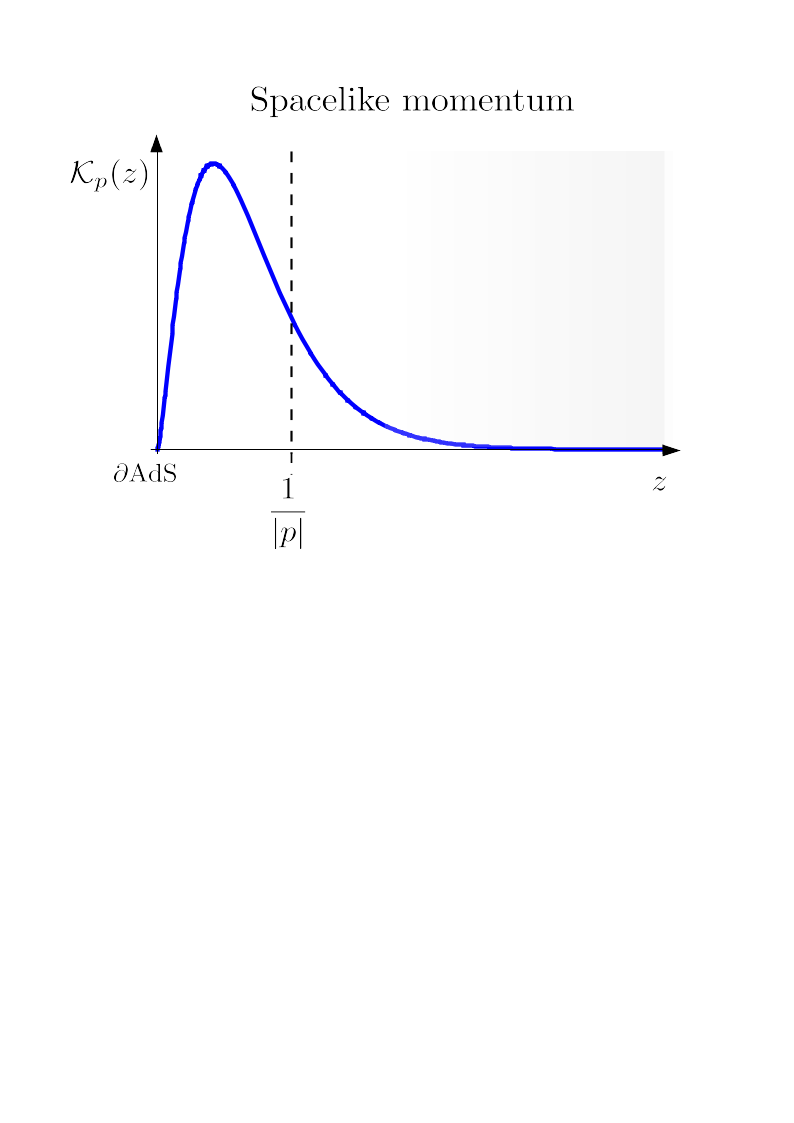}
\includegraphics[angle=0,width=0.48\linewidth,trim={1cm 11cm 2cm 1cm},clip]{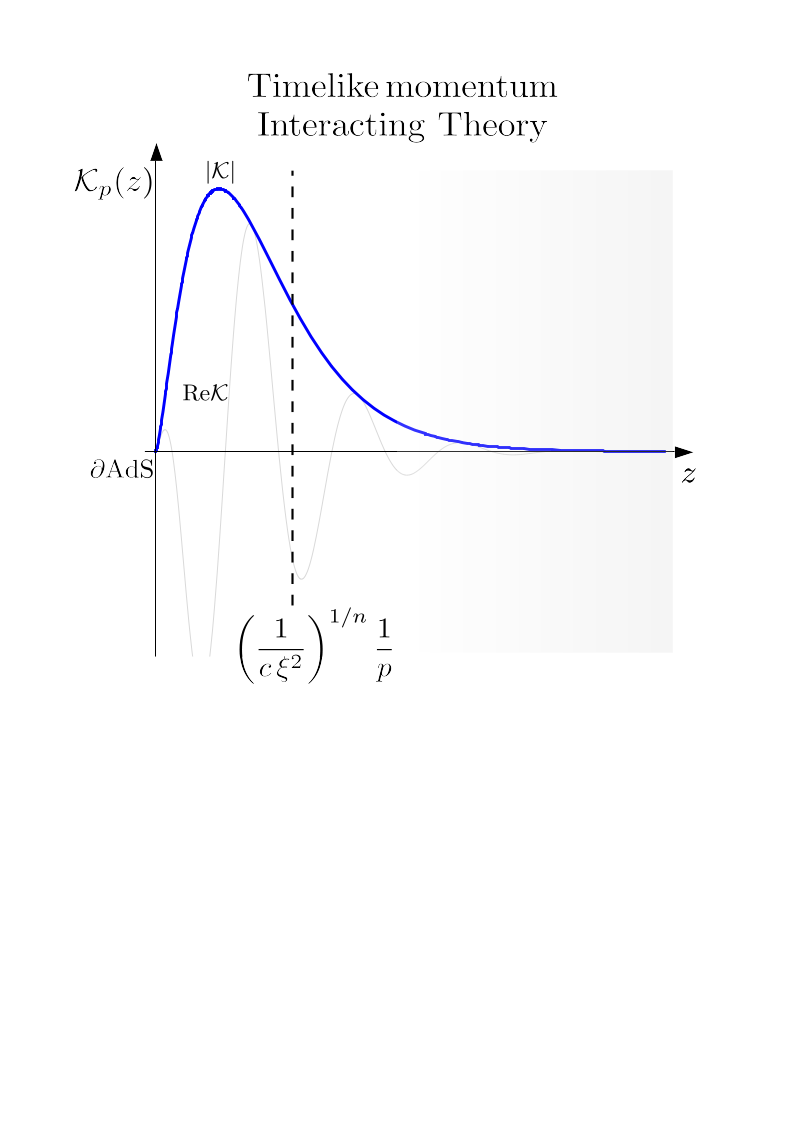}
\caption{
AdS boundary-to-bulk propagators in Poincar\'e position-momentum space $(p^\mu,z)$.  For spacelike momentum the propagator decays exponentially for $|p|z \gg 1$ (left). For timelike momentum, exponential decay is induced by interactions at the quantum level  for 
$c\,\xi^2(pz)^n \gg 1$. The dimensionless coupling $\xi$ parameterizes some cubic interaction and $c\ll 1$ encodes a loop factor suppression. In both cases, AdS becomes ``opaque'' to propagation at large enough $z$. 
\label{fig:CartoonBo2B}}
\end{figure}

\begin{itemize}

\item We derive the imaginary part of the simplest scalar bubble in the three representations mentioned above. In the conformal spectral representation, AdS/CFT arises and we recover various results from the literature.\,\footnote{
In the conformal spectral representation, we show that ${\rm Im}\Pi$ takes the form of a sum over double-trace propagators with coefficients that exactly match those from \cite{Dusedau:1985ue,Fitzpatrick:2011hu}.
 When summing the Born series in the conformal spectral representation, we recover the anomalous dimension found in  \cite{Giombi:2017hpr}.}  We prove the equivalence of all three representations of $\text{Im}\Pi$ and show some of its elementary and asymptotic properties.

\item We solve the dressed equation of motion for timelike momentum $p^2>0$ in the conformally flat  region $pz\gg1$ using a WKB-type approximation. We work in the momentum spectral representation of ${\rm Im }\,\Pi$ and the canonical representation of the propagator. To render the calculations analytically tractable and obtain a closed-form result, we employ kinematic and saddle-point approximations. This is a new method of studying the dressed propagator.

\item The dressing induces an exponential damping of the propagator in the $pz\gg1$ region. This damping behaviour occurs even if a single point is in the conformally flat region, \textit{i.e.} $pz_< \ll1$, $pz_> \gg 1$, 
which includes the  boundary-to-bulk propagator with $ pz\gg 1$ as a particular case. This regime has no flat-space equivalent---the $pz\ll1 $ region vanishes if one Weyl-transforms to flat space.
Renormalizable interactions may not give rise to an exponential damping, but operators of sufficiently high dimension---as present in an EFT---will induce it. This is pictured in Fig.\,\ref{fig:CartoonBo2B}.

\item In the EFT paradigm, the EFT breaks down when higher dimensional operators give contributions of the same order. In AdS this occurs at sufficiently large $pz$, \textit{i.e.} in the IR region of the Poincar\'e patch. Working in the EFT paradigm, we find that the exponential damping censors the region of EFT breakdown.

\item In case of a scalar field theory in AdS$_5$, the leading damping occurs from the $\Phi^3-\Phi \partial_M \Phi \partial^M \Phi$ bubble. The timelike propagator behaves schematically as
\be 
G(p;z,z') \propto e^{
- c \frac{k}{\Lambda} (pz_>)^2
}\,
\ee
where the coefficient $c = O(10^{-2})$ (see Sec.\,\ref{se:AdS5}) includes suppression by a loop factor and  estimates from 5D dimensional analysis. 
Here $k$ is the AdS curvature and $\Lambda$ characterizes the strength of couplings in the EFT. 

\item A partial contribution from bulk gravity in AdS$_5$ leads to the suppression
\be 
G(p;z,z') \propto e^{
- c' \kappa^2 (pz_>)^4}\,
\ee
with $c'=O(10^{-6})$ and $\kappa = \frac{k}{M_{\rm Pl}}$. This effect depends only on the strength of gravity $\kappa$.

\end{itemize}

\subsection{Outline}

Our investigation takes the following steps. 
We lay down the basic formalism and derive the scalar propagator in various representations in Sec.\,\ref{se:AdS}.
Asymptotic properties of Poincar\'e position-momentum space are also discussed. 
Using the conformal spectral representation,
 we evaluate a bubble diagram 
and sum the Born series in Sec.\,\ref{se:AdSCFT}. This involves AdS/CFT---the CFT elements required for the calculations are also given. 
In Sec.\,\ref{se:Properties} we derive
the expressions of ${\rm Im }\Pi$ in the various representations, as well as their equivalence proofs. We proceed to prove some general properties of ${\rm Im }\Pi$. 

In  Sec.\,\ref{se:Opacity} we adopt the EFT viewpoint and study the dressed propagator in the conformally flat region. We present a WKB-like approach and justify some related assumptions needed to tackle the calculations analytically. We use this approach to derive the behavior of the dressed propagator in a given regime. Aspects of EFT validity, arbitrary dimensions, higher order diagrams and  deformed AdS backgrounds are also discussed. A universal contribution from gravity to opacity is then calculated in Sec.\,\ref{se:grav}. 
Closing remarks and possible future directions (including implications for field-theoretical AdS model-building) are given in Sec.\,\ref{se:Conclusion}.

\section{A Scalar Field in $\text{AdS}_{d+1}$ }
\label{se:AdS}

We focus on a scalar quantum field theory in $(d+1)$-dimensional Anti-de Sitter (AdS) spacetime with $d> 2$. The action has the form
\begin{align}
	S &= S_{\rm EH}+
	\int d^{d+1}X \sqrt{g}
	\left(
		\frac{1}{2} \partial_M\Phi \partial^M\Phi 
		- \frac{1}{2} m^2_\Phi \Phi^2   + \ldots
	\right)
	 \, \label{eq:SPhi}
\end{align}
 where $S_{\rm EH}$ is the Einstein-Hilbert action. 
The metric of the AdS background is denoted as $\gamma_{MN}$ such that $g_{MN}=\gamma_{MN}+\ldots$ where the ellipses represent  fluctuations of the metric. 
The background metric in Poincar\'e coordinates is \begin{align}
	ds^2	=\gamma_{MN} \, dX^{M}dX^{N}
			=\frac{1}{(kz)^2}\left( \eta_{\mu\nu}dx^\mu dx^\nu-dz^2 \right)  
\label{eq:metric}
\end{align}
with $k$ the AdS curvature and $z\in [0, \infty ]$.\,\footnote{Poincar\'e coordinates render manifest the $SO(1,1)\times SO(1,d-1)$ subgroup of the $SO(2,d)$  isometry group of AdS$_{d+1}$, which encodes dilatation and  $d$-dimensional Poincar\'e isometries.  }
The AdS boundary is at $z=0$, the Poincar\'e horizon at $z \rightarrow\infty$.   We use a mostly-minus metric such that $\eta=\text{diag}(1,-1,-1, \ldots)$.

The ellipses in Eq.\,\eqref{eq:SPhi} represent operators with more fields and/or more derivatives, including interactions such as $\frac{1}{3!}\Phi^3$, $\frac{1}{3!}\Phi \partial_M\Phi \partial^M\Phi $. Interactions play a central role in this work and will be specified further on. 
The Einstein-Hilbert action  is expanded in Sec.\,\ref{se:grav}, which considers gravity-scalar interactions. In the other sections, we focus on scalar interactions.

Regarding the free part of the scalar action, it is often useful to parameterize the scalar mass as
\begin{align}
	m_\Phi^2 \equiv \left(\alpha^2-\frac{d^2}{4}\right)k^2 \,. 
	\label{eq:mPhi}
\end{align}
The Breitenlohner-Freedman bound is satisfied for $\alpha^2 \ge0$ in any dimension.
 In general we  have $\alpha \in \mathbb{R} $. Throughout this work we restrict to $\alpha \in \mathbb{R}_+ $ without loss of generality. 
In any dimension the  $\alpha=1/2$ value corresponds to a conformally massless scalar, see Sec.\,\ref{se:asymptotics} for details.

In this work we focus on exact AdS, with no departure or truncation of the metric in the UV (towards the boundary) or in the IR (towards the Poincar\'e horizon).  However  our results will also be relevant in the context of deformed AdS backgrounds. This is discussed in Sec.\,\ref{se:deformed_ads}.

\subsection{Free and Dressed Propagator}
\label{se:Dressed_Prop}
 The equation of motion for the free field---when all interactions are neglected---can be obtained by extremizing the fundamental action. This gives
\begin{align}
{\cal D}\Phi_{\rm cl}\equiv\frac{1}{\sqrt{\gamma}}\partial_M(\gamma^{MN}\sqrt{\gamma}\partial_N\Phi_{\rm cl})+m^2_\Phi\Phi_{\rm cl}=0 \label{eq:eom}
\end{align}
where with have introduced the  differential operator ${\cal D}$. 
The Green's function of $\cal D$  is the propagator of the free field,  $\Gf(X,X')=\langle \Phi(X)\Phi(X')\rangle_{\rm free}$, satisfying
\begin{align}
	{\cal D}_X \Gf(X,X')
	= \frac{-i}{\sqrt{\gamma}} \delta^{(d+1)}(X-X') \ . 
	\label{eq:DDelta_free}
\end{align}

In the presence of interactions, the propagator  is dressed by self-energy insertions, \textit{i.e.} by bilinear operators resulting from the quantum dynamics. This is described using the partition function and derived quantities such as the quantum effective action.  

The dressed equation of motion can be obtained from the partition function
\be Z=\int {\cal D}[{\rm fields}]\, e^{i S} \ee
by using invariance under an infinitesimal change of the field variable $\Phi(X)\rightarrow \Phi(X)+\epsilon(X)$. An explicit  derivation is given in App.\,\ref{app:dressed_EOM}.  The result for the  propagator dressed by a generic self-energy $i \Pi (X,X')$ is given by
\begin{align}
	{\cal D}_X G(X,X') - \frac{1}{\sqrt{\gamma}} \Pi * G(X,X')
	= \frac{-i}{\sqrt{\gamma}} \delta^{(d+1)}(X-X') \,
	\label{eq:DDelta}
\end{align}
where $*$ is the convolution product, $A*B(X,X')=\int d^{d+1}Y A(X,Y)B(Y,X')$.

Treating the self-energy operator perturbatively,  one can verify that
Eq.\,\eqref{eq:DDelta} implies the well-known Born series representation of the dressed propagator $ G=\sum^\infty_{n=0} G^{(n)}$ with
\be
G(X,X')= \Gf(X,X')+G^{(1)}(X,X')+\ldots =  \Gf(X,X')+G^{(0)} * i \Pi * \Gf (X,X')+\ldots \label{eq:geom}
\ee
For example, from the perturbative solving of Eq.\,\eqref{eq:DDelta}, the first nontrivial term satisfies \be
{\cal D}_X G^{(1)}(X,X') = \frac{1}{\sqrt{\gamma}} \Pi * G^{(0)}(X,X')\,,
\ee
which gives the contribution Eq.\,\eqref{eq:geom} using that the solution to ${\cal D} \Phi = J$ is given by  $\Phi(X) = i \int d^{d+1}Y \sqrt{|\gamma|} \Gf(X,Y)  J(Y) $. The higher order terms are obtained recursively, generating the Born series representation of $G$.

\subsection{Three Representations   in Poincar\'e Coordinates}
\label{se:prop_reps}

The free Feynman propagator in the  Poincar\'e coordinates $(x^\mu, z)$ defined in Eq.\,\eqref{eq:DDelta_free} is derived in Ref.\,\cite{Burges:1985qq}. Here we rather work in momentum space along the Minkowski slices, using Fourier transform \be \Phi(p^\mu, z)=\int d^d x \Phi(x^\mu, z) e^{i \eta_{\mu\nu} p^\mu x^\nu}\label{eq:FT1}\,.\ee
When working in Fourier-transformed Poincar\'e coordinates, 
the $d$-dimensional Poincar\'e isometries $SO(1,d-1)$ remains manifest, and the dilatation isometry becomes $(p^\mu, z)\rightarrow (p^\mu /\lambda, \lambda z)$.\,\footnote{This follows from requiring invariance of the Fourier transform Eq.\,\eqref{eq:FT1} under these symmetries. } One discrete symmetry is lost: the inversion  $X^M\rightarrow \frac{X^M}{(X_N X^N)} $. 
As a counterpart to position space, position-momentum space offers supplemental insights. These are discussed throughout this section.

We define $p^2=\sqrt{\eta_{\mu\nu}p^\mu p^\nu}$. Notice that the quantity $p z $ is invariant under dilatations in addition to $SO(1,d-1)$ transformations. It is thus a good quantity to characterize points in the position-momentum space. The $pz$ invariant appears throughout this work. 

In position-momentum space the EOM operator becomes  
\begin{align}
{\cal D}=\square+m_\Phi^2= k^2\left(-z^2p^2-z^{d+1}\partial_z\left(z^{1-d}\partial_z\right)+\left(\alpha^2-\frac{d^2}{4}\right)\right) \label{eq:D} \,.
\end{align}
The EOM of the propagator is then given by
\be
{\cal D}_zG^{(0)}(p;z,z')= -i (kz)^{d+1} \delta(z-z') \,. \label{eq:eom_G}
\ee
$G^{(0)}(p;z,z')$ corresponds to the reduced 2-point function defined by factoring out a Dirac delta function associated with overall momentum conservation,
 \be \langle \Phi(p^\mu,z)\Phi(p'^{\mu},z')\rangle_{\rm free} = G^{(0)}(p;z,z') (2\pi)^d \delta^{(d)}(p^\mu-p'^{\mu})\,. \ee
In the following we often use the shortcut
$ G(p;z,z')=G_p(z,z') $.

In Lorentzian space, the physical $p^2$ takes both signs. Since we have chosen the mostly minus metric, we have $p^2<0$ for spacelike momentum, $p^2>0$ for timelike momentum.
In the free theory, $p^2$ is made slightly complex to resolve the non-analyticities arising  for timelike momentum. This corresponds to the inclusion of an infinitesimal imaginary shift $p^2+i\epsilon$, $\epsilon\rightarrow 0$. 
$\epsilon>0$, the ``Feynman prescription'',
is consistent with causality and  defines  the Feynman propagator. 
The $i\epsilon$ shift will often be left implicit in our notations.

 In Fourier space, the homogeneous solutions to the EOM are linear combinations of
\be
z^{d/2} J_\alpha \left(pz\right)\,,\quad z^{d/2} Y_\alpha \left(pz\right) \,, \label{eq:sols}
\ee
using the bulk mass parameter $\alpha$ introduced in Eq.\,\eqref{eq:mPhi}.
It is also useful to use the basis 
\be
z^{d/2} I_\alpha \left(\sqrt{-p^2}z\right)\,,\quad z^{d/2} K_\alpha \left(\sqrt{-p^2}z\right) \,, \label{eq:sols2}
\ee
which shows explicitly the occurence of a branch cut  for timelike momentum $p^2>0$. In this case one has the  identities
\be
I_\alpha \left(\sqrt{-p^2}z\right)K_\alpha \left(\sqrt{-p^2}z\right) =
\begin{dcases}
~~\frac{i \pi}{2}
J_\alpha \left(pz\right)H^{(1)}_\alpha \left(pz\right)
\quad{\rm if}\quad \epsilon>0 \\
-\frac{i\pi}{2}
J_\alpha \left(pz\right)H^{(2)}_\alpha \left(pz\right)
\quad{\rm if}\quad \epsilon<0
\end{dcases}
\ee

We now derive three different representations of the propagator.

\subsubsection{Canonical Representation}

\label{se:prop_closed}

A direct solving of Eq.\,\eqref{eq:eom_G} is possible using standard ODE techniques (see App.\,A of \cite{Fichet:2019owx}). In this reference a solving has been done for $d=4$, but the generalization to arbitrary dimension is straightforward.

The propagator takes the general form \be
G_p(z,z')=\frac{i}{C}F_<(z_<)F_>(z_>) \,
\label{eq:prop_closed}
\ee
with $z_<=\min(z,z')$, $z_>=\max(z,z')$. 
The $F_{<,>}$ functions are linear combinations of the  solutions Eq.\,\eqref{eq:sols} and are determined such that the propagator decays at $z\rightarrow 0, \infty$.  In the timelike regime,  this decay is ensured by the $i\epsilon$ prescription.\,\footnote{The $i\epsilon$ is presumably replaced by a physical effect in the interacting theory. We explicitly show how this occurs in Sec.\,\ref{se:Opacity}.  }
An equivalent method is to assume boundary conditions on branes placed at $z \neq 0, \infty$, and then send those branes to $z\rightarrow 0$ and $z\rightarrow \infty$. 
The overall coefficient $C$ is related to the Wronskian $W=F_<F_>'-F_<'F_>$  such that $C=W(z)/(kz)^{d-1}$ \,\cite{Fichet:2019owx}.

Since we have chosen $\alpha>0$, the condition that Eq.\,\eqref{eq:prop_closed} does not diverge as $z\rightarrow \{0,\infty\}$ dictates that $F_<(z)=z^2 I_\alpha\left(\sqrt{-p^2}z\right)$ and $F_>(z)=z^2 K_\alpha\left(\sqrt{-p^2}z\right)$.
We thus find 
\be \label{eq:CFprop2}
\Gf_p(z,z')= -\frac{i}{k}(kz)^{d/2}(kz')^{d/2}   I_\alpha\left(\sqrt{-p^2}z_<\right) K_\alpha\left(\sqrt{-p^2}z_>\right)
\ee
with $C =-   k $.
In the timelike regime we obtain
\be \label{eq:CFprop}
\Gf_p(z,z')=\begin{dcases}
~~\frac{\pi}{2k}(kz)^{d/2}(kz')^{d/2}   J_\alpha(pz_<)H^{(1)}_\alpha(pz_>)\quad{\rm if}\quad \epsilon>0 \\
~~ - \frac{\pi}{2k}(kz)^{d/2}(kz')^{d/2} J_\alpha(pz_<)H^{(2)}_\alpha(pz_>)\quad {\rm if}\quad  \epsilon<0.
\end{dcases}
\ee

\subsubsection{Spectral Representations}

The homogeneous solutions Eq.\,\eqref{eq:sols} depends on two external continuous parameters, $\alpha$ and $p$. The physical value of $\alpha$ is real and the physical value of $p$ can be either purely real or imaginary, but both of these parameters can be analytically continued everywhere into the complex plane. 

Each of these parameters can be used to develop a spectral representation of the propagator. What is required is a spectral function $\Omega_X(z,z')$ solution of the homogeneous EOM ${\cal D}\Omega_X(z,z')=0$ and satisfying a completeness relation of the form
\be
\int dX \Omega_X(z,z') \propto \delta(z-z')\,
\ee
where the integration is over some specified domain and 
where $X$ is either $\alpha$ or $p$ in our case. As we will see below, an integral representation of $\Gf$ can be easily obtained whenever $\Omega_X$ is known. 
The completeness relation can be difficult to guess directly. However it can be built starting from the propagator, possibly with an appropriate analytic continuation of the relevant parameter.

\subsubsection{Conformal Spectral Representation}
\label{se:alpharep}

Here we consider  the spectral representation based on the bulk mass parameter $\alpha$. In this subsection we  indicate explicitly the $\alpha$ dependence of $\Gf$ and of $\cal D$ via subscripts. 
 The $\alpha$ parameter plays a central role in the AdS/CFT and CFT literature.  $\alpha$ is directly related to the conformal dimension of the operators in the conformal field theory, see Sec.\,\ref{se:AdSCFT} for further details.

A spectral function in $\alpha$ is found to be 
\be \label{eq:Omega_alpha}
\Omega_\alpha(z,z') = \frac{i\alpha \sin(\pi\alpha) (kz)^\frac{d}{2}(kz')^\frac{d}{2}}{\pi^2}K_\alpha\left(\sqrt{-p^2} z \right)K_{\alpha}\left(\sqrt{-p^2} z' \right) \,.
\ee
We find it satisfies the completeness relation
\be
\int^{i\infty}_{-i \infty} d\hat\alpha\,\Omega_{\hat\alpha}(z,z') =(kz)^{d+1} \frac{ \delta(z-z')}{k}\,. \label{eq:comp_alpha}
\ee
The direct proof of Eq.\,\eqref{eq:comp_alpha} is not trivial, we give it in App.\,\ref{se:comp_alpha}.
The spectral function also satisfies the homogeneous EOM ${\cal D}_{z;\alpha} \Omega_\alpha(z,z')=0$.

The propagator with bulk mass parameter $ \alpha$  (\textit{i.e.} with bulk mass $(\alpha^2-d^2/4)k^2$) expressed in the conformal spectral representation takes the form
\be
\Gf_{p, \alpha}(z,z')= \frac{i}{k} \int^{i\infty}_{-i\infty} d\hat\alpha \, P(\hat \alpha,\alpha) \, \Omega_{\hat \alpha}(z,z')\,,\quad \quad P(\hat \alpha,\alpha)=\frac{1}{\hat\alpha^2-\alpha^2}\,.
\label{eq:Delta_alpha}
\ee
One can notice that ${\cal D}_{z;\alpha} \Omega_\alpha(z,z')=0$ implies \be\label{eq:domegaalpha}{\cal D}_{z; \alpha}\Omega_{\hat\alpha}(z,z')=( \alpha^2- \hat\alpha^2)k^2\Omega_{\hat\alpha}(z,z')\,.\ee  
By using Eq.\,\eqref{eq:domegaalpha} and the completeness relation Eq.\,\eqref{eq:comp_alpha}, one can show that Eq.\,\eqref{eq:Delta_alpha} obeys the EOM.

One can also verify that the spectral function satisfies
\be
\Omega_\alpha(z,z')= \frac{\alpha}{2\pi}\left(\Gf_{p,\alpha}(z,z') -\Gf_{p,-\alpha}(z,z') \right)\,. \label{eq:diff_alpha}
\ee
This can be used to prove Eq.\,\eqref{eq:Delta_alpha}. 
Substituting Eq.\,\eqref{eq:diff_alpha} into Eq.\,\eqref{eq:Delta_alpha}, and using that  $\Gf_{p,\alpha}(z,z')\propto \left(\frac{z_<}{z_>}\right)^\alpha$ for large $\text{Re}(\alpha)>0$ as discussed in and below  Eq.\,\eqref{eq:Delta_confconf}, we can close the contour of the $\int d\hat\alpha$ integral of $\Gf_{p,\hat \alpha}(z,z')$ clockwise towards the positive reals and the contour integral of $\Gf_{p,-\hat \alpha}(z,z')$ counterclockwise towards the negative reals. This picks respectively the  $\hat\alpha= \alpha$ and $\hat\alpha=- \alpha$ poles of the $P(\hat\alpha, \alpha)$ measure. The residues combine to prove  Eq.\,\eqref{eq:Delta_alpha}.

The conformal spectral representation obtained here amounts to the Fourier transform of the so-called split representation of the AdS propagator, usually taken in position space in the AdS/CFT literature. The split representation in position space involves a convolution of two boundary-to-bulk propagators \cite{Leonhardt:2003qu}. 
In our position-momentum space formalism, such  boundary convolution integral changes into a product. The ``split'' feature simply corresponds to the fact that $\Omega_\alpha(z,z')$ factors  into the product of the  $z^2 K_\alpha\left(\sqrt{-p^2} z \right)$, $z'^2 K_\alpha\left(\sqrt{-p^2} z' \right)$ solutions. This is not surprising since $\Omega_\alpha(z,z')$ has to satisfy the homogeneous EOM for both $z$ and $z'$. This  dictates that $\Omega_\alpha(z,z')$ must factor into a product of  solutions of the free EOM.

\subsubsection{Momentum Spectral Representation}\label{se:MSrepprop}

Here we consider  the spectral representation based on the absolute 4-momentum $p$. In this subsection we need to  indicate explicitly the $p$ dependence of $\cal D$ in subscript. 

A spectral function in $p$ is found to be 
\be
\Omega_p(z,z') = \frac{p}{k} (kz)^{d/2}(kz')^{d/2}  J_\alpha(pz)J_\alpha(pz') \,.
\ee
It satisfies the homogeneous EOM. Moreover $\Omega_p(z,z')$ satisfies the completeness relation
\be \int^{\infty}_{0} dp  \,\Omega_p(z,z') = 
(kz)^{d-1} \delta(z-z')
\,. \label{eq:comp_p}
\ee
This follows directly from the identity \be \int^\infty_0 d\hat p \hat p J_\alpha(\hat pz)J_\alpha(\hat pz')= z^{-1}\delta(z-z')
\label{eq:JJ}
\,.\ee

The propagator with momentum $ p$ expressed in the momentum spectral representation takes the form
\be
\Gf_{ p}(z,z')= i \int^{\infty}_{0} d\hat p \, P(\hat p, p)\, \Omega_{\hat p}(z,z')\,,\quad \quad P(\hat p, p)=\frac{1}{ p^2-\hat p^2}\,.
\label{eq:Delta_p}
\ee
One can notice that ${\cal D}_{z;p} \Omega_p(z,z')=0$ implies \be\label{eq:domegap}{\cal D}_{z; p}\Omega_{\hat p}(z,z')=( \hat p^2-  p^2)(kz)^2\Omega_{\hat p}(z,z')\,.\ee By using Eq.\,\eqref{eq:domegap} and the completeness relation Eq.\,\eqref{eq:comp_p}, one can show that Eq.\,\eqref{eq:Delta_p} obeys the EOM.
We note that for timelike $\hat p$  the spectral function is given by
\be
\Omega_{\hat p}(z,z') = \frac{\hat p   }{\pi}\left(
\Gf_{ \hat p  + i \epsilon }(z,z')-\Gf_{-\hat p  + i \epsilon }(z,z')
\right) \label{eq:diff_p}
\ee
using that 
\begin{align}\label{eq:JHidentity}
    J(x)H^{(1)}(y)=-(J(-x^*)H^{(1)}(-y^*))^*
\end{align}
for ${\rm Im}(x,y)>0$.

We see that the structure of these expressions is similar to those of the conformal spectral representation. For instance $\Omega_p$ has a split structure just like $\Omega_\alpha$, since it satis\-fies the homogeneous EOM for both $z$ and $z'$. 
The analogy is completed 
by extending the integration  over $\hat p\in[-\infty,+\infty]$ ---and including an overall factor of $1/2$. This then reproduces a structure similar  to the integral Eq.\,\eqref{eq:Delta_alpha} when taking spacelike momentum $ p^2<0$ such that the poles are imaginary.
It follows that Eq.\,\eqref{eq:Delta_p} can be derived from Eq.\,\eqref{eq:diff_p} by closing the contour of the $\hat p$ integral.\,\footnote{ One has 
$\Gf_{\hat p }\propto e^{-{\rm Im} \,\hat p|z-z'| }$ at large $|\hat p|$. One closes the contour upward for both terms. The residues from each term have  opposite signs and add up. }  
Alternatively one could use the known result, Eq.\,(10.22.69) of \cite{NIST:DLMF}.

The form of the spectral function Eq.\,\eqref{eq:diff_p} is reminiscent of the K\"all\'en-Lehmann spectral representation from flat space QFT (see e.g. \cite{Zwicky:2016lka,Peskin:257493}). 
This is because using Eq.\,\eqref{eq:diff_p} with Eq.\,\eqref{eq:JHidentity} amounts to picking the discontinuity of $\Gf$ across the branch cut in $ p$,  
\be\label{eq:discontinuity}
{\rm Disc} [\Gf_p(z,z')]_{p\in R^+}= \frac{\pi }{k} (kz)^{d/2}(kz')^{d/2}  J_\alpha(pz)J_\alpha(pz') \,.
\ee
 Eq.\,\eqref{eq:Delta_p} is then equivalent to  
\be\label{eq:discrep}
\Gf_{ p}(z,z') = \frac{i}{2\pi} \int^\infty_0 d\hat p^2 \frac{{\rm Disc} [\Gf_{\hat p}(z,z')]_{\hat p\in R^+}}{ p^2- \hat p^2}\,,
\ee
which follows from Cauchy's integral formula.

\subsection{Asymptotics in Poincar\'e Coordinates}

\label{se:asymptotics}

In Poincar\'e momentum space, the factorized structure of the propagator implies that the asymptotic behaviour of the propagator can be independently understood for the $z$ and $z'$ endpoints.
The behaviour is dictated by Bessel functions asymptotics, where
the relevant quantities to expand about are either the $pz$, $pz'$ invariants or the $\alpha$ parameter.

In contrast, notice that in full position space, the propagator admits asymptotic behaviours as a function of the chordal distance $ \xi=\frac{(z-z')^2-x^\mu x_\mu}{zz'}$. Hence the asymptotic behaviour in position space involves information from both endpoints, the endpoints are not disentangled like in position-momentum space.

A given propagator has three distinct regimes: ($pz_<\ll 1$, $pz_>\ll 1$), ($pz_<\gg 1$,  $pz_>\gg 1$), and ($pz_<\ll 1$, $pz_>\gg 1$). Expanding the Bessel functions at fixed complex $\alpha$,  the three asymptotic regimes are\,\footnote{The exact criterion for Bessel asymptotics depends on $\alpha$,  throughout the paper we write simply $pz\ll 1$, $pz\gg 1$ for convenience.  }
\be
\Gf_p(z,z')\approx -i \left(k^2zz'\right)^{\frac{d-1}{2}} \frac{ \sin\left(pz_<-\varphi\right)e^{i\left(pz_>-\varphi\right)}}{ p} \quad {\rm if} \quad pz,pz'\gg 1
\label{eq:Delta_flatflat}
\ee

\be
\Gf_p(z,z')\approx
-i \frac{\sqrt{\pi}}{\Gamma(\alpha+1)} \left(k^2zz'\right)^{\frac{d-1}{2}}
\left(\frac{p z_<}{2}\right)^{\alpha+1/2}\frac{e^{i\left(pz_>-\varphi\right)}}{p } 
\quad {\rm if} \quad pz_<\ll 1\,,~pz_>\gg 1\,.
\label{eq:Delta_confflat}
\ee

\be
\Gf_p(z,z')\approx
-\frac{i}{2k \Gamma(\alpha+1)} \left(k^2zz'\right)^{d/2}
\left(
\Gamma(\alpha)\left(\frac{z_<}{z_>} \right)^\alpha+\Gamma(-\alpha) e^{-i\pi\alpha}\left(\frac{p^2z_<z_>}{4}\right)^{\alpha}
\right)
\quad {\rm if} \quad pz,pz'\ll 1
\label{eq:Delta_confconf}
\ee
for $\alpha\not\in \mathbb{Z}$ with $\varphi=\frac{\pi (2\alpha-1)}{4}$. Note that for spacelike momentum $p^2<0$, the expressions Eq.\,\eqref{eq:Delta_flatflat} and Eq.\,\eqref{eq:Delta_confflat} are exponentially suppressed for $|p|z_> \gg 1$.
Some other features of these expressions are discussed further below.

The limit of large bulk mass/conformal dimension at fixed $pz$, $pz'$ is equivalent to the limit taken in Eq.\,\eqref{eq:Delta_confconf} when $\alpha\not\in \mathbb{Z}$. When $\text{Re}(\alpha)$ is large and positive, the first term in Eq.\,\eqref{eq:Delta_confconf} dominates. When $\text{Re}(\alpha)$ is large and negative, it is the second term that dominates.
This limit is useful for  Mellin-Barnes-type integrals appearing  in the conformal spectral representation.

\subsubsection{Conformally Massless Scalar}

Here we further discuss the behaviour of the solutions at $pz \gg 1$.

Consider a scalar in  $d+1$-dimensional flat space with mass $m^2_0$.  A conformal Weyl transform from flat space (${\cal R}=0$) to curved space (${\cal R}\neq 0$) gives  an additional mass contribution uniquely fixed by the geometry, 
\be
m^2_\Phi= \frac{d-1}{4d} {\cal R} +m^2_0\,. \label{eq:m2Phi_weyl}
\ee
We have  ${\cal R}=-d(d+1)k^2$ in AdS. Setting $m_0=0$ in Eq.\,\eqref{eq:m2Phi_weyl} then gives \be
m^2_{\Phi}=\frac{1-d^2}{4}k^2\,. \label{eq:conformal_mass}
\ee
We refer to a scalar with such mass as a \textit{conformally massless} scalar---since it becomes massless when Weyl-transforming from AdS to flat space.

In position-momentum space, 
a scalar field in the $pz\gg 1$ regime \textit{behaves asymptotically as a conformally massless scalar}, \textit{i.e.} its mass is set to Eq.\,\eqref{eq:conformal_mass},
or equivalently  has  $\alpha=1/2$ in any dimension. 
This property can be noted by inspecting the solutions to the EOM for arbitrary $\alpha$ at $pz\gg 1$. The $\alpha$ dependence remains only in phases which are irrelevant for solutions to the homogeneous EOM. An $\alpha$-dependence also remains in higher order terms $O\left(\frac{1}{pz}\right)$ of the asymptotic expansion of the Bessel functions.  

This implies that  $\Gf_p(z,z')$ in the $pz,pz'\gg1$ regime (Eq.\,\eqref{eq:Delta_confconf}) is equivalent to a \textit{massless} propagator in $d+1$-dimensional flat  space under a Weyl transformation. This is shown in next section
where we study the flat space limit in more details.
In contrast, for $\Gf_p(z,z')$ in the $pz_<\ll1$ and $pz_>\gg1$ (Eq.\,\eqref{eq:Delta_confflat}), only ``one half'' of the propagator has such conformally massless behaviour. 
In next section we make clear the propagators in the $pz_<\ll1$, $pz_>\ll1$ and $pz_<\ll1$, $pz_>\gg1$ regimes have no equivalent in flat space. 

\subsubsection{The Flat Space Limit}

\label{se:flat}

Here we  study the flat space limit in order
to bring perspective to the asymptotic behaviours Eqs.\,\eqref{eq:Delta_flatflat}-\eqref{eq:Delta_confconf}.
This limit is defined by the  Weyl transform \be \gamma_{MN}\rightarrow (kz)^2 \gamma_{MN}\ee which takes  AdS space to flat space with a boundary at $z=0$ \cite{birrell1984quantum}.

Since the Weyl transform  amounts to undoing the overall $1/(kz)^2$ factor of the metric Eq.\,\eqref{eq:metric}, such transform also amounts to take the AdS curvature to zero, $k\rightarrow 0$ ---keeping  track of the $k$-dependence of the bulk mass. Such zero-curvature limit is best taken by first switching from Poincar\'e coordinates to the $y$-coordinates
\be
z=\frac{e^{ky}}{k}\,,\quad \gamma_{MN}dX^M dX^N= e^{-2ky} \eta_{\mu\nu}dx^\mu dx^\nu -dy^2\,
\ee
in which it is manifest that $k\rightarrow 0$ gives the $(d+1)$ Minkowski metric. The AdS boundary is  at $y=-\infty$.

For any point which is not on the boundary, taking $k\rightarrow 0$ gives $z-z'\rightarrow y-y'$, $kz\rightarrow 1$, while sending $z, z'$ to infinity. 
This last feature implies that, for  any fixed $p$, there is no $pz\ll 1$ regime in the flat space limit. Instead, any $(p^\mu,z)$ point of the position-momentum space ends up satisfying  $pz\gg 1$. We refer to $pz\gg 1$ as the \textit{conformally flat region}. 

This feature illuminates how the propagator behaves under the flat space limit. For $k\rightarrow 0$ both endpoints have $pz\gg 1$, $pz'\gg 1$, such that the  propagator is in the conformally  flat regime of Eq.\,\eqref{eq:Delta_flatflat}.
Taking $kz\rightarrow 1$ and $\alpha\rightarrow 1/2$ in this expression reproduces the massless scalar propagator in $d+1$-dimensional flat space with a boundary. 

Conversely, the flat space limit makes clear that the $pz\ll 1$, $pz'\ll 1$ and $pz_<\ll 1$, $pz_>\gg 1$ asymptotic regimes shown in Eqs.\,\eqref{eq:Delta_confconf}, \eqref{eq:Delta_confflat} have \textit{no} flat space equivalent. Instead these regimes vanish in the flat space limit. 
These non-conformally flat regimes include the cases of one or two endpoints on the AdS boundary and   are thus key for AdS/CFT features. 
Our study of dressing will focus on the $pz_<\ll 1$, $pz_>\gg 1$ regime in Sec.\,\ref{se:Opacity}.

\section{Bubble and Dressing in the Conformal Spectral Representation \label{se:AdSCFT}}

This section focuses on the self-energy and the dressed propagator in the 
conformal spectral representation. In this representation, the AdS/CFT correspondence\,\footnote{ For AdS/CFT see \cite{Maldacena:1997re,
Gubser:1998bc,
Witten:1998qj}, 
subsequent early works 
\cite{Freedman:1998bj,
Liu:1998ty,
Freedman:1998tz,
DHoker:1999mqo,
DHoker:1999kzh}, 
and some yet unmentioned recent developments \cite{Paulos:2011ie,
Mack:2009gy,
Kharel:2013mka,
Fitzpatrick:2011dm,
Costa:2014kfa,
Paulos:2016fap,
Jepsen:2018ajn,
Jepsen:2018dqp,
Gubser:2018cha,
Hijano:2015zsa,
Parikh:2019ygo,
Jepsen:2019svc,
Penedones:2019tng,
Ponomarev:2019ofr,
Caron-Huot:2018kta}. 
Some lecture notes and reviews are \cite{Aharony:1999ti,Zaffaroni:2000vh,Nastase:2007kj,Kap:lecture}. } naturally arises\,\cite{Fitzpatrick:2011ia}.

The evaluation of the dressed propagator is the primary goal of this work as a whole. The advantage of the conformal spectral representation is that CFT objects naturally appear, which greatly simplifies the evaluation and allows us to use known CFT results. This approach allow us to obtain the dressed propagator written as a spectral integral. Though exact, this representation of the dressed propagator is not convenient for the purpose of obtaining a simple and intuitive idea of its behaviour in position-momentum space. This aspect will instead be tackled in Sec.\,\ref{se:Properties} and Sec.\,\ref{se:Opacity} with the help of the momentum spectral and canonical representations. 

From the dressed propagator in the conformal spectral representation, we can use a simple trick to obtain a form of the self-energy, $\Pi$, itself. We will use this form in the calculations of Sec.\,\ref{se:Properties}. The self-energy is a necessary component of our analysis in Sec.\,\ref{se:Opacity}.

Some results of this section connect to earlier results  from AdS/CFT works \cite{Giombi:2017hpr, Carmi:2018qzm}. The calculations we present here in momentum space offers various cross-checks with these works, and perhaps a different perspective.

\subsection{Bubble Diagram}

We consider a cubic interaction of the form\,\footnote{ \label{foot:int}
The results can be trivially extended to the cubic self-interaction of a scalar field via the interaction $\frac{\lambda}{3!} \Phi^3$.
A symmetry factor of $1/2$ has then to be taken in the  bubble amplitude. 
}
\be
S \supset \int dX^M \sqrt{|\gamma|} \lambda \Phi \Phi_1 \Phi_2 \,.
\ee
The coupling has dimension $[\lambda]=\frac{5-d}{2}$, we introduce the dimensionless coupling 
\be\hat\lambda^2=
\lambda^2 k^{d-5}\,.
\ee

The external propagating field is chosen to be $\Phi$. The $\Phi_1$, $\Phi_2$ fields form the self-energy bubble. In the conformal representation calculations it is convenient to use $\bar\alpha$ for the external field and no bar for the dummy variable $\alpha$ to be integrated over. 
The bulk masses for $\Phi$, $\Phi_{1,2}$ are parametrized as 
\be \label{eq:330}
m^2_\Phi=\left(\bar \alpha^2-\frac{d^2}{4}\right)k^2\,,\quad 
m_{\Phi_{1,2}}^2=\left(\bar \alpha_{1,2}^2-\frac{d^2}{4}\right)k^2\,.
\ee
The respective free propagators are noted $\Gf$ and $\Gf_{1,2}$.

This cubic interaction gives rise to dressing by bubble self-energy diagrams as shown in Fig.\,\ref{fig:Dressed_propa_Plane} (left). The bubble amplitude is  given by
\be
i \Pi(p;u,v) = -\frac{\lambda^2}{(ku )^{d+1}(kv )^{d+1} }  \int\frac{d^dq}{(2\pi)^d} \Gf_1(q+p;u,v)\Gf_2(q;v,u)
\ee
where we have included the metric factors coming from the vertices. This  $\Pi$ is the self-energy entering in the dressed equation of motion  Eq.\,\eqref{eq:DDelta}.

\subsection{A Bit of AdS/CFT in Momentum Space}

The elements of AdS and CFT needed for calculating the dressed propagator in the conformal spectral representation are collected in App.\,\ref{se:CFT}. Here we give a brief description of these AdS/CFT ingredients together with  an outline of the upcoming calculation.  
 
\begin{figure}
\centering
	\includegraphics[angle=0,width=0.23\linewidth,trim={0cm 3cm 0cm 1cm},clip]{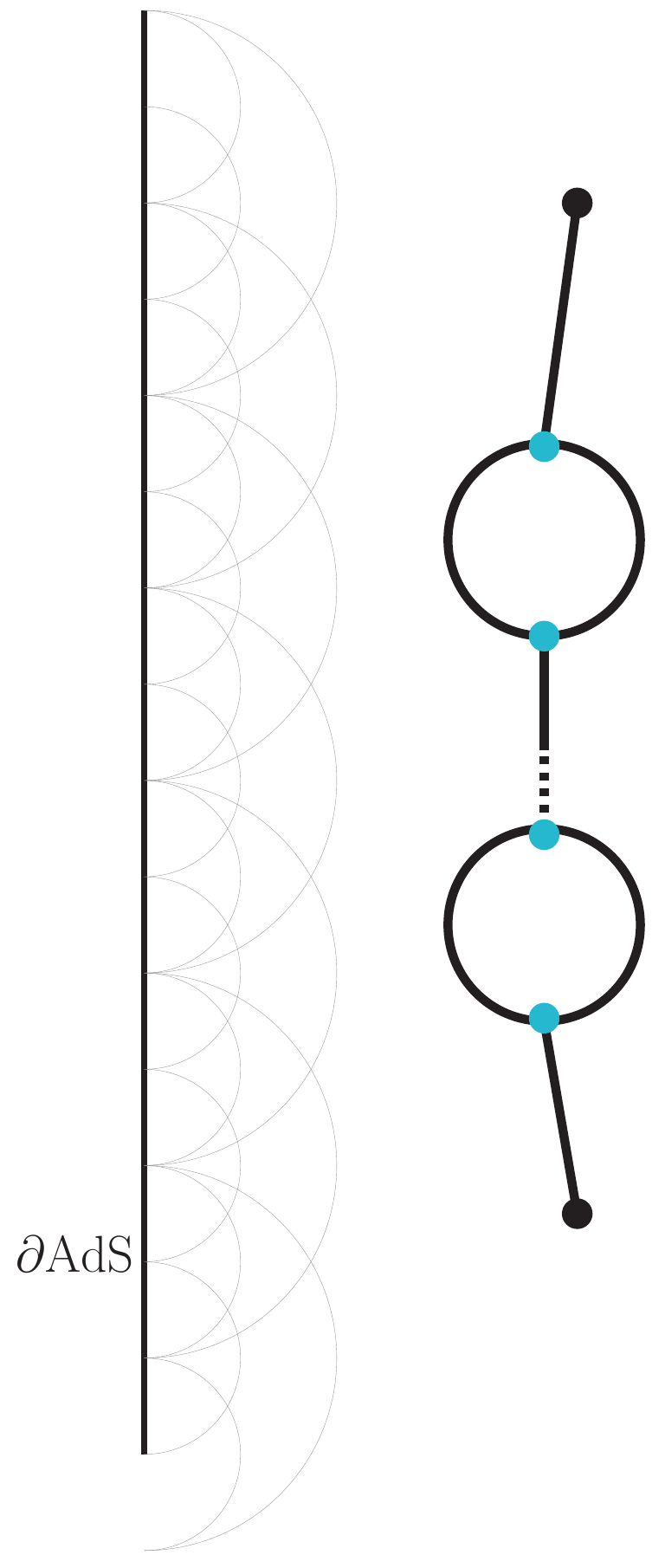}
\quad\quad\quad\quad\quad\quad
\includegraphics[angle=0,width=0.20\linewidth,trim={0cm 3cm 0cm 1cm},clip]{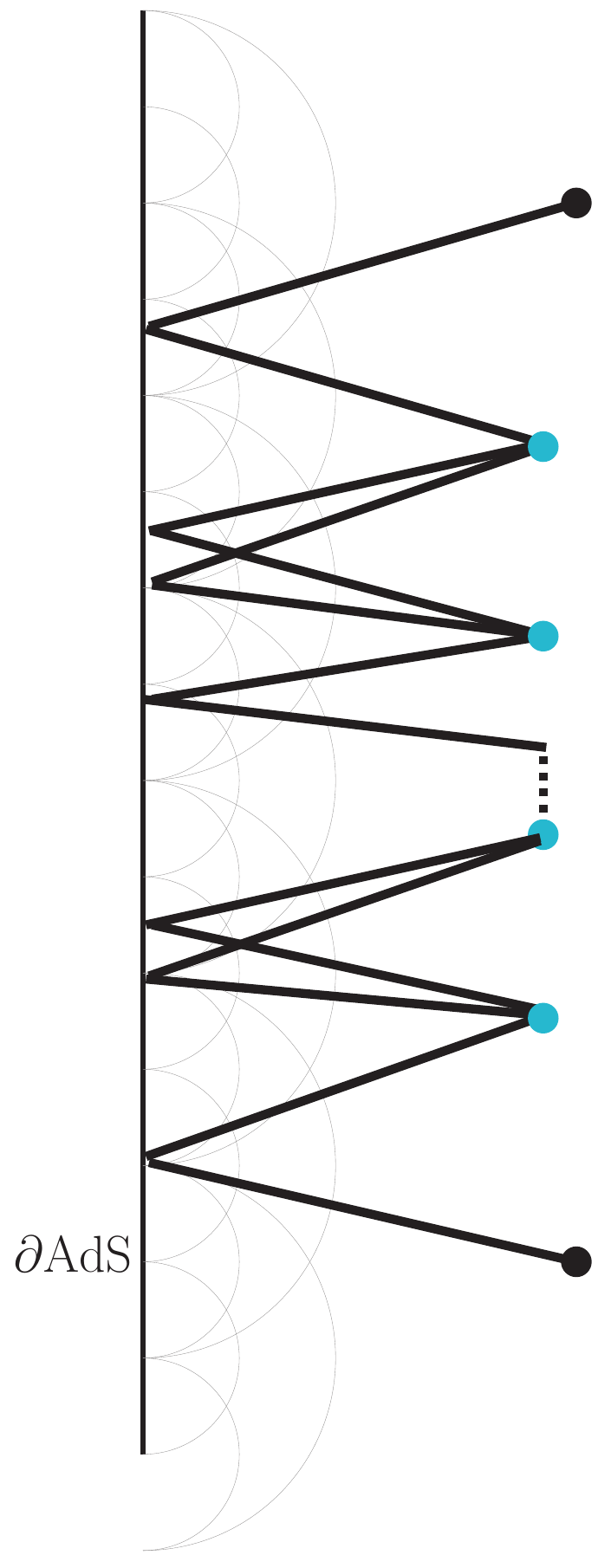}
\quad\quad\quad\quad\quad\quad
\includegraphics[angle=0,width=0.20\linewidth,trim={0cm 3cm 0cm 1cm},clip]{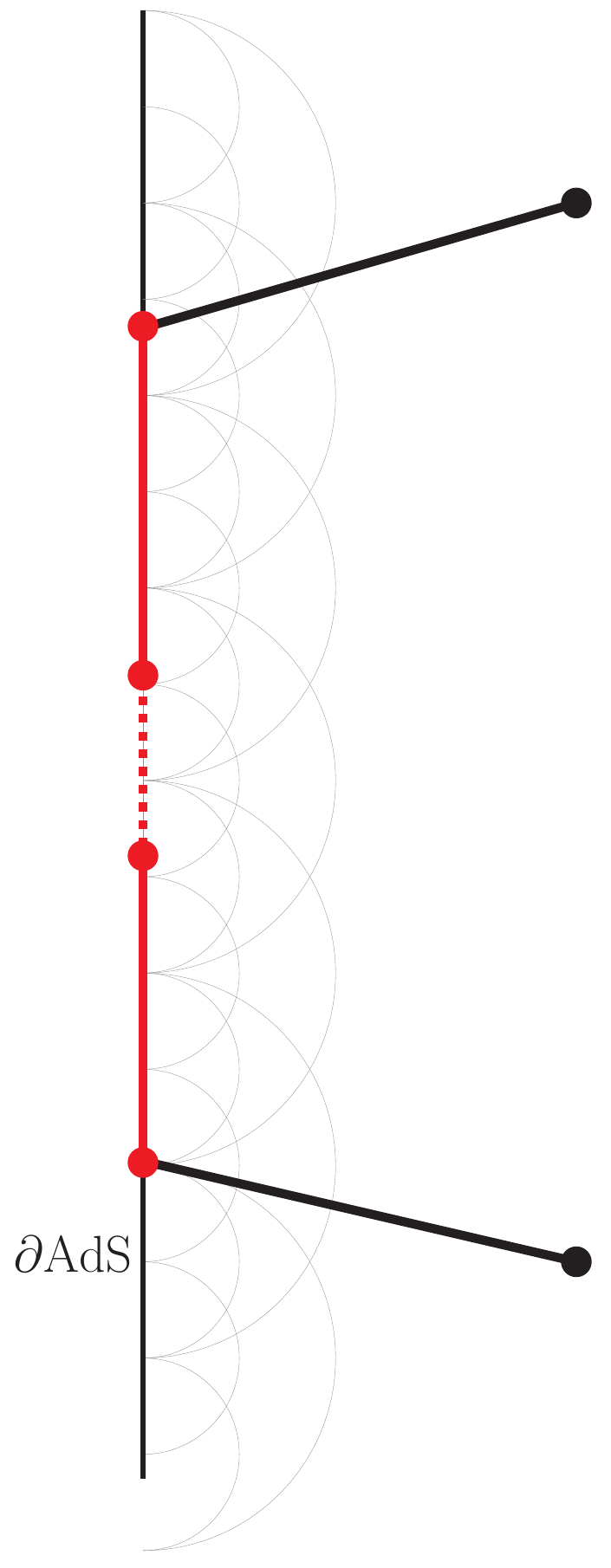}
\caption{
Left: A bulk line dressed by  bubble self-energy diagrams.
Center: Rewriting using the conformal spectral (\textit{i.e.} split)  representation. 
Right: Structure after integrating intermediate bulk points. Each solid red line represents a conformal bubble integral. 
\label{fig:Dressed_propa_Plane}}
\end{figure}

In momentum space, the spectral function $\Omega_\alpha$ can be decomposed as the product
\be
\Omega_\alpha(z,z')\sim {\cal K}_\alpha^+(p,z){\cal K}_\alpha^-(p,z')
\ee
 where ${\cal K}_\alpha^\pm(p,z)$ are boundary-to-bulk propagators defined in App.\,\ref{se:Bo2B}. 
 Writing all the propagators in this form for a given term of the Born series gives the diagram in Fig.\,\ref{fig:Dressed_propa_Plane} (center), where each line represents a   ${\cal K}$ propagator. 
 
The $z$-coordinate of each internal vertex (blue points in Fig.\,\ref{fig:Dressed_propa_Plane}) are  integrated over in each term of the Born series. Each cubic vertex gives rise to convolutions of three ${\cal K}_{\alpha_i}^\pm(p_i,z)$ with distinct $\alpha_i$ and $d$-momentum $p_i$. Each of these triple-K integrals give rise to a CFT three-point correlator in $d$-dimensional Minkowski space, as given by
\be
\int dz \sqrt{|\gamma|} {\cal K}_\alpha^\pm(p_1,z) {\cal K}_\alpha^\pm(p_2,z) {\cal K}_\alpha^\pm(p_3,z) \sim \llangle  {{\cal O }}_1(p_1) {\cal O}_2(p_2) {\cal O}_3(p_3)   \rrangle \,  .
\ee
This is  AdS/CFT in action. The coefficient is given in  Eq.\,\eqref{eq:3K_full}. Only $\alpha$ and $d$-momentum integrals remain. 
This is pictured in Fig.\,\ref{fig:Dressed_propa_Plane} (right). 

Each bulk bubble gives rise to two CFT three-point correlators  connected together via the pairing described in App.\,\ref{se:CFT_shadow}, forming a (non-amputated) CFT bubble. 
The evaluation of the CFT bubble in momentum space is given in  App.\,\ref{se:CFT_bubble} 
\be 
\int \frac{d^dq}{(2\pi)^d}
\llangle  {{\cal O }}_a(p) {\cal O}_1(q) {\cal O}_2(-p-q)   \rrangle
\llangle  \tilde{{\cal O }_1}(-q) \tilde{\cal O}_2(p+q) \tilde {\cal O}_b(-p)   \rrangle  \sim \delta_{ab} \delta(\alpha-\alpha')+s.t.\, 
\ee
where $\tilde{ \cal O}$ denotes the shadow transform of $ \cal O$ (see App.\,\ref{se:CFT_shadow}) and $s.t$ is short for shadow transform. 

Finally, the $\delta(\alpha\pm \alpha')$ arising from each CFT bubble  eliminates a $\int d\alpha $ integral from one of the $G^{(0)}$  connected to it. This implies that for any term of the Born series, a single $\int d\alpha $ integral ultimately remains. This allows the Born series to be summed.

In the next sections we proceed with the calculation of the dressed propagator per se.

\subsection{One Bubble Insertion}\label{se:G1}

In this subsection, we work out the details of the $G^{(1)}$ term of the Born series $G=\sum^\infty_{n=0} G^{(n)}$ which is the term with one insertion, see  Eq.\,\eqref{eq:geom}. Computation of higher terms is similar, and 
the full series  will be summed in next subsection.

The $G^{(1)}$ term of the series reads 
\begin{align} 
G^{(1)}\, (z,z')&=\Gf*i \Pi*\Gf\, (z,z') \label{eq:G1_1}
\\ \nn &=
\int_{-i \infty}^{i \infty} d \alpha \int_{-i \infty}^{i \infty} d \alpha' \,
\left(\frac{i}{k}\right)^2 P(\alpha,\bar \alpha)P(\alpha',\bar \alpha) \,\, \Omega^{(0)}_{ \alpha} *
i \Pi * \Omega^{(0)}_{ \alpha'}\,(z,z')
\end{align}
with the measures
\be\label{eq:alphaprops}
P(\alpha,\bar \alpha)=\frac{1}{ \alpha^2-\bar\alpha^2}\,,\quad
P(\alpha',\bar \alpha)=\frac{1}{ \alpha'^2-\bar\alpha^2} \,.
\ee
We single out the  convolution appearing in Eq.\,\eqref{eq:G1_1}, 
\begin{align}
& \Omega^{(0)}_{\alpha}* i \Pi * \Omega^{(0)}_{\alpha'}(z,z') = \int du \int dv \,\, \Omega^{(0)}_{\alpha}(z,u) \, i \Pi(u,v) \, \Omega^{(0)}_{\alpha'}(v,z')  
\end{align}
and introduce the conformal spectral representation  for the bubble and the external legs. We write all spectral functions in terms of the boundary-to-bulk propagators as shown in Eq.\,\eqref{eq:Omega_KK}, giving
\begin{align} \label{eq:G1_2}
& 
\int du \int dv \,\, \Omega_{\alpha}^{(0)}(z,u) \, i \Pi(u,v) \, \Omega_{\alpha'}^{(0)}(v,z')
\\ & \nn = \left(\frac{1}{4 \pi}\right)^4 \left(\frac{\lambda}{k}\right)^2
\int \frac{ du}{(ku)^{d+1}} \int \frac{ dv}{(kv)^{d+1}}  \,
\int_{-i \infty}^{i \infty} d \alpha_1 \int_{-i \infty}^{i \infty} d \alpha_2
\,P(\alpha_{1},\bar\alpha_{1})P(\alpha_{2},\bar\alpha_{2})\, \times \\ & \nn 
  \int\frac{d^dq}{(2\pi)^d} 
 {\cal K}^+_{\alpha }( p; z)
 {\cal K}^-_{\alpha }( p; u)
 {\cal K}^+_{\alpha_1 }( q+p; u)
    {\cal K}^-_{\alpha_1 }( q+p;  v)
 {\cal K}^+_{\alpha_2 }( q;  v) 
   {\cal K}^-_{\alpha_2 }( q;  u)
   {\cal K}^+_{\alpha' }( p;  v)
  {\cal K}^-_{\alpha'}(p;  z').
\end{align}
The diagram involves products of boundary-to-bulk propagators---it corresponds to Fig.\,\ref{fig:Dressed_propa_Plane} (center) with a single bubble. 

In $G^{(1)}$ there are two triple-K integrals, one corresponding to each vertex. Using results from App.\,\ref{se:Bo2B}, we have
\begin{align}
& \int \frac{du}{(ku)^{d+1}}  {\cal K}^-_{\alpha }( p; u){\cal K}^+_{\alpha_1 }( q+p; u){\cal K}^-_{\alpha_2 }( q;  u) 
 =  c_{\alpha,-\alpha_1,\alpha_2}k^{d/2-1}
\llangle  \tilde{{\cal O }}(p) {\cal O}_1(-p-q) \tilde {\cal O}_2(q)   \rrangle 
\\
& \int \frac{dv}{(kv)^{d+1}}  {\cal K}^+_{\alpha' }( p; v){\cal K}^-_{\alpha_1 }( q+p; v){\cal K}^+_{\alpha_2 }( q;  v) 
 =  c_{-\alpha',\alpha_1,-\alpha_2}k^{d/2-1}
\llangle  {{\cal O }}(-p) \tilde {\cal O}_1(p+q)  {\cal O}_2(-q)   \rrangle
\,\,.
\end{align}
We see that a CFT bubble integral arises in the last line of Eq.\,\eqref{eq:G1_2} when performing the intermediate bulk point integrations. This is shown in Fig.\,\ref{fig:Dressed_propa_Plane} (right). The evaluation of the CFT bubble   in momentum space is given in Eq.\,\eqref{eq:CFT_bubble}. Both terms of the right-hand side of Eq.\,\eqref{eq:CFT_bubble}  give identical contribution by shadow symmetry ($\alpha\leftrightarrow-\alpha$).

Putting the pieces together, we have
\begin{align} \label{eq:G1_3}
&\Omega^{(0)}_\alpha* i \Pi * \Omega^{(0)}_{\alpha'}(z,z') =  2 \delta(\alpha-\alpha')  \frac{\lambda^2 k^{d-4} }{(4\pi)^4}     \,2\pi {\cal B}_{\cal O} \,
{\cal K}^+_{\alpha}(pz){\cal K}^-_{\alpha'}(pz)
\\ & \nn 
\quad\quad\quad
\times  \int d\alpha_1  d\alpha_2 P(\alpha_1,\bar\alpha_1) P(\alpha_2,\bar\alpha_2)
 c_{\alpha,-\alpha_1,\alpha_2}c_{-\alpha',\alpha_1,-\alpha_2},
\end{align}
where ${\cal B}_{\cal O}$ is given in Eq.\,\eqref{eq:BO}. In Eq.\,\eqref{eq:G1_3} we are left with two boundary-to-bulk propagators associated with the two endpoints of $G^{(1)}$. Since $\alpha=\alpha'$ from the Dirac delta,  these two boundary-to-bulk propagators can be combined into a spectral function using Eq.\,\eqref{eq:Omega_KK}. We thus obtain
\begin{align} \label{eq:G1_4}
\Omega^{(0)}_\alpha* i \Pi * \Omega^{(0)}_{\alpha'}(z,z')& = 
\delta(\alpha-\alpha') 
 \lambda^2k^{d-4}\, \frac{1}{4\pi} {\cal K}^+_{\alpha}(pz){\cal K}^-_{\alpha'}(pz) \, {\cal B}(\alpha)
\\ & \nn
=\delta(\alpha-\alpha') 
\lambda^2k^{d-4}  \,i {\cal B}(\alpha)\, \Omega^{(0)}_{\alpha}(z,z'),
\end{align}
where we introduced the dimensionless bubble function 
\begin{align}\label{eq:bubblefunction}
&{\cal B}(\alpha)= - 2
\frac{1}{(4\pi)^3}\,
 2\pi {\cal B}_{\cal O}
\times \int d\alpha_1 d\alpha_2  \,P(\alpha_{1},\bar\alpha_{1})P(\alpha_{2},\bar\alpha_{2})\,
 c_{\alpha,-\alpha_1,\alpha_2}c_{-\alpha,\alpha_1,-\alpha_2}
 \\ & \nn
 = -   
 \frac{1}{2^5 \pi^{4+d/2}}
 \frac{1}{\Gamma(d/2)\Gamma(d/2-\alpha)\Gamma(d/2+\alpha)}
 \int d\alpha_1 d\alpha_2 \frac{\alpha_1 \alpha_2 \sin(\pi \alpha_1)\sin(\pi \alpha_2)}{(\alpha^2_1- \bar\alpha^2_{1})(\alpha^2_2- \bar\alpha^2_{2})}
  \\ & \nn \times  
\Gamma\left( { \begin{Smaller11} \frac{\alpha+\alpha_1+\alpha_2+d/2}{2} \end{Smaller11} }  \right)
\Gamma\left({\begin{Smaller11} \frac{\alpha-\alpha_1+\alpha_2+d/2}{2} \end{Smaller11} }\right)
\Gamma\left({\begin{Smaller11} \frac{\alpha+\alpha_1-\alpha_2+d/2}{2}
\end{Smaller11} }  \right)
\Gamma\left({\begin{Smaller11}  \frac{-\alpha+\alpha_1+\alpha_2+d/2}{2} \end{Smaller11} }\right)
  \\ & \nn \times   
\Gamma\left({\begin{Smaller11}  \frac{-\alpha-\alpha_1-\alpha_2+d/2}{2} \end{Smaller11} }\right)
\Gamma\left({\begin{Smaller11}  \frac{-\alpha+\alpha_1-\alpha_2+d/2}{2} \end{Smaller11} }\right)
\Gamma\left({\begin{Smaller11}  \frac{-\alpha-\alpha_1+\alpha_2+d/2}{2} \end{Smaller11} }\right)
\Gamma\left({\begin{Smaller11}  \frac{\alpha-\alpha_1-\alpha_2+d/2}{2} \end{Smaller11} }\right)
\end{align}
This bubble function is proportional to the one obtained in \cite{Giombi:2017hpr}. An additional cross-check is obtained further below.

In Eq.\,\eqref{eq:G1_4} we have reduced the convolutions in $G^{(1)}$ into a nontrivial algebraic factor. 
We complete the evaluation by putting Eq.\,\eqref{eq:G1_4} back into Eq.\,\eqref{eq:G1_1}, which gives
\begin{align}
&G^{(1)}\, (z,z')=\Gf*i \Pi*\Gf\, (z,z') \label{eq:G1}
\\ \nn &=
\int_{-i \infty}^{i \infty} d\alpha \int_{-i \infty}^{i \infty} d\alpha' \,
\left(\frac{i}{k}\right)^2 P(\alpha,\bar \alpha) P(\alpha',\bar \alpha)  \,\, \delta(\alpha-\alpha') 
\lambda^2 k^{d-4} i {\cal B}(\alpha) \Omega^{(0)}_{\alpha}\,(z,z')
\\ \nn &=
\frac{i}{k}
\int_{-i \infty}^{i \infty} d\alpha \,
 P^2(\alpha,\bar \alpha)   \, \left[-\hat\lambda^2 {\cal B}(\alpha)\right] \Omega^{(0)}_{\alpha}\,(z,z').
\end{align}

\subsection{The Spectral Born Series}\label{se:series}

The above steps can be reproduced for an arbitrary number of bubble insertions. One finds the $n$-bubble contribution to be
\be
G^{(n)}\, (z,z') =
\frac{i}{k}
\int_{-i \infty}^{i \infty} d\alpha \,
 P(\alpha,\bar \alpha)  \,\left[ -  \hat\lambda^2 P(\alpha,\bar \alpha) {\cal B}(\alpha) \right]^n \Omega^{(0)}_{\alpha}\,(z,z').
\ee
We see that a geometric series appears and hence the full Born sum can be performed. This is because spectral transforms turn convolutions into products, as discussed in \cite{Carmi:2018qzm}. Here we have recovered this feature via direct calculation. There is no analog to this property in the momentum spectral representation, because unlike in the conformal representation the bubble integral does not conserve the spectral variable $\hat p$. 

Summing the Born series, we find that the complete dressed propagator in the conformal spectral representation is
\be
\sum_{n=0}^\infty G^{(n)}\, (z,z') = G\, (z,z') = 
\left(\frac{i}{k}\right)
\int_{-i \infty}^{i \infty} d\alpha \,
\frac{1}{P(\alpha,\bar \alpha)^{-1} +  \hat\lambda^2 {\cal B}(\alpha)}
 \Omega^{(0)}_{\alpha}\,(z,z')\,.
 \label{eq:G_dressed_alpha}
\ee
Thus it turns out that in the conformal spectral representation the dressing amounts to a deformation of the $P(\alpha,\bar\alpha)$ measure and $\hat\lambda^2 {\cal B}(\alpha)$ effectively has the role of a ``spectral self-energy''. In particular we can see that the $\alpha=\pm\bar\alpha$ poles of the free spectral function $P(\alpha,\bar \alpha)$ get shifted by the $\hat\lambda^2 {\cal B}(\alpha)$ term. 

 \subsection{Aside: Anomalous Dimension from the Dressed Propagator}

For sufficiently small coupling $\hat \lambda \ll 1$ we can approximate ${\cal B}(\alpha)\approx {\cal B}(\bar \alpha)  $ near the $\alpha=\pm \bar \alpha$ poles.
Other poles exist whenever $\hat \lambda \neq 0$, but the corresponding residues are expected to be small for  $\hat \lambda \ll 1$ and are not our focus.\,\footnote{
From direct evaluation \cite{Giombi:2017hpr,Carmi:2018qzm}, ${\cal B}( \alpha)$ features an infinite set of simple poles. ${\cal B}( \alpha)$ goes to $\pm\infty$ when approaching each pole from either side. Hence for any  $\hat \lambda\neq 0$, $1/(P(\alpha,\bar\alpha)^{-1}+\hat\lambda^2{\cal B}(\alpha))$ has an infinite set of poles contributing to Eq.\,\eqref{eq:G_dressed_alpha}.
For $\hat \lambda \ll 1$ these extra residues are expected to vanish with $\hat \lambda $. Otherwise, there would be a discontinuity in $G(z,z')$ (and thus in the spectrum of operators obtained from related Witten diagrams) when turning on  $\hat \lambda$. 
See \cite{Carmi:2018qzm} for a related study of  ${\cal B}(\bar \alpha)$  in the $O(N)$ model.
}
The ${\cal B}(\bar \alpha)$ constant can be seen as a correction to  the bulk mass,
\be
m^2_\Phi|_{\rm dressed}=\left(\bar\alpha^2 -\frac{d^2}{4}-{\cal B}(\bar \alpha)\right)k^2 \,.
\ee
In the dual CFT, the correction amounts to an anomalous dimension $\gamma$ shifting the conformal dimensions $\Delta_\pm$. We define $\gamma$ such that \be \Delta_\pm|_{\rm dressed}=\Delta_\pm\pm\gamma=\frac{d}{2}\pm\bar\alpha\pm\gamma 
\label{eq:an_dim_def}
\,,\ee with 
\be
\gamma=-\frac{{\cal B}(\bar \alpha)}{2 \bar \alpha}\,. \label{eq:an_dim}
\ee
The fact that the sign of the correction  flips between $\Delta_+$ and $\Delta_-$ is enforced from the AdS side since it is a correction on $\bar \alpha$ and since $\Delta_\pm=\frac{d}{2}\pm \bar\alpha$. The sign flip is also clear from the CFT side in order for  shadow symmetry to be respected. 

We find that the imaginary part of ${\cal B}(\bar \alpha)$ vanishes. To note this, close the  contours in Eq.\,\eqref{eq:bubblefunction} and use the residue theorem. The residues are real and hence by counting factors of $i$, we determine that ${\cal B}(\bar \alpha)$ is purely real. This is consistent with the conclusions in \cite{Giombi:2017hpr}. Thus the leading correction to $\bar\alpha$ is purely real. 

We find that our result Eq.\,\eqref{eq:an_dim} precisely matches the anomalous dimension found in \cite{Giombi:2017hpr} (Eq.\,(2.37) in that reference). In that reference the anomalous dimension was evaluated by looking at the $\log(\Delta X)$  term in a on-shell  bubble amplitude in position space. Here we have confirmed their result with a  different approach. From Eq.\,\eqref{eq:G_dressed_alpha} other  higher order effects such as wave-function renormalization could be studied.

\subsection{The Self-Energy as a Spectral Transform}

To obtain a useful form of $\Pi$ in the conformal spectral representation, we apply the EOM operator $\cal D$ on both sides of $G^{(1)}(z,z')=\Gf*i\Pi * \Gf (z,z')$. Using the free EOM Eq.\,\eqref{eq:eom_G}, we obtain
\begin{align}
 {\cal D}_z{\cal D}_{z'} G^{(1)}(z,z')=-i \frac{1}{\sqrt{|\gamma|}_z\sqrt{|\gamma|}_{z'}} \Pi(z,z')  =
-i (kz)^{d+1}  (kz')^{d+1} \Pi(z,z'). \label{eq:DDG1}
\end{align}
$G^{(1)}$ in the conformal spectral representation has been computed in Eq.\,\eqref{eq:G1}. Using it in Eq.\,\eqref{eq:DDG1} 
gives 
\begin{align}
 {\cal D}_z{\cal D}_{z'} G^{(1)}(z,z') &=
 \frac{i}{k}
\int_{-i \infty}^{i \infty} d\alpha \,
 P^2(\alpha,\bar \alpha)  \, \left[-\hat\lambda^2 {\cal B}(\alpha) \right]  {\cal D}_z{\cal D}_{z'}\Omega^{(0)}_{\alpha}\,(z,z')\\ \nn & = -   ik^3
\int_{-i \infty}^{i \infty} d\alpha \,
 \, \hat\lambda^2 {\cal B}(\alpha) \Omega^{(0)}_{\alpha}\,(z,z')\,.
\end{align}
In doing so we have thus obtained
\be \label{eq:SEidentity}
i (kz)^{d+1}  (kz')^{d+1} \Pi(z,z')  = i \hat\lambda^2 k^3
\int_{-i \infty}^{i \infty} d\alpha \,
    {\cal B}(\alpha)  \Omega^{(0)}_{\alpha}\,(z,z') \,.
\ee
This establishes that $\hat \lambda^2 k^3 {\cal B}(\alpha)$ is the spectral transform of $ (k^2zz')^{d+1} \Pi$. This is the form of the self-energy that we will use in the conformal spectral calculation of Sec.\,\ref{se:CSresult}.

\section{Representations and Properties of ${\rm Im}\Pi$}
\label{se:Properties}

In this section we evaluate  a self-energy bubble diagram $i \Pi$ in the three different representation  shown in Sec.~\ref{se:AdS}. Our focus is on the imaginary part ${\rm Im}\Pi$. 
The equivalence of the expressions in various representations is nontrivial, hence we demonstrate it explicitly in Sec.~\ref{se:equality}.
The various representations render manifest different properties of the self-energy. These properties are discussed in Sec.~\ref{se:PropertiesSE}. 

Throughout this section the interaction considered is a cubic coupling between three inequivalent real scalar fields
\be
S \supset \int dX^M \sqrt{|\gamma|} \lambda \Phi \Phi_1 \Phi_2 \,.
\ee
We introduce $\alpha_{1,2}$ to parameterize the bulk masses of the $\Phi_{1,2}$ fields. When working in the conformal spectral representation, we change our convention $\alpha_{1,2}\rightarrow\bar\alpha_{1,2}$ as done in Sec.\,\ref{se:AdSCFT}. Our results are trivially extended to a cubic self-interaction  via the operator $\frac{\lambda}{3!} \Phi^3$ and  taking into account a $ \frac{1}{2}$  symmetry factor in the loop.

In our calculations, we focus on $p^2$ close to the real line. For $p^2\in\mathbb{R_-}$, we are permitted to stay exactly on the real line. The propagator has a branch cut for timelike momentum however, so taking $p^2\in\mathbb{R_+}$ is not well defined (see Sec.~\ref{se:prop_reps}). Instead, one must give $p^2$ a small imaginary part $p^2\rightarrow p^2+i\epsilon$ to resolve the branch ambiguity. We allow for either $\epsilon>0$ or $\epsilon<0$ in our calculations. We show in Sec.~\ref{se:PropertiesSE} that the sign of $\epsilon$ controls the sign of ${\rm Im}\Pi$.

The self-energy of $\Phi$ is given by
\begin{align}\label{eq:SEstart}
\Pi_p(z,z')=\frac{i\lambda^2}{\left(k^2zz'\right)^{d+1}}\int\frac{d^dq}{(2\pi)^d}\Gf_1(p+q,z,z')\Gf_2(q,z',z).
\end{align}
Eq.\,\eqref{eq:SEstart} serves as the starting point for the following calculations.

\subsection{${\rm Im}\Pi$ in the Canonical Representation}
\label{se:CFresult}

We substitute the canonical propagator Eq.\,\eqref{eq:CFprop2} into Eq.\,\eqref{eq:SEstart} and use the infinite series-representations of the Bessel functions,
\begin{align}
I_\alpha(\sqrt{-p^2}z)=&\sum_{m=0}^{\infty}\frac{1}{m!\Gamma(m+\alpha+1)}\left(\frac{\sqrt{-p^2}z}{2}\right)^{2m+\alpha}\label{eq:Iseries}
\\
\label{eq:Kseries}
K_\alpha(\sqrt{-p^2}z) =&\frac{\pi}{2\sin(\pi\alpha)}\sum_{m=0}\frac{1}{m!}\left(\frac{\sqrt{-p^2}z}{2}\right)^{2m}\times\\&\left[\frac{1}{\Gamma(m-\alpha+1)}\left(\frac{\sqrt{-p^2}z}{2}\right)^{-\alpha}-\frac{1}{\Gamma(m+\alpha+1)}\left(\frac{\sqrt{-p^2}z}{2}\right)^{\alpha}\right].\nonumber
\end{align}
In this representation, the self-energy is
\begin{align}\label{eq:45}
& \Pi_p   (z,z') = 
\frac{-i\pi^2\lambda^2}{4k^4zz'\sin(\pi\alpha_1)\sin(\pi\alpha_2)}\int\frac{d^dq}{(2\pi)^d}\sum_{s,t=0}^{\infty}\sum_{n,m=0}^{\infty}
\frac{(-1)^{n+m+s+t}}{n!m!s!t!}\,\times\\&
\frac{1}{\Gamma(n+\alpha_2+1)}\frac{1}{\Gamma(s+\alpha_1+1)}\left(\frac{qz_<}{2}\right)^{2s+\alpha_1}\left(\frac{(p+q)z_<}{2}\right)^{2n+\alpha_2}
\left(\frac{qz_>}{2}\right)^{2t}\left(\frac{(p+q)z_>}{2}\right)^{2m}\nonumber~\times\\&\left[\frac{\left(
\frac{ \vphantom{(} qz_>}{2}\right)^{-\alpha_1}}{\Gamma(t-\alpha_1+1)}-
\frac{\left( \frac{ \vphantom{(} 
qz_>}{2}\right)^{\alpha_1}e^{-i\alpha_1\pi\text{Sign}(\epsilon)}}{\Gamma(t+\alpha_1+1)}\right] 
\left[\frac{\left(\frac{(p+q)z_>}{2}\right)^{-\alpha_2}}{\Gamma(m-\alpha_2+1)}-\frac{\left(\frac{(p+q)z_>}{2}\right)^{\alpha_2}e^{-i\alpha_2\pi\text{Sign}(\epsilon)}}{\Gamma( m+\alpha_2+1  )}\right]\nonumber .
\end{align}

$\Pi_p(z,z')$ contains many terms, each of which go like 
\begin{align}
    \Pi_p(z,z') \sim \int\frac{d^dq}{(2\pi)^d} \left(\left(p+q\right)^2\right)^a \left(q^2\right)^b
\end{align}
for some $a,b$. To each term, we apply the identity
\begin{align}\label{eq:FIidentity}
\left(\left(p+q\right)^2\right)^a \left(q^2\right)^b=\int_0^1 dx\frac{(x\left(p+q\right)^2+(1-x)q^2)^{a+b}}{x^{a+1}(1-x)^{b+1}}\frac{\Gamma(-a-b)}{\Gamma(-a)\Gamma(-b)}\,.
\end{align}
The integral on the right-hand side converges for  ${\rm Re}(a),{\rm Re} (b)<0$.
However, provided the final result of the calculation is analytic in $a,b$, the result can be extended by analytical continuation such that restrictions on $a,b$ are ultimately lifted.\,\footnote{Although the details of analytic continuation are usually left implicit, it is interesting to know how it concretely happens in  the intermediate steps. The Feynman parametrization Eq.\,\eqref{eq:FIidentity} follows from  the integral representation of Gamma functions valid for  ${\rm Re}(a),{\rm Re} (b)<0$, which involves integrals along the real line. The analytical continuation of the Feynman parametrization  relies on  analytically continued Gamma functions, whose integral representation involves Hankel contours in the complex plane. It turns out that the $x$ parameter must follow a \textit{Pochhammer} contour appropriately wrapping the $0$ and $1$ points in the complex plane. 
The Pochhammer contour in $x$  gives rise to the analytically continued integral representation of the hypergeometric function, lifting the ${\rm Re}(a),{\rm Re} (b)<0$ restriction in Eq.\,\eqref{eq:FIidentity} and ultimately
 giving rise to the analytically continued 
Beta function in \textit{e.g.} Eq.\,\eqref{eq:Pi_beta}.
}

Shifting the loop momentum $l\equiv q+px$, we obtain
\begin{align}
    \Pi_p(z,z') \sim \int_0^1 dx\int\frac{d^dl}{(2\pi)^d} \frac{(l^2+x(1-x)p^2)^{a+b}}{x^{a+1}(1-x)^{b+1}}\frac{\Gamma(-a-b)}{\Gamma(-a)\Gamma(-b)}.
\end{align}
We evaluate the loop integral with
\begin{align}
\int\frac{d^dl}{(2\pi)^d}\left(l^2+\Delta\right)^c=\frac{i(-i\text{Sign}(\epsilon))^{d}}{(4\pi)^{\frac{d}{2}}}\frac{\Gamma\left(-c-\frac{d}{2}\right)}{\Gamma(-c)}\Delta^{c+\frac{d}{2}}\,.\label{eq:loopintegral}
\end{align}
Again, the loop integrals are performed in the domain of $(c,d)$ where the integral on the left-hand-side converges. The functions on the right-hand-side are analytic in $c$ anywhere away from $c$ integer, hence the final result will be ultimately analytically continued in $c$. 
The particular points where a divergence appear require renormalization. However such divergences are in the real part of $\Pi$ and are irrelevant for the study of ${\rm Im}\Pi$.

Putting Eqs.\,\eqref{eq:FIidentity} and \eqref{eq:loopintegral} together yields
\begin{align}
\label{eq:Pi_beta}
    \Pi_p(z,z') \sim  \frac{i(-i\text{Sign}(\epsilon))^{d}}{(4\pi)^{\frac{d}{2}}}\left(p^2\right)^{a+b+\frac{d}{2}}\frac{\Gamma\left(-a-b-\frac{d}{2}\right)}{\Gamma(-a)\Gamma(-b)}\int_0^1 dxx^{b+\frac{d}{2}-1}(1-x)^{a+\frac{d}{2}-1}.
\end{align}
We identify the remaining integral as being the integral representation of the Beta function. Evaluating the integral, we obtain\,\footnote{S.F. thanks M.\,Quiros for providing insight on this loop integral calculation in an early unpublished work \cite{Loop_Quiros}.
}
\begin{align}\label{eq:eq411}
    \Pi_p(z,z') \sim  \frac{i(-i\text{Sign}(\epsilon))^{d}}{(4\pi)^{d/2}}\left(p^2\right)^{a+b+d/2}\frac{\Gamma(-a-b-d/2)}{\Gamma(-a)\Gamma(-b)}\frac{\Gamma(a+d/2)\Gamma(b+d/2)}{\Gamma(a+b+d)}.
\end{align}

There are four terms in Eq.\,\eqref{eq:45} corresponding to the four sign pairings of $\alpha_1,\alpha_2$ in the exponents. All but the $z_>^{\alpha_1+\alpha_2}$ term have either $a\in\mathbb{N}$ or $b\in\mathbb{N}$ or both. These terms vanish trivially on account of the $\Gamma(-a)\Gamma(-b)$ function in the denominator of Eq.\,\eqref{eq:eq411}. 
The disappearance of the $z_>^{-\alpha_1-\alpha_2}$ term can also be shown by taking the imaginary part of $\Pi$. In particular for even $d$ when $\Gamma(-a-b-d/2)$ diverges, we find that the $z_>^{-\alpha_1-\alpha_2}$ term is purely real. 

Making everything explicit again and taking the imaginary part, we have
\begin{align}\label{eq:eq412}
& \text{Im}\Pi_p  (z,z')=
\\ \nn &\text{Im}\Bigg[\frac{\pi^2\lambda^2p^de^{-i\pi\left(\alpha_1+\alpha_2+d/2\right)\text{Sign}(\epsilon)}}{4k^4zz'\sin(\pi\alpha_1)\sin(\pi\alpha_2)(4\pi)^{d/2}}\sum_{s,t=0}^{\infty}\sum_{n,m=0}^{\infty}\left(\frac{pz}{2}\right)^{2n+2s+\alpha_1+\alpha_2}\left(\frac{pz'}{2}\right)^{2m+2t+\alpha_1+\alpha_2}\times\\&
\quad\quad
\frac{(-1)^s}{s!\Gamma(s+\alpha_1+1)}
\frac{(-1)^t}{t!\Gamma(t+\alpha_1+1)}\frac{(-1)^n}{n!\Gamma(n+\alpha_2+1)}
\frac{(-1)^m}{m!\Gamma(m+\alpha_2+1)}\nonumber\,\times\\&
\quad\quad
\frac{\Gamma(-n-m-s-t-\alpha_1-\alpha_2-d/2)}{\Gamma(-n-m-\alpha_2)\Gamma(-s-t-\alpha_1)}
\frac{\Gamma (s+t+\alpha_1+d/2) \Gamma (n+m+\alpha_2+d/2)}{\Gamma (s+t+n+m+d+\alpha_1+\alpha_2)} \Bigg] \nonumber \,,
\end{align}
where we removed the $<,>$ subscripts because the expression is symmetric upon relabeling. 
We regroup the complex exponential with the factors of $p$ in Eq.\,\eqref{eq:eq412} to make the $\sqrt{-p^2}$ factor appear explicitly. This involves reabsorbing the $i\epsilon$ into $p^2$, where it remains implicitly, yielding
\begin{align}\label{eq:415}
\text{Im}  \Pi_p  &(z,z')= \sum_{n,m,s,t=0}^{\infty}\frac{\lambda^2 \pi^2}{4\pi^{d/2}
k^4\left(zz'\right)^{1+d/2}\sin(\pi\alpha_1)\sin(\pi\alpha_2)
}\times  \\ \nn &\frac{(-1)^s}{s!\Gamma(s+\alpha_1+1)}
\frac{(-1)^t}{t!\Gamma(t+\alpha_1+1)}\frac{(-1)^n}{n!\Gamma(n+\alpha_2+1)}
\frac{(-1)^m}{m!\Gamma(m+\alpha_2+1)} \times\\&\frac{\Gamma(-n-m-s-t-\alpha_1-\alpha_2-d/2)}{\Gamma(-n-m-\alpha_2)\Gamma(-s-t-\alpha_1)}
\frac{\Gamma (s+t+\alpha_1+d/2) \Gamma (n+m+\alpha_2+d/2)}{\Gamma (s+t+n+m+d+\alpha_1+\alpha_2)}
\nonumber~\times\\&
\text{Im}\Bigg[\left(\begin{Smaller11}\frac{\sqrt{-p^2}z}{2}\end{Smaller11}\right)^{2n+2s+\alpha_1+\alpha_2+d/2}\left(\begin{Smaller11}\frac{\sqrt{-p^2}z'}{2}\end{Smaller11}\right)^{2m+2t+\alpha_1+\alpha_2+d/2}\Bigg]\nonumber .
\end{align}

In this form, $\text{Im}\Pi_p(z,z')$ explicitly vanishes for spacelike momentum. Using Euler's reflection formula
\begin{align}\label{eq:reflection}
\Gamma(x)\Gamma(1-x)=\frac{\pi}{\sin(\pi x)}\quad\text{for}\quad x\not\in\mathbb{Z},
\end{align}
and taking the imaginary part, Eq.\,\eqref{eq:415} can be written as
\begin{align}\label{eq:seriesresult}
\text{Im}\Pi_p & (z,z')=
\\ \nn
&\frac{\text{Sign}(\epsilon)\pi\lambda^2p^d}{4k^4zz'(4\pi)^{d/2}}\sum_{n,m,s,t=0}^{\infty}\left(\frac{pz}{2}\right)^{2n+2s+\alpha_1+\alpha_2}\left(\frac{pz'}{2}\right)^{2m+2t+\alpha_1+\alpha_2}\times\\&
\frac{(-1)^s}{s!\Gamma(s+\alpha_1+1)}
\frac{(-1)^t}{t!\Gamma(t+\alpha_1+1)}\frac{(-1)^n}{n!\Gamma(n+\alpha_2+1)}
\frac{(-1)^m}{m!\Gamma(m+\alpha_2+1)}\nonumber ~ \times\\&
\frac{\Gamma(n+m+\alpha_2+1)\Gamma(s+t+\alpha_1+1)}{\Gamma(n+m+s+t+\alpha_1+\alpha_2+d/2+1)}
\frac{\Gamma (s+t+\alpha_1+d/2) \Gamma (n+m+\alpha_2+d/2)}{\Gamma (s+t+n+m+d+\alpha_1+\alpha_2)}\nonumber .
\end{align}
Eq.\,\eqref{eq:seriesresult} represents our final result for $\text{Im}\Pi_p(z,z')$ in the canonical representation.

\subsection{${\rm Im}\Pi$ in the Conformal Spectral Representation} \label{se:CSresult}

Some preliminary work was done in Sec.\,\ref{se:AdSCFT} to calculate ${\rm Im}\Pi$ in the conformal spectral representation. Here we start from the identity Eq.\,\eqref{eq:SEidentity}, repeated here:
\begin{align}
& {\rm Im} (k^2 zz')^{d+1} \Pi(z,z')  = {\rm Im}\left[ \hat\lambda^2 k^3
\int_{-i \infty}^{i \infty} d\alpha \,
    {\cal B}(\alpha)  \Omega^{(0)}_{\alpha}\,(z,z')\right]. 
    \end{align}

In this subsection only, we follow the convention of Sec.\,\ref{se:AdSCFT} and use $\alpha,\alpha_1,\alpha_2$ to denote integration variables. We use $\bar\alpha_1,\bar\alpha_2$ for the bulk mass parameters, Eq.\,\eqref{eq:mPhi}. We expand the Bessel functions in a series using Eq.\,\eqref{eq:Kseries} and use the $\alpha\leftrightarrow-\alpha$ shadow symmetry of the integrand to write the four terms in $ \Omega^{(0)}_{\alpha}\,(z,z')$ as two. We have
\begin{align}\label{eq:417}
&  {\rm Im}\Pi(z,z')  = 
\\ & \nn
{\rm Im}\Bigg[\frac{i\hat\lambda^2 k^3}{\left(k^2 zz'\right)^{d/2+1}}
\int_{-i \infty}^{i \infty} d\alpha \,
{\cal B}(\alpha)\frac{\alpha }{2\sin(\pi \alpha)} \sum_{m,n=0}^\infty\frac{1}{m!n!}\left(\begin{Smaller11}\frac{\sqrt{-p^2}z'}{2}\end{Smaller11}\right)^{2n}\left(\begin{Smaller11}\frac{\sqrt{-p^2}z}{2}\end{Smaller11}\right)^{2m}
\times\\&\nonumber
\frac{1}{\Gamma(n-\alpha+1)}\left(\begin{Smaller11}\frac{\sqrt{-p^2}z'}{2}\end{Smaller11}\right)^{-\alpha}
\left(\frac{1}{\Gamma(m-\alpha+1)}\left(\begin{Smaller11}\frac{\sqrt{-p^2}z}{2}\end{Smaller11}\right)^{-\alpha}-\frac{1}{\Gamma(m+\alpha+1)}\left(\begin{Smaller11}\frac{\sqrt{-p^2}z}{2}\end{Smaller11}\right)^{\alpha}\right)\Bigg].
\end{align}

The imaginary part of the above expression will ultimately come from $\text{Im}\left(-p^2\right)^a$ for some $a$. That is, all other factors of $i$ from applications of the residue theorem conspire to produce a real prefactor. In the second term of Eq.\,\eqref{eq:417} above, the two powers of $\alpha$ cancel, and hence $a$ is necessarily an integer. Thus we find that this second term vanishes upon taking the imaginary part, as it is purely real. The poles from the $1/\sin(\pi\alpha)$ factor sit at integer $\alpha$, which again yields integer $a$ upon applying the residue theorem. Hence these residues do not contribute to  ${\rm Im}\Pi(z,z')$.

Following this preliminary result, only the first term in the bracket in Eq.\,\eqref{eq:417} 
contributes to the imaginary part,
and only the poles within the bubble function ${\cal B}(\alpha)$ are relevant. We close the $\alpha$-contour towards the negative reals, where the integrand vanishes exponentially on account of the Gamma functions. This selects an infinite number of residues from the four relevant Gamma functions within the bubble function, Eq.\,\eqref{eq:bubblefunction}. 
The sets of poles that we enclose are located at
\be
\alpha=\pm\alpha_1\pm\alpha_2-\frac{d}{2}-2l \quad\quad\quad\text{for}\quad\quad\quad l\in\mathbb{N}
\,. \label{eq:poles_B}
\ee
The corresponding four sets of residues all take the same form, but with relative sign flips ($\alpha_1\leftrightarrow-\alpha_1$ and/or $\alpha_2\leftrightarrow-\alpha_2$). We have
\begin{align}
& {\rm Im}\Pi(z,z')  = {\rm Im}\sum_{m,n,l=0}^\infty\int d\alpha_1 d\alpha_2\left(\mathcal{I}_{\alpha_1,\alpha_2}+\mathcal{I}_{-\alpha_1,\alpha_2}+\mathcal{I}_{\alpha_1,-\alpha_2}+\mathcal{I}_{-\alpha_1,-\alpha_2}\right)
\end{align}
for
\begin{align}
&\mathcal{I}_{\alpha_1,\alpha_2}\equiv\\&\nonumber \frac{\hat\lambda^2 k^3}{\left(k^2zz'\right)^{d/2+1}}
\left(\frac{\sqrt{-p^2}z}{2}\right)^{2m+2l+\alpha_1+\alpha_2+d/2}\left(\frac{\sqrt{-p^2}z'}{2}\right)^{2n+2l+\alpha_1+\alpha_2+d/2}\times\\&\nonumber
\frac{1}{2^4 \pi^{1+d/2}}\frac{1}{l!m!n!}
\frac{(2l+\alpha_1+\alpha_2+d/2)\csc(\pi (\alpha_1+\alpha_2+d/2))}{\Gamma(d/2)\Gamma(2l+\alpha_1+\alpha_2+d)\Gamma(l+\alpha_1+\alpha_2+1)}
 \frac{\alpha_1\alpha_2 }{(\alpha^2_1-\bar \alpha^2_{1})(\alpha^2_2-\bar \alpha^2_{2})}\times 
\\ & \nonumber 
\frac{\Gamma\left(l+d/2\right)
\Gamma\left(l+\alpha_1+d/2\right)
\Gamma\left(l+\alpha_2+d/2\right)
\Gamma\left(2l+\alpha_1+\alpha_2+1\right)
	\Gamma\left(l+\alpha_1+\alpha_2+d/2\right)}{\Gamma(n+2l+\alpha_1+\alpha_2+d/2+1)\Gamma(m+2l+\alpha_1+\alpha_2+d/2+1)\Gamma\left(l+\alpha_1+1\right)
	\Gamma\left(l+\alpha_2+1\right)},
\end{align}
where we have simplified the expression by using the reflection formula Eq.\,\eqref{eq:reflection}. All four terms are equivalent---as can be seen by $\alpha_j\rightarrow-\alpha_j$ relabelings---and hence the integrand amounts to $4\mathcal{I}_{\alpha_1,\alpha_2}$. 

We then perform the $\alpha_1$ and $\alpha_2$ integrals.
We close the integrals towards the positive reals, where the integrand vanishes exponentially on account of the Gamma functions. Within the contour, there are poles at $\alpha_i=\bar\alpha_i$ from the $P\left(\bar \alpha_i,\alpha_i \right)$ measures (see Eq.\,\eqref{eq:alphaprops}) and poles at $\alpha_1+\alpha_2+d/2\in\mathbb{Z}_+$ from $\csc(\pi (\alpha_1+\alpha_2+d/2))$.
The residues from the cosecant only give a contribution to ${\rm Re}\Pi(z,z')$ because they give rise to even powers of $\sqrt{-p^2}$.  
The residues from the $P\left(\bar \alpha_i,\alpha_i \right)$ measures yield a contribution to ${\rm Im}\Pi(z,z') $.

  We obtain \begin{align}
{\rm Im}\Pi(z,z')  & =-\sum_{m,n,l=0}^\infty\frac{\hat\lambda^2 k^3}{\left(k^2zz'\right)^{d/2+1}}\frac{1}{2^2 \pi^{d/2-1}}\frac{1}{l!m!n!}\times \\&\nonumber 
\frac{\Gamma\left(l+\bar\alpha_1+\bar\alpha_2+d/2\right)(2l+\bar\alpha_1+\bar\alpha_2+d/2) \csc(\pi (\bar\alpha_1+\bar\alpha_2+d/2))}{\Gamma(d/2)\Gamma(n+2l+\bar\alpha_1+\alpha_{2,0}+d/2+1)\Gamma(m+2l+\bar\alpha_1+\bar\alpha_2+d/2+1)}\times  
\\ &\frac{\Gamma\left(l+d/2\right)
\Gamma\left(l+\bar\alpha_1+d/2\right)
\Gamma\left(l+\bar\alpha_2+d/2\right)
\Gamma\left(2l+\bar\alpha_1+\bar\alpha_2+1\right)}{\Gamma(2l+\bar\alpha_1+\bar\alpha_2+d)\Gamma(l+\bar\alpha_1+\bar\alpha_2+1)\Gamma\left(l+\bar\alpha_1+1\right)
	\Gamma\left(l+\bar\alpha_2+1\right)} 
\nonumber\times\\&
	{\rm Im}\left[
\left(\begin{Smaller11}\frac{\sqrt{-p^2}z}{2} \end{Smaller11}\right)^{2m+2l+\bar\alpha_1+\bar\alpha_2+d/2}\left(\begin{Smaller11}\frac{\sqrt{-p^2}z'}{2} \end{Smaller11} \right)^{2n+2l+\bar\alpha_1+\bar\alpha_2+d/2}\right]\nonumber 
\,.
\end{align}

The self-energy becomes real for spacelike momenta, which is straightforward to note in this form. For timelike momenta, the imaginary part is given by
\begin{align}
& {\rm Im}\Pi(z,z')  =\\&\nonumber\text{Sign}\left(\epsilon\right) \sum_{m,n,l=0}^\infty\frac{(-1)^{n+m}}{m!n!l!} \left(\frac{pz}{2}\right)^{2m+2l+\bar\alpha_1+\bar\alpha_2+d/2}\left(\frac{pz'}{2}\right)^{2n+2l+\bar\alpha_1+\bar\alpha_2+d/2}\times\\&\nonumber
\frac{\hat\lambda^2 k^3}{2^2 \pi^{d/2-1}\left(k^2zz'\right)^{d/2+1}}
\frac{\Gamma\left(l+d/2\right)
\Gamma\left(l+\bar\alpha_1+d/2\right)
\Gamma\left(l+\bar\alpha_2+d/2\right)
\Gamma\left(2l+\bar\alpha_1+\bar\alpha_2+1\right)}{\Gamma(d/2)\Gamma(2l+\bar\alpha_1+\bar\alpha_2+d)\Gamma(l+\bar\alpha_1+\bar\alpha_2+1)}\times
\\ & \nonumber   
\frac{(2l+\bar\alpha_1+\bar\alpha_2+d/2) \Gamma\left(l+\bar\alpha_1+\bar\alpha_2+d/2\right)}{\Gamma\left(l+\bar\alpha_1+1\right)
	\Gamma\left(l+\bar\alpha_2+1\right)\Gamma(n+2l+\bar\alpha_1+\bar\alpha_2+d/2+1)\Gamma(m+2l+\bar\alpha_1+\bar\alpha_2+d/2+1)}\,.
\end{align}
We recognize the sums over $n,m$ as the series representation of the Bessel function $J$,
\be\label{eq:BesselJ}
J_{2l+\bar\alpha_1+\bar\alpha_2+d/2}\left(x\right)=\sum_{j=0}^{\infty}\frac{(-1)^j\left(\frac{x}{2}\right)^{2j+2l+\bar\alpha_1+\bar\alpha_2+d/2}}{j!\Gamma(j+2l+\bar\alpha_1+\bar\alpha_2+d/2+1)}.
\ee
This gives 
\begin{align}\label{eq:CSresult}
& {\rm Im}\Pi(z,z')  =\\&\nonumber\frac{\text{Sign}\left(\epsilon\right)\hat\lambda^2 k^3}{2^2 \pi^{d/2-1}\left(k^2zz'\right)^{d/2+1}} \sum_{l=0}^\infty(2l+\bar\alpha_1+\bar\alpha_2+d/2) J_{2l+\bar\alpha_1+\bar\alpha_2+d/2}\left(pz\right)J_{2l+\bar\alpha_1+\bar\alpha_2+d/2}\left(pz'\right)\times\\&\nonumber
\frac{\Gamma\left(l+d/2\right)
\Gamma\left(l+\bar\alpha_1+d/2\right)
\Gamma\left(l+\bar\alpha_2+d/2\right)
\Gamma\left(2l+\bar\alpha_1+\bar\alpha_2+1\right)\Gamma\left(l+\bar\alpha_1+\bar\alpha_2+d/2\right)}{\Gamma(d/2)\Gamma(l+1)\Gamma(2l+\bar\alpha_1+\bar\alpha_2+d)\Gamma(l+\bar\alpha_1+\bar\alpha_2+1)\Gamma\left(l+\bar\alpha_1+1\right)
	\Gamma\left(l+\bar\alpha_2+1\right)}\,.
\end{align}
Eq.\,\eqref{eq:CSresult} represents our final result for $\text{Im}\Pi_p(z,z')$ in the conformal spectral representation. In Sec.\,\ref{se:DT}, we identify this result as being a sum over the imaginary part of propagators.

\subsection{${\rm Im}\Pi$ in the Momentum Spectral Representation}\label{se:MSresult}

We again start from Eq.\,\eqref{eq:SEstart} and express the propagator in the $p$-spectral representation, Eq.\,\eqref{eq:Delta_p}. We introduce a Feynman parameter and shift the loop momentum, $l\equiv q+px$. This yields 
\begin{align}
\Pi_p(z,z')=\frac{-i\lambda^2}{\left(k^2zz'\right)^{d+1}}\int_0^1dx\int_0^\infty dm_1\Omega_{m_1}(z,z')\int_0^\infty dm_2\Omega_{m_2}(z,z')\int\frac{d^dl}{(2\pi)^d}\frac{1}{\left(l^2-\Delta\right)^2}
\end{align}
for 
\be
\Delta\equiv xm_2^2+(1-x)m_1^2-x(1-x)p^2.
\ee

We evaluate the loop integral in arbitrary dimension with
\begin{align} 
  \int \frac{d^dl}{(2\pi)^d}\frac{1}{(l^2 - \Delta)^2} &=
    \frac{i\Gamma\left(2-\frac{d}{2}\right)}{(4\pi)^{d/2}}\Delta^{d/2-2} \,.
\end{align}
The left-hand side integral converges for $d<4$. However it can be analytically continued in $d$. For odd $d$, we write the square root as
\begin{align}
\left.\sqrt{\Delta}\right|_{\text{Re}\Delta\gg\text{Im}\Delta} \approx i\sqrt{\left|\Delta\right|}\Theta\left(-\text{Re}\Delta\right)\text{Sign}\left(\text{Im}\Delta\right)+\sqrt{\left|\Delta\right|}\Theta\left(\text{Re}\Delta\right)\,,
\end{align}
valid for $\Delta$ near the real line. For even $d$, we take $d=d_0\pm\varepsilon$ for $\varepsilon\ll 1$ and use the reflection formula Eq.\,\eqref{eq:reflection}. We obtain
\begin{align} \label{eq:evenDstuff}
  \left.\int \frac{d^dl}{(2\pi)^d}\frac{1}{(l^2 - \Delta)^2}\right|_{d_0 \text{ even}} &=
    \frac{i\pi(-1)^{d_0/2-1}}{(4\pi)^{d_0/2}\sin\left(\pm\frac{\pi\varepsilon}{2}\right)\Gamma\left(\frac{d_0}{2}-1\right)}\Delta^{d_0/2-2}\Delta^{\pm\varepsilon/2}\,.
\end{align}
We expand Eq.\,\eqref{eq:evenDstuff} for $\varepsilon\ll1$ using
\begin{align} 
\Delta^{\pm\varepsilon/2}\approx1+\pm\frac{\varepsilon}{2}\log\Delta\,,
\quad\quad\sin\left(\pm\frac{\pi\varepsilon}{2}\right)\approx\pm\frac{\pi\varepsilon}{2}.
\end{align}
We note that the $\log$ can be described in terms of $\arctan$ and a step function,
\begin{align}
\log\left(\Delta\right)=\log\left|\Delta\right|+i\arctan\left(\frac{\text{Im}\Delta}{\text{Re}\Delta}\right)+i\pi\Theta\left(-\text{Re}\Delta\right)\text{Sign}\left(\text{Im}\Delta\right)\,.
\end{align}
For $\Delta$ near the real line, the $\arctan$ term is small and can be dropped. The $1$ and $\log\left|\Delta\right|$ terms yield contributions to the self-energy that are purely real and hence vanishes when taking the imaginary part. 

With these simplifications, we can combine results for both even and odd $d$ to obtain
\begin{equation}
\text{Im}\left[i\int \frac{d^dl}{(2\pi)^d}\frac{1}{(l^2 - \Delta)^2}\right]_{\text{Re}\Delta\gg\text{Im}\Delta}  \approx \frac{\pi(-\Delta)^{d/2-2}}{(4\pi)^{d/2}\Gamma\left(\frac{d}{2}-1\right)}\Theta\left(-\text{Re}\Delta\right)\text{Sign}\left(\text{Im}\Delta\right)\,,\label{eq:effectiveloop}
\end{equation}
valid for $d>2$. Taking the imaginary part of the self-energy, we now have
\begin{align}\label{eq:se433}
\text{Im}\Pi_p(z,z')=&\frac{-\pi}{(4\pi)^{d/2}\Gamma\left(\frac{d}{2}-1\right)}\frac{\lambda^2}{\left(k^2zz'\right)^{d+1}}\int_0^1dx ~~\times
\\&\int_0^\infty dm_1\Omega_{m_1}\int_0^\infty dm_2\Omega_{m_2}(-\Delta)^{d/2-2}\Theta\left(-\text{Re}\Delta\right)\text{Sign}\left(\text{Im}\Delta\right)\,.\nonumber
\end{align}
We find that for spacelike momenta $p^2<0$, $\text{Re}\Delta>0$ and hence $\text{Im}\Pi=0$. Specializing to timelike momenta $p^2>0$, we have
\be
\text{Sign}\left(\text{Im}\Delta\right)=-\text{Sign}\left(\epsilon\right),
\ee
where $\epsilon>0$ is the Feynman prescription for the propagator. The step function truncates the integrals over $m_1,m_2$ and $ x$. We now have
\begin{align}
\text{Im}\Pi_p(z,z')=\frac{\text{Sign}\left(\epsilon\right)\pi}{(4\pi)^{d/2}\Gamma\left(\frac{d}{2}-1\right)}\frac{\lambda^2}{\left(k^2zz'\right)^{d+1}}\int_0^{p} dm_1\Omega_{m_1}\int_0^{p-m_1} dm_2\Omega_{m_2}\int_{x_-}^{x_+}dx(-\Delta)^{d/2-2}\label{eq:se434}
\end{align}
for
\be
x_\pm \equiv \frac{p^2+\left(m_1^2-m_2^2\right)\pm\sqrt{\left(p^2-(m_1-m_2)^2\right)\left(p^2-(m_1+m_2)^2\right)}}{2p^2}\,.
\ee
Notice that $-\Delta=(x_+-x)(x-x_-)p^2$. To evaluate the integral, we change variables to $y= \left(x-x_-\right)/\left(x_+-x_-\right)$ and recognize the integral representation of the Beta function,
\begin{align}\label{eq:betafucntion}
\int_0^1 dy y^{n_1} (1-y)^{n_2}=\frac{\Gamma (n_1+1) \Gamma (n_2+1)}{\Gamma (n_1+n_2+2)}\equiv \text{B}(n_1+1,n_2+1)\,.
\end{align}
$n_1=n_2=\frac{d}{2}-2$ in the current case, and thus we obtain
\begin{align}\label{eq:xint}
\Xi(d,p,m_1,m_2)& \equiv\frac{1}{\Gamma\left(\frac{d}{2}-1\right)}\int_{x_-}^{x_+}dx(-\Delta)^{d/2-2}=2p^{2-d}K^{d-3}\frac{\Gamma\left(\frac{d}{2}\right)}{\Gamma\left(d-1\right)}
\end{align}
Here $K=K(p,m_1,m_2)$ is the standard four-dimensional two-body kinematic factor,
\begin{align}
	K(p,m_1,m_2)
	\equiv
	\sqrt
	{\left(
		p^2-\left(m_1+m_2\right)^2
	\right)
	\left(
		p^2-\left(m_1-m_2\right)^2
	\right)} \ .
\end{align}
The $\Xi(d,p,m_1,m_2)$ function 
 is the two-body kinematic factor in arbitrary dimension $d$.
Thus we obtain
\begin{align}
\text{Im}\Pi_p(z,z')=\text{Sign}\left(\epsilon\right)&\frac{2\pi}{(4\pi)^{d/2}}\frac{\Gamma\left(\frac{d}{2}\right)}{\Gamma\left(d-1\right)}\frac{\lambda^2}{\left(k^2zz'\right)^{d+1}}\,\times
\\ \nn &\int_0^{p} dm_1\Omega_{m_1}\int_0^{p-m_1} dm_2\Omega_{m_2}
p^{2-d}K(p,m_1,m_2)^{d-3}\,.\label{eq:MSresult}
\end{align}
Eq.\,\eqref{eq:MSresult} represents our final result for $\text{Im}\Pi_p(z,z')$ in the momentum spectral representation.

\subsection{Proofs of Equality}\label{se:equality}

 The equivalence of the representations of ${ \rm Im}\Pi$ obtained in Secs.\,\ref{se:CFresult}, \ref{se:CSresult}, \ref{se:MSresult} is not manifest.
 In the canonical representation, a key piece of the calculation was a loop integral with non-integer powers. In the conformal spectral representation, AdS/CFT naturally arises: CFT correlators appear and combine to form a CFT bubble diagram. Finally in the momentum spectral representation, a generalized two-body kinematic threshold  emerges---with no intrinsic dependence on $\alpha$.
 These three evaluations thus involve rather different objects and methods.  

In this section, we prove the equality of these various representations of ${\rm Im}\Pi$.

\subsubsection{Canonical---Momentum Spectral Equivalence}

Here we relate the self-energy from the calculation in the momentum spectral representation to the calculation in the canonical representation. We start from Eq.\,\eqref{eq:se433} in the momentum spectral representation and substitute in the spectral functions $\Omega_m$ in terms of the Bessel functions explicitly
\begin{align}
\text{Im}\Pi_p(z,z')=&\text{Sign}(\epsilon)\frac{\pi^2}{(4\pi)^{d/2+1}\Gamma\left(\frac{d}{2}-1\right)}\frac{\lambda^2}{k^4zz'}\int_0^1dx\int_0^\infty dm_1^2\int_0^\infty dm_2^2~\times\\&
J_{\alpha_1}(m_1z)
J_{\alpha_1}(m_1z')J_{\alpha_2}(m_2z)
J_{\alpha_2}(m_2z')(-\Delta)^{d/2-2}\Theta\left(-\text{Re}\Delta\right)\,.\nonumber
\end{align}
We use the step function to cut the $m_1,m_2$ integrals instead of the $x$ integral. We also expand out the Bessel functions in terms of the series 
\begin{align}
J_\alpha(z)=\sum_{m=0}^{\infty}\frac{(-1)^m}{m!\Gamma(m+\alpha+1)}\left(\frac{z}{2}\right)^{2m+\alpha}.\label{eq:Jseries}
\end{align}
We obtain
\begin{align}
\text{Im}& \Pi_p(z,z')=
\\ & \nn 
\frac{\text{Sign}(\epsilon)\pi^2}{(4\pi)^{d/2+1}\Gamma\left(\frac{d}{2}-1\right)}\frac{\lambda^2}{k^4zz'}\int_0^1dx\int_0^{xp^2} dm_1^2\int_0^{(1-x)(xp^2-m_1^2)/x} dm_2^2~\times\\&\nonumber
\sum_{s,t,m,n=0}^{\infty}\frac{(-1)^{s+t+n+m}}{s!t!n!m!\Gamma(s+\alpha_1+1)\Gamma(t+\alpha_1+1)\Gamma(m+\alpha_2+1)\Gamma(n+\alpha_2+1)}
~\times\\&
\left(\frac{m_1z}{2}\right)^{2s+\alpha_1}\left(\frac{m_1z'}{2}\right)^{2t+\alpha_1}\left(\frac{m_2z}{2}\right)^{2n+\alpha_2}\left(\frac{m_2z'}{2}\right)^{2m+\alpha_2}(-\Delta)^{d/2-2}\,.\nonumber
\end{align}
We perform the integral over $m_2$, then the integral over $m_1$, and lastly the integral over $x$. Each of these integrals can be expressed as the integral representation of the Beta function, Eq.\,\eqref{eq:betafucntion}.
Thus we obtain
\begin{align}
\text{Im} & \Pi_p(z,z')=
\\ & \nn 
\frac{\text{Sign}(\epsilon)p^d\pi^2}{(4\pi)^{d/2+1}}\frac{\lambda^2}{k^4zz'}\sum_{s,t,m,n=0}^{\infty}\left(\frac{pz}{2}\right)^{2s+2n+\alpha_1+\alpha_2}\left(\frac{pz'}{2}\right)^{2t+2m+\alpha_1+\alpha_2}\times\\&\nonumber
\frac{(-1)^{s+t+n+m}}{s!t!n!m!\Gamma(s+\alpha_1+1)\Gamma(t+\alpha_1+1)\Gamma(m+\alpha_2+1)\Gamma(n+\alpha_2+1)}
\times\\&
\frac{\Gamma(1+s+t+\alpha_1)\Gamma(1+m+n+\alpha_2)}{\Gamma\left(1+s+t+n+m+\alpha_1+\alpha_2+\frac{d}{2}\right)}\frac{\Gamma\left(\frac{d}{2}+s+t+\alpha_1\right)\Gamma\left(\frac{d}{2}+m+n+\alpha_2\right)}{\Gamma(d+s+t+n+m+\alpha_1+\alpha_2)}\,,\nonumber
\end{align}
which is exactly Eq.\,\eqref{eq:seriesresult}.

\subsubsection{Equivalence with the Conformal Spectral Representation}

It is difficult to show the exact equivalence between the conformal spectral representation and either the canonical or momentum spectral representation. The proof involves sums over generalized hypergeometric functions $_4F_3$ which are nontrivial to perform. It is not as difficult to show the equivalence order by order in $pz_<$. Here we present a proof of equivalence for $pz_<\ll 1$ with arbitrary $pz_>$. 

We take the self-energy in the canonical representation Eq.\,\eqref{eq:seriesresult} to leading order in $pz_<$,

\begin{align}
&\text{Im}\Pi_p (z,z')=\frac{\text{Sign}(\epsilon)\pi\lambda^2p^d}{4k^4zz'(4\pi)^{d/2}}\frac{1}{\Gamma(\alpha_1+1)}
\frac{1}{\Gamma(\alpha_2+1)}\left(\frac{pz_<}{2}\right)^{\alpha_1+\alpha_2}\times
\\ \nn
&\sum_{n,s=0}^{\infty}\left(\frac{pz_>}{2}\right)^{2n+2s+\alpha_1+\alpha_2}\frac{(-1)^{s+n}}{s!n!}
\frac{\Gamma (s+\alpha_1+d/2) \Gamma (n+\alpha_2+d/2)}{\Gamma (s+n+d+\alpha_1+\alpha_2)\Gamma(n+s+\alpha_1+\alpha_2+d/2+1)}\nonumber .
\end{align}
We introduce $j\equiv s+n$ and reorganize the sums
\begin{align}
&\text{Im}\Pi_p (z,z')=\frac{\text{Sign}(\epsilon)\pi\lambda^2p^d}{4k^4zz'(4\pi)^{d/2}}\frac{1}{\Gamma(\alpha_1+1)}
\frac{1}{\Gamma(\alpha_2+1)}\left(\frac{pz_<}{2}\right)^{\alpha_1+\alpha_2}\times
\\ \nn
&\sum_{j=0}^{\infty}\frac{(-1)^{j}\left(\frac{pz_>}{2}\right)^{2j+\alpha_1+\alpha_2}}{\Gamma (j+d+\alpha_1+\alpha_2)\Gamma(j+\alpha_1+\alpha_2+d/2+1)}\sum_{s=0}^{j}
\frac{\Gamma (s+\alpha_1+d/2) \Gamma (j-s+\alpha_2+d/2)}{s!(j-s)!}\nonumber .
\end{align}
We directly evaluate the sum over $s$ with Mathematica \cite{Mathematica}. We obtain
\begin{align}
&\text{Im}\Pi_p (z,z')=\frac{\text{Sign}(\epsilon)\pi\lambda^2p^d}{4k^4zz'(4\pi)^{d/2}}\frac{1}{\Gamma(\alpha_1+1)}
\frac{1}{\Gamma(\alpha_2+1)}\left(\frac{pz_<}{2}\right)^{\alpha_1+\alpha_2}\times
\\ \nn
&\sum_{j=0}^{\infty}\frac{(-1)^{j}\left(\frac{pz_>}{2}\right)^{2j+\alpha_1+\alpha_2}\Gamma \left(\alpha_1+\frac{d}{2}\right) \Gamma \left(\alpha_2+\frac{d}{2}+j\right) \, _2F_1\left(\alpha_1+\frac{d}{2},-j;-\alpha_2-j-\frac{d}{2}+1;1\right)}{j!\Gamma (j+d+\alpha_1+\alpha_2)\Gamma(j+\alpha_1+\alpha_2+d/2+1)}\nonumber .
\end{align}
We now use a known result for this special case of the hypergeometric function---Eq.\,(15.4.20) of \cite{NIST:DLMF}---reprinted here:
\be
_2F_1(a,b;c;1)=\frac{\Gamma(c)\Gamma(c-a-b)}{\Gamma(c-a)\Gamma(c-b)}.
\ee
We obtain
\begin{align}
&\text{Im}\Pi_p (z,z')=\frac{\text{Sign}(\epsilon)\pi\lambda^2p^d}{4k^4zz'(4\pi)^{d/2}}\frac{1}{\Gamma(\alpha_1+1)}
\frac{1}{\Gamma(\alpha_2+1)}\left(\frac{pz_<}{2}\right)^{\alpha_1+\alpha_2}\times
\\ \nn
&\sum_{j=0}^{\infty}\frac{(-1)^{j}\left(\frac{pz_>}{2}\right)^{2j+\alpha_1+\alpha_2}\Gamma \left(\alpha_1+\frac{d}{2}\right) \Gamma \left(\alpha_2+\frac{d}{2}+j\right) \Gamma(-\alpha_1-\alpha_2-d+1)\Gamma(-\alpha_2-j-\frac{d}{2}+1)}{j!\Gamma (j+d+\alpha_1+\alpha_2)\Gamma(j+\alpha_1+\alpha_2+d/2+1)\Gamma(-\alpha_1-\alpha_2-j-d+1)\Gamma(-\alpha_2-\frac{d}{2}+1)}\nonumber .
\end{align}
We simplify further by using Euler's reflection formula, Eq.\,\eqref{eq:reflection}, and trigonometric identities. We obtain
\begin{align}
\text{Im}\Pi_p  (z,z')=&\frac{\text{Sign}(\epsilon)\pi\lambda^2}{4k^4(zz')^{d/2+1}\pi^{d/2}}\frac{\Gamma \left(\alpha_1+\frac{d}{2}\right) \Gamma \left(\alpha_2+\frac{d}{2}\right)}{\Gamma(\alpha_1+1)\Gamma(\alpha_2+1)}\frac{1}{\Gamma (d+\alpha_1+\alpha_2)}\times
\\ \nn
&\left(\frac{pz_<}{2}\right)^{\alpha_1+\alpha_2+d/2}\sum_{j=0}^{\infty}\frac{(-1)^j\left(\frac{pz_>}{2}\right)^{2j+\alpha_1+\alpha_2+d/2}}{j!\Gamma(j+\alpha_1+\alpha_2+d/2+1)} .
\end{align}
We recognize the remaining sum as the series representation of the Bessel function $J$ as given in Eq.\,\eqref{eq:BesselJ}. Thus we obtain
\begin{align}\label{eq:CSCFmatchresult}
\text{Im}\Pi_p & (z,z')=\frac{\text{Sign}(\epsilon)\pi\lambda^2}{4k^4(zz')^{d/2+1}\pi^{d/2}}\frac{\Gamma \left(\alpha_1+\frac{d}{2}\right) \Gamma \left(\alpha_2+\frac{d}{2}\right)}{\Gamma(\alpha_1+1)\Gamma(\alpha_2+1)}\frac{J_{\alpha_1+\alpha_2+d/2}\left(pz_>\right)}{\Gamma (d+\alpha_1+\alpha_2)}\left(\frac{pz_<}{2}\right)^{\alpha_1+\alpha_2+d/2}.
\end{align}

The conformal spectral result is given by Eq.\,\eqref{eq:CSresult}. To leading order in $pz_<$, $l=0$ and we take the small argument limit of the $pz_<\ll 1$ Bessel function.
Upon application of the recurrence formula for the Gamma functions, we immediately obtain Eq.\,\eqref{eq:CSCFmatchresult}. 

We have checked the equality of the next-to-leading order terms in $pz_<$. This calculation goes just as above, but with more terms that nontrivially sum. When attempting to show equality generally (for any $pz_<$), sums over generalized hypergeometric functions $_4F_3$ emerge which are difficult to evaluate. We expect equality to hold to all orders in $pz_<$.

\subsection{Properties of the Self-Energy}\label{se:PropertiesSE}

Based on our results, here we discuss some properties of the imaginary part of the bubble diagram ${ \rm Im}\Pi$.

\subsubsection{Elementary Features}
\label{se:ImPi_el}

\begin{itemize}
    \item {\bf The self-energy becomes real  for spacelike momentum $p^2<0$, \textit{i.e.} } \be{ \rm Im}\Pi|_{p^2<0}=0\,.\ee  
    \noindent This is easily shown from the conformal spectral or canonical representations, where we note that ${\rm Im}\Pi\propto{\rm Im}\left(-p^2\right)^{\gamma}$ for some power $\gamma$. When the momenta becomes spacelike $-p^2>0$, ${\rm Im}\Pi$ subsequently vanishes. In the momentum spectral representation, we note that the Heaviside $\Theta$-function in Eq.\,\eqref{eq:effectiveloop} cannot be satisfied for spacelike momenta, and hence ${\rm Im}\Pi=0$.
    
      \item {\bf ${\rm Im}\Pi$ is finite for  nonzero $z,z'$ and finite timelike momentum $p$.}
      
      This can be shown from any of the representations. Consider for instance  the canonical representation Eq.\,\eqref{eq:seriesresult}. At fixed $p$ and $z,z'$, when going to high enough order in a given sum,  the Gamma functions in the denominator imply that the absolute value of the ratios of two subsequent terms  is strictly smaller than $1$. This implies that the series (absolutely) converges, therefore ${\rm Im}\Pi$ is finite.

        \item {\bf If $z=z'$, ${\rm Im}\Pi$ carries the same sign as $\epsilon$, } \textit{i.e.} \be{\rm Sign}\left({\rm Im}\Pi(z,z)\right)= {\rm Sign} (\epsilon)\,.\ee
        This property can be shown from  the momentum spectral representation Eq.\,\eqref{eq:MSresult}. For $z=z'$, both spectral functions are positive since $\propto \left( J_\alpha(m_i z)\right)^2$.  Hence the integrals are positive and $\text{Sign}\left({\rm Im}\Pi\right)=\text{Sign}\left(\epsilon\right)$. This property can also be shown in the conformal spectral representation by noting that the self-energy can be written as a sum over squares when $z=z'$.
        Along the same lines, we show that ${\rm Sign}\left({\rm Im}\Pi(z,z')\right)= {\rm Sign} (\epsilon)$ if $p|z-z'|\ll 1$, \textit{i.e.} when $z$ and $z'$ are close enough (see below).  
        
       \subsubsection{Asymptotic Features} 
        \label{se:ImPi_lims}

  \item {\bf At large   $pz\rightarrow \infty$ with timelike momentum $p$, ${\rm Im}\Pi$ becomes asymptotically local, with}
    \be
   { \rm Im}\Pi_p(z,z')\Big|_{p\rightarrow \infty} \propto 
\frac{ p^{d-3}}{z^4}
  \frac{\sin(p(z-z'))}{z-z'}\Big|_{p\rightarrow \infty}\sim \frac{ p^{d-3}}{z^4}
  \delta(z-z')\,. \label{eq:ImPiLargep}
    \ee 
    We show this in the momentum spectral representation. Details are given in App.\,\ref{app:large_pz}. In short, at large $pz$ the spectral functions rapidly oscillate such that their average can be pull out from the mass integrals. The averages asymptotically give rise to a nascent Dirac delta in $z$. The remaining mass integral depends only on $p$ and must scale as $p^{d-3}$. 
  This limit makes clear that Im$\Pi_p(z,z')$ is peaked on $z\sim z'$, and that it becomes increasingly peaky with $pz\gg 1$ since in this limit  it  approaches a  Dirac delta.
  This asymptotic feature should not be seen as a literal divergence. 
  These properties are more difficult to show in the conformal spectral and canonical representations.

    \item {\bf At small  $pz$, $pz'$ with timelike momentum $p$, the behaviour of ${\rm Im}\Pi$ is} 
    \be
   { \rm Im}\Pi_p(z,z')\Big|_{pz \ll 1, pz' \ll 1} \propto (zz')^{{\alpha_1}+{\alpha_2}-1} p^{d+2\alpha_1+2\alpha_2}=(zz')^{\Delta_1+\Delta_2-d-1} p^{2\Delta_1+2\Delta_2-d}
\,. \label{eq:ImPiSmallp}
    \ee   
    This behaviour can be shown from any of the representations, and can be directly read from \textit{e.g.} Eq.\,\eqref{eq:seriesresult} by taking the first term of the series.  The scaling in $p$ is consistent with the one found in \cite{Meltzer:2020qbr}.  For the last equality we used $\Delta_i=\frac{d}{2}+\alpha_i$. 
    If  $z$ approaches zero, Im$\Pi_p(z,z')$ goes as $z^{{\alpha_1}+{\alpha_2}-1}$, which diverges if ${\alpha_1}+{\alpha_2}<1$. In physical expressions, however, Im$\Pi_p(z,z')$ is always multiplied by extra metric factors which remove this possibly divergent behavior near the boundary. 
    Such $z,z'$ factors appear for instance in Eq.\,\eqref{eq:S1}, thus justifying the saddle-point expansion.

\end{itemize}

\subsubsection{Double-Trace Formula from Conformal Spectral Representation}
\label{se:DT}

Starting from the conformal spectral representation of ${\rm Im}\Pi$ Eq.\,\eqref{eq:CSresult}, we notice that each term of the sum can be rewritten as the discontinuity of a free propagator with bulk mass parameter $\alpha_1+\alpha_2+d/2$ along the timelike branch cut (see Eq.\,\eqref{eq:discontinuity}). This is  equivalently given by the 
imaginary part of $i \Gf$,   \be 
{\rm Im}\left[ 
i\Gf_{\alpha_1+\alpha_2+d/2+2l}(z,z') \right]= \text{Sign}\left(\epsilon\right) \frac{\pi}{2k} (kz)^{d/2}(kz')^{d/2} J_{\alpha_1+\alpha_2+d/2+2l}(pz)J_{\alpha_1+\alpha_2+d/2+2l}(pz')
\label{eq:ImG}
\ee

In terms of conformal dimension, using $\Delta_i=\alpha_i+d/2$, this propagator amounts to the exchange of operators with dimension $\Delta_1+\Delta_2+2l$.  We recognize the structure of the well-known ``double-trace'' formula \cite{Dusedau:1985ue,Fitzpatrick:2011hu} which gives the product of position-space propagators $\Gf_{\alpha_1}(X,Y)\Gf_{\alpha_2}(X,Y)$ as an infinite sum over  $\Gf_{\alpha_1+\alpha_2+d/2+2l}(X,Y)$. 

Making the imaginary part of $i\Gf$ explicit in Eq.\,\eqref{eq:CSresult}, we obtain
\begin{align}\label{eq:CS_DT}
& {\rm Im}\Pi_p(z,z')  = \frac{\hat\lambda^2 k^3}{\left(k^2zz'\right)^{d+1}} \sum_{l=0}^\infty a_{\Delta_1,\Delta_2,l}\,
{\rm Im}\left[ 
i\Gf_{\alpha_1+\alpha_2+d/2+2l}(z,z') \right]
\end{align}
with
\begin{align}
&a_{\Delta_1,\Delta_2,l}=      \frac{2l+\alpha_1+\alpha_2+d/2}{2 \pi^{d/2}} \, \times \\ &
\frac{\Gamma\left(l+d/2\right)
\Gamma\left(l+\alpha_1+d/2\right)
\Gamma\left(l+\alpha_2+d/2\right)
\Gamma\left(2l+\alpha_1+\alpha_2+1\right)\Gamma\left(l+\alpha_1+\alpha_2+d/2\right)}{\Gamma(d/2)\Gamma(l+1)\Gamma(2l+\alpha_1+\alpha_2+d)\Gamma(l+\alpha_1+\alpha_2+1)\Gamma\left(l+\alpha_1+1\right)
	\Gamma\left(l+\alpha_2+1\right)}\,.\nn
\label{eq:DT}
\end{align}

Using Legendre's duplication formula \be
\frac{\Gamma(2z)}{\Gamma(z)} = \frac{2^{2z-1}}{\sqrt{\pi}} \Gamma\left(z+\frac{1}{2}\right)
\ee
it turns out that one can re-write $a_{\Delta_1,\Delta_2,l}$ in terms of Pochammer functions $(x)_n=\Gamma(x+n)/\Gamma(x)$, obtaining
\be
a_{\Delta_1,\Delta_2,l}= \frac{\left(\frac{d}{2}\right)_l}{2 \pi^{d/2 }l!}\frac{\left(\Delta_1+\Delta_2+2l\right)_{1-d/2} \left(\Delta_1+\Delta_2+l-d+1\right)_{l}}{
\left(\Delta_1+l\right)_{1-d/2}\left(\Delta_2+l\right)_{1-d/2} \left(\Delta_1+\Delta_2+l-d/2\right)_{l} }  \,. 
\ee
This matches precisely the coefficients  of the double-trace formula \cite{Fitzpatrick:2011hu}. 
With Eq.\,\eqref{eq:CS_DT}, we have thus recovered the imaginary part of the double-trace formula by direct calculation  starting from   the  conformal spectral representation.\,\footnote{Upon Fourier-transform and appropriate translation into our Lorentzian  conventions.}

\subsubsection{Relation to AdS Unitarity Methods}
\label{ref:unitarity}

Unitarity methods in AdS have recently been investigated in \cite{Meltzer:2019nbs,Meltzer:2020qbr}. These methods typically aim to compute the double discontinuity of the dual CFT correlators  from the AdS side \cite{Caron-Huot:2017vep}, and therefore focus on full Witten diagrams. However amputated diagram such as our bubble $i\Pi$  encapsulate essential information about the  Witten diagram they are a part of. Let us consider the interplay between our results for ${\rm Im}\Pi$  and  existing AdS unitarity methods. 

In \cite{Meltzer:2019nbs}, a $\mathrm{\mathbf{Cut}}$ operation was introduced in the space of  conformal dimensions.\,\footnote{A $\widehat{\mathrm{\mathbf{Cut}}}$ operation is also introduced, which   involves an extra projection and applies only to full Witten diagrams. Here our focus is on the $\mathrm{\mathbf{Cut}}$ operation only.}
This $\mathrm{\mathbf{Cut}}$ applies in  the conformal spectral representation:  It picks the poles of the $P(\alpha,\bar\alpha)$ measure
enclosed by the corresponding $\alpha$-contour integral.
One can apply this operation to the bubble diagram in the conformal spectral representation (Sec.\,\ref{se:CSresult}).  Consider the  $\mathrm{\mathbf{Cut}}_{12}$ operation, which cuts both lines of the bubble in the sense defined above. This cut amounts to selecting  the poles of $P(\alpha_1,\bar\alpha_1)$ and $P(\alpha_2,\bar\alpha_2)$ in the  $\alpha_1$, $\alpha_2$ integrals. This is intuitively what the ${\rm Im}$ operation would do in flat space. What is the effect of ${\rm Im}$  on the AdS bubble as compared to $\mathrm{\mathbf{Cut}}_{12}$?     
We have seen by direct calculation in Sec.\,\ref{se:CSresult} that the ${\rm Im}$ does the same work as $\mathrm{\mathbf{Cut}}_{12}$ because other residues are real and are thus projected out. We  conclude that 
\be
{\rm Im}(
\mathrm{\mathbf{Cut}}_{12} \Pi) = \mathrm{\mathbf{Cut}}_{12}( {\rm Im} \Pi ) =   {\rm Im} \Pi \,.
\ee
These identities imply that the ${\rm Im}$ projection is at least as strong as $\mathrm{\mathbf{Cut}}_{12}$, \textit{i.e.} ${\rm Im} \Pi \subset \mathrm{\mathbf{Cut}}_{12} \Pi$. 
The equality of the operations is not guaranteed.  The evaluation of $\mathrm{\mathbf{Cut}}_{12} \Pi$ in our formalism would require further investigation. 

In \cite{Meltzer:2020qbr}, unitarity cuts in the momentum spectral representation have been explored. 
 Following this reference, 
 we can evaluate a cut of the $\Pi$ bubble instead of evaluating its imaginary part.  To do so, we implement Cutkosky rules in the propagator Eq.\,\eqref{eq:Delta_p}, cutting the $P(\hat p , p)$  using the substitution
\be
P(\hat p , p )\propto \delta\left(
p^2-\hat p^2
\right)\,.
\ee
With such substitution in Eq.\,\eqref{eq:Delta_p}, one can evaluate the integral in $\hat p$, which gives rise to the AdS Wightman propagator \cite{Meltzer:2020qbr} 
given schematically by\,\footnote{We thank the authors of \cite{Meltzer:2020qbr} for pointing this out in private correspondence.}
\be G^W_p(z,z') \propto 
(z z')^{d/2}J_\alpha(pz)J_\alpha(pz')\,.
\ee 
Pursuing the evaluation of the cut of $\Pi$ along these lines,  we obtain a product of four Bessel $J$ functions. Expanding the Bessel functions as a series and performing the loop integrals as in Sec.\,\ref{se:CFresult}, we ultimately find that the cut diagram reproduces the result of ${\rm Im}\Pi$ in the canonical representation Eq.\,\eqref{eq:CSresult}.

\section{Opacity of AdS}
\label{se:Opacity}

In this section we solve the dressed equation of motion in the timelike regime. 
This is done using an improved version of the WKB method\,\footnote{Also known as the Carlini--Liouville--Green--Rayleigh--Gans--Jeffreys--Wentzel--Kramers--Brillouin approximation}  and other approximations. 
Our focus is on the effect of the imaginary contribution ${\rm Im}\Pi_p(z,z')$. 

As discussed in Sec.\,\ref{se:asymptotics},  three distinct regimes can be distinguished in position-momentum space, depending on whether the invariants $pz$, $pz'$ are smaller or larger than $1$. 
In the $pz,pz'\ll 1$ regime, the effects of dressing are expected to be small because the self-energy gets suppressed by higher powers of $pz$ with respect to the terms in the free EOM operator ${\cal D}_z$ (see \textit{e.g.} Eq.\,\eqref{eq:seriesresult} or \eqref{eq:ImPiSmallp}). The behaviour of the dressed propagator for  $pz,pz'\gg 1$ could be understood from a flat space viewpoint by using a Weyl transform (see Sec.\,\ref{se:flat}). Hence we do not focus on these cases. 
Our focus in this section is rather on the non-trivial $pz_<\ll 1$, $pz_>\gg 1$ regime, which has no flat space analog since the $pz \ll 1$  vanishes in the flat space limit. 
Note this regime includes the boundary-to-bulk propagators (Sec.\,\ref{se:Bo2B}) as a particular case.

Recall that for spacelike momentum, the propagator in
the $|p|z_<\ll 1$, $|p|z_>\gg 1$ regime
 decays as
\be
G_p(z,z')\propto \exp\left(-|p| z_>\right)
\ee
irrespective of $z_<$. This can be seen from Eq.\,\eqref{eq:Delta_confflat} using that $p$ is imaginary in the spacelike region. 
The purpose of this section is to show that a similar exponential decay happens for \textit{timelike} momentum in the \textit{interacting} theory. This exponential fall-off directly results  from the imaginary part of $\Pi$, which introduces a damping in the EOM solutions, which would otherwise be oscillating. 

We have provided various representations of ${\rm Im }\Pi$ in Sec.\,\ref{se:Properties}. The momentum spectral representation---made available by working in position-momentum space---turns out to be the most convenient to pursue the calculation and make intuitive approximations.

In this section we will work within the effective field theory (EFT) paradigm. 
Before delving into the dressing calculations, some details about EFT are given in Sec.\,\ref{eq:EFT_intro}.

\subsection{Interactions and Effective Field Theory}
\label{eq:EFT_intro}

 Matter interactions have been left implicit in the action  Eq.\,\eqref{eq:SPhi}. In Secs.\,\ref{se:AdSCFT},\,\ref{se:Properties}, we have   restricted the calculation to  a non-derivative scalar cubic coupling for simplicity.
 
Here we are interested more broadly in the low-energy effective field theory (EFT) viewpoint, in which the action $S$ encodes local interactions with arbitrary high number of fields and derivatives. In addition to scalar cubic interactions with no derivatives---which are renormalizable for $d\leq 6$ ---cubic interactions with an arbitrary number of extra derivatives are  also present in principle.  
We consider the  cubic couplings\,\footnote{
In the EFT, other two-derivative cubic interactions such as $\Phi_1 \partial_M\Phi \partial^M\Phi_2$ can be reduced to the ones in Eq.\,\eqref{eq:int_opacity} using integration by parts and the equation of motion.
}
 \be
S \supset \int dX^M \sqrt{|\gamma|} \left( \lambda \Phi \Phi_1 \Phi_2 -
\zeta \Phi \partial_M\Phi_1 \partial^M\Phi_2+\ldots
\right)\, \label{eq:int_opacity}
\ee
where the ellipses denote scalar cubic couplings  with higher number of derivatives. 
In Sec.\,\ref{se:grav}, cubic matter-gravity interactions will be considered. 

Following the EFT paradigm, these interactions of higher dimension are suppressed by a typical energy scale $\Lambda$,  the EFT cutoff. The natural relation for the couplings in Eq.\,\eqref{eq:int_opacity} is expected to be $\zeta/\lambda\sim 1/\Lambda^2 $, \textit{i.e.} each derivative brings an extra  $1/\Lambda$ factor. 
Conversely, the  EFT action is valid only up to a proper distance scale of order $\Delta X \sim 1/\Lambda$.  

How does the cutoff appear in our Poincar\'e position-momentum space? To see it, one can compare the effects of operators of different order, \textit{e.g.} the two operators in Eq.\,\eqref{eq:int_opacity}, or the scalar kinetic term and a bilinear four-derivative operator $\Lambda^{-2} \square(\partial_M \Phi \partial^M \Phi) $, in some physical amplitude.\,\footnote{ One can  \textit{e.g.} examine the correction to the free propagator from $\Lambda^{-2} \square(\partial_M \Phi \partial^M \Phi) $. }  
This shows that the cutoff of the EFT is reached for 
\be
pz \sim \frac{\Lambda}{k}\,. \label{eq:EFT_cutoff} 
\ee
This happens because higher derivatives operators necessarily come together with extra powers of $kz$. More generally, at the value set by Eq.\,\eqref{eq:EFT_cutoff}, operators with arbitrary number of derivatives become equally important in the amplitudes, signalling that the low-energy EFT has reached its limit and that a deeper UV-completion should be used instead. 
We in fact recover the scaling of Eq.\,\eqref{eq:EFT_cutoff} in our results Eq.\,\eqref{eq:pz_timelike}.
This feature is well-known, see \cite{ArkaniHamed:2000ds,Goldberger:2002cz,Goldberger:2002hb,Ponton:2012bi,Fichet:2019hkg,Costantino:2020msc}  for further details.

\subsection{Solving the Dressed EOM}
\label{se:WKB}

The dressed EOM can be treated with standard solving techniques just like the free case (see Sec.\,\ref{se:prop_closed} and Eq.\,\eqref{eq:prop_closed}), such that the  dressed canonical propagator admits the structure
\be
G(p;z,z')=\frac{i}{\tilde C} \tilde F_<(z_<)\tilde F_>(z_>)\,.
\ee
The $\tilde F_i$ satisfies the homogeneous dressed EOM
\be
{\cal D} \tilde F_i(z)-\frac{1}{\sqrt{|\gamma|}} \Pi * \tilde  F_i(z)=0 \,. \label{eq:EOM_F}
\ee
The quantum dressing affects both solutions $\tilde F_<$ and $\tilde F_>$. Our focus is on the $pz_<\ll 1$, $pz_>\gg 1$ regime, and we are interested in the effect of dressing on the $\tilde F_>$ part of the propagator. 
As discussed above,  the effects of dressing on $\tilde F_<$ and $\tilde C$ are small---this follows from \textit{e.g.}  Eq.\,\eqref{eq:ImPiSmallp}---and are neglected for our purposes.

\subsubsection{A WKB-like Approximation }
In the $p z_> \gg 1$ regime we know  from Sec.\,\ref{se:asymptotics} that $\tilde F_>$ is conformally equivalent to a massless  flat space solution, which is a   $e^{i pz}$ exponential---the asymptotic AdS  solution is $z^{\frac{d-1}{2}}e^{i pz}$. This feature implies that we can use developed WKB-type methods to find the solution $\tilde F_>$ as a suitable perturbation of the free solution $F_>$.
The WKB ansatz of $\tilde F_>$ for timelike momentum, assuming 
$\epsilon>0$ and taking the $pz \gg 1$ limit, is the asymptotic form\,\footnote{
We dropped an irrelevant multiplicative constant as compared to the convention of the free case of Sec.\,\ref{se:prop_closed}. By construction such factors cancel in the full expression, see Sec.\,\ref{se:prop_closed}. }
\be
\tilde F_>(z)\Big|_{pz \gg 1} \approx \tilde F^{\rm c.m.}_>(z) = z^{\frac{d-1}{2}} e^{i pz +S(z)}\,.
\label{eq:F_dressed}
\ee
The $S(z)$ function is determined perturbatively. 
Setting $S\rightarrow 0$ gives the asymptotic conformally massless solution of the free EOM, noted $ F^{\rm c.m.}_>(z)$. In our analysis, we show that $S(z)<0$. Thus we find that loop corrections are consistent with the $\epsilon>0$ prescription of the free theory.

Plugging $\tilde F^{\rm c.f.}_>(z)$ into the dressed equation of motion gives
\be
 S'(z) 
= - \frac{e ^{- i p  z}}{2p}  \int^{\infty}_0 du\, {\rm Im}\Pi(z,u)  (k^2 \, z)^{\frac{d-1}{2}} F_>(u)  \,\label{eq:Sprime1}
\ee 
at first order, assuming $|S'|\ll p$, $|S''|\ll p |S'|$ and $|S|$ not arbitrarily large.\,\footnote{
More precisely, taking $ e^{-S(z)+S(u)} \sim 1$ (with $S<0$) on the right-hand side of Eq.\,\eqref{eq:Sprime1} is valid as long as the rest of the integrand is sufficiently peaked. 
In Sec.\,\ref{se:saddle}, we use a saddle-point approximation for the rest of the integrand which gives $e^{-\frac{1}{2C}p^2(z-u)^2}$.
So long as $ e^{-S(z)+S(u)}$ does not compete with $e^{-\frac{1}{2C}p^2(z-u)^2}$, our approximation is valid. The breakdown occurs at values of $|S|$ sensibly larger than $1$. 
} 
These conditions are satisfied for $pz \gg 1$.

Upon extracting the $u^{\frac{d-1}{2}}$ prefactor from $F_>(u)$, we note that the resulting integrand is highly peaked at $z\sim u$ ---as expected from locality. Upon inspecting $\left(k^2\,  z\,u\right)^{\frac{d-1}{2}}{\rm Im} \Pi(z,u)$, one can also confirm that the dominant contributions to $S(z)$ do indeed come from the $pz\gg 1$ region. Further, the steepness of the peak increases with $pz$. This can be checked  in the amplitudes of Sec.\,\eqref{se:AdS5}, see \textit{e.g.} Eq.\,\eqref{eq:ImPi_ll_int}. Additional supporting details are discussed in Sec.\,\ref{se:PropertiesSE}.

Having a peak at $z\sim u$, and having $pz\gg 1$ by assumption, we can safely use the $pu\gg 1$ approximation for the convoluted propagator such that
$F_>(u)\approx F^{\rm c.m.}_>(u) $ in
Eq.\,\eqref{eq:Sprime1}. This gives 
\be
 S(z) 
= - \frac{1}{2p} \int_z d\tilde z\int^{\infty}_0 du\, {\rm Im}\Pi(\tilde z,u)  \left(k^2\, \tilde z\,u\right)^{\frac{d-1}{2}} e^{+ i p (u- \tilde z)}\, \label{eq:S1}
\ee 
where we have integrated $S'$ to obtain $S$.

The integration constant for the last $d\tilde z$ integral would be determined by a matching condition at $z=z'$ in the propagator. Such condition is not accessible analytically in the asymptotic regime considered, since $z_>\gg z_<$. However we will see that the integration constant  is essentially irrelevant for our analysis.

\subsubsection{Aside on an Elementary Approximation}

For large $pz$, the integrand in Eq.\,\eqref{eq:S1} is highly peaked. The most basic approximation that we can perform is to take $\left(k^2\, z\,u\right)^{\frac{d-1}{2}}{\rm Im} \Pi(z,u)$ to be proportional to a Dirac distribution, \textit{i.e.} assuming exact locality along the $z$ direction. 
Under this approximation we have
\be\label{eq:localapprox}
\left( z\,u\right)^{\frac{d-1}{2}}\,{\rm Im}\Pi(z,u) \approx \FPi(z) \delta(z-u), \,\quad  \FPi(z)=\int^\infty_0 du\,\left( z\,u\right)^{\frac{d-1}{2}}\, {\rm Im}\Pi(z,u)\,. 
\ee
In this limit, the convolution $\Pi* F_>(z)$  becomes a product $\FPi(z) F_>(z)$. The convolution being gone, the problem reduces to a  standard WKB one in which the EOM is  perturbed by a potential $\FPi(z)$. This approximation reproduces  the large $p$ limit of the self-energy established in Sec.\,\ref{se:ImPi_lims}. Even though the integrand in Eq.\,\eqref{eq:S1} is very peaked at large $pz$, it multiplies an  exponential which oscillates in $pz$  and thus cannot be treated as a function varying slowly with respect to ${\rm Im} \Pi$. 
Thus this approximation scheme is potentially inaccurate and a more refined approach is necessary.  

\subsubsection{Saddle-Point Approximation}
\label{se:saddle}

As an improved approximation, we take $\left(k^2\, z\,u\right)^{\frac{d-1}{2}}{\rm Im} \Pi(z,u)$ as a Gaussian centered on $\tilde z=u$, whose width is controlled by the second derivative at this point.\,\footnote{While performing this approximation, ${\rm Im} \Pi(z,u)>0$ is required. While this is not strictly the case for ${\rm Im} \Pi(z,u)$ globally, it is true on the $z\sim u$ peak. A side effect of the kinematic approximation discussed in Sec.\,\ref{se:ImPi_approxes} is to smooth the oscillations of ${\rm Im} \Pi(z,u)$. This smoothed version satisfies ${\rm Im} \Pi(z,u)>0$ everywhere.} This is the saddle-point approximation. It allows us to treat $\left(k^2\, z\,u\right)^{\frac{d-1}{2}}{\rm Im} \Pi(z,u)$ as peaked while also accounting for the complex exponential. The saddle-point approximation will allow us to perform the integrals analytically. 

We write Eq.\,\eqref{eq:S1} as
\be
 S(z) 
= - \frac{1}{2p} \int_z d\tilde z\int^{\infty}_0 du\, e^{-f(\tilde z,u)+i p (u- \tilde z)}
\ee 
for
\be
f( z,u)=-\log\left(\left(k^2\, z\,u\right)^{\frac{d-1}{2}}{\rm Im}\Pi( z,u)\right).
\ee 
The maximum of $\left(k^2\, z\,u\right)^{\frac{d-1}{2}}{\rm Im} \Pi(z,u)$ being at $z=u$, $f(z,u)$ has a minimum at $z\sim u$ up to a small $O(1/pz)$ suppressed shift. 
Expanding $f(z,u)$ about this minimum, we have
\be
 S(z) 
= - \frac{1}{2p} \int_z d\tilde z\int^{\infty}_0 du\, e^{-f( \tilde z,\tilde z)+i p (u- \tilde z)-\frac{1}{2}\left(  \left.\partial_u^2f(\tilde z,u)\right|_{u=\tilde z}\right) (u-\tilde z)^2}
\ee 
where 
$\left.\partial_u^2f(z,u)\right|_{u=z}$ is given by
\be
\left. \partial_u^2f( z,u)\right|_{u= z}\approx-\left.\frac{\partial^2_u\left(\left(k^2\,  z\,u\right)^{\frac{d-1}{2}}{\rm Im}\Pi( z,u)\right)}{\left(\left(k^2\,  z\,u\right)^{\frac{d-1}{2}}{\rm Im}\Pi( z,u)\right)}\right|_{u= z} +\left. \left(\partial_u f( z,u)\right|_{u= z}\right)^2
\equiv C_p \left( z\right)p^2.
\ee
Note that $\left. \partial_u f( z,u)\right|_{u= z}=0$ by construction.
The $p^2$ factor we have extracted in our convention will naturally appear in the upcoming explicit calculations. 
We now have
\be
 S(z) 
= - \frac{1}{2p} \int_z d\tilde z \left(k \tilde z\right)^{d-1}{\rm Im}\Pi(\tilde z,\tilde z)\int^{\infty}_0 du\, e^{i p (u- \tilde z)-\frac{1}{2}C_p \left(\tilde z\right)p^2(u-\tilde z)^2},
\ee
which is a Gaussian integral. Evaluating this integral gives 
\be
S(z) 
\approx - \frac{1}{2p^2} \int_z d\tilde z\, {\rm Im} \Pi_p(\tilde z,\tilde z)   \,  (k\tilde z)^{d-1} 
\sqrt{\frac{2\pi}{ C_p(\tilde z) }}e^{-\frac{1}{2C_p(\tilde z)}}\,.
\label{eq:S_final}
\ee

Compared to the delta-function approximation shown in Eq.\,\eqref{eq:localapprox}, it turns out that
an extra  factor $e^{-\frac{1}{2C_p(z)}}$ appears above in Eq.\,\eqref{eq:S_final} ---inducing an extra suppression of the $S$ function. 
This extra effect reflects the departure from the standard WKB problem taken into account by the saddle-point approximation. It  encodes the fact that the perturbation to the EOM  is a convolution $\Pi * G$, and not a mere potential multiplying $G$ as would be approximated by Eq.\,\eqref{eq:localapprox}. We will use this improved WKB-like approximation in our analysis.

\subsection{Kinematic Approximation for Im$\Pi$ in Any Dimension }

\label{se:ImPi_approxes}

We turn to the imaginary part of the self-energy itself. In Sec.\,\ref{se:Properties}, we gave the exact form of ${\rm Im}\Pi$ from a $\Phi^3$ interaction in various forms. In this section, we allow for a more general form that could arise from the derivative cubic interactions in Eq.\,\eqref{eq:int_opacity}.

Using the momentum spectral representation established in Sec.\,\ref{se:MSresult}, the imaginary part of the bubble diagram takes the structure
\begin{align}
{\rm Im}&\Pi_p(z,z')=  \label{eq:ImPi_gen_p}
\\
& \frac{c_a c_b\pi}{(4\pi)^{d/2} k^4 zz'}\int^{p}_0 dq_1 \int^{p-q_1}_0 d q_2 q_1q_2
{\cal V}_a \, 
J_{\alpha_1}(q_1z) J_{\alpha_2}(q_2z)
{\cal V}_b \,
J_{\alpha_1}(q_1z')
J_{\alpha_2}(q_2z')
\K(d,p,q_1,q_2)
\,. \nn
\end{align}
The ${\cal V}_i\equiv{\cal V}_i(q_1,q_2,\partial_z,\partial_{z'})$ are operators encoding the Lorentz structure of the vertices. The $c_i$ are the overall dimensionful coefficient of the vertices. Dimensions are such that  $[c_i{\cal V}_i]=\frac{5-d}{2}$. 
The $\K(d,p,q_1,q_2)$ function is the $d$-dimensional 2-body kinematic function appearing in the momentum spectral representation (see Sec.\,\ref{se:MSresult}). Its dimension is $[\K]=d-4$. 

Eq.\,\eqref{eq:ImPi_gen_p} is exact, but is in general difficult to evaluate. In any dimension the $\K(d,p,q_1,q_2)$ function drops to zero at the $p=q_1+q_2$ threshold. Away from the threshold $p\gg q_1+q_2$, $K$ tends to $p^2$ and the kinematic function simplifies to 
\be
   \K(d,p,q_1,q_2) \approx \K(d,p,0,0)=  \frac{2p^{d-4} 
\Gamma\left(\frac{d}{2} \right)}   
   {\Gamma\left(d-1  
   \right)} \,
   \ee
   giving for instance
\begin{center} 
\begin{tabular}{c|cccc}
    $d$ & 3 & 4 & 5 & 6 \\
    \hline
   $ \K(d,p,0,0)$  & $ \frac{\sqrt{\pi}}{p}$ & $1$ & $\frac{\sqrt{\pi} }{4}p$ & $\frac{ p^2}{6}$
\end{tabular}
\end{center}

Following these observations, we introduce a kinematic threshold approximation for which $\K(d,p,q_1,q_2)$ is replaced by $\Theta\left(q_1<\frac{p}{2}\right)\Theta\left(q_2<\frac{p}{2}\right) \K(d,p,0,0)$. We have checked this approximation gives a typical error of $O(10\%)$ upon $z$ integration. Numerical examples are shown in App.\,\ref{app:kin_approx}.
The approximate expression for ${\rm Im}\Pi_p$ becomes
\begin{align}
{\rm Im} & \Pi_p(z,z') \label{eq:ImPi_gen_p_approx}  \approx \\ & \nn 
\frac{c_a c_b\pi}{(4\pi)^{d/2} k^4 zz'}\int^{p/2}_0 dq_1 \int^{p/2}_0 d q_2 q_1q_2
{\cal V}_a \, 
J_{\alpha_1}(q_1z) J_{\alpha_2}(q_2z)
{\cal V}_b \,
J_{\alpha_1}(q_1z')
J_{\alpha_2}(q_2z')
\K(d,p,0,0)
\,. 
\end{align}

Additionally, when working  in the $pz, pz' \gg 1$ regimes, it turns out that the propagators can be accurately approximated as the conformally massless ones, such that one can also take $\alpha_{1,2}=1/2$ in Eq.\,\eqref{eq:ImPi_gen_p_approx}. This last approximation is not strictly necessary to render the evaluation of the WKB function $S$ analytically tractable, but it will simplify the following expressions and integrals.

\subsection{Bubbles in AdS$_5$}

\label{se:AdS5}

In the following subsections, we calculate the various contributions to the self-energy coming from the cubic vertices in Eq.\,\eqref{eq:int_opacity}. A detailed calculation of the $\Phi^3-\Phi^3$ bubble was performed in Sec.\,\ref{se:Properties}. Abridged versions of all bubble calculations are given in App.\,\ref{app:bubble}.

Here we specialize to the $d=4$ case, \textit{i.e} five-dimensional AdS. Results in other dimension are qualitatively similar, this is discussed in Sec.\,\ref{se:generaldim}.
As described in the previous subsection, to leading order we can set ${\alpha_{1,2}}=1/2$ to simplify our expressions in the $pz\gg1$ region.

\subsubsection{$\Phi^3-\Phi^3$ Bubble}

The contribution to ${\rm Im}\Pi$ from  the bubble induced by two $\lambda \Phi\Phi_1\Phi_2$ vertices is given by
\be \label{eq:ImPi_ll}
{\rm Im}\Pi_{\lambda\lambda}(z_1,z_2)=
\frac{\lambda^2}{16\pi k^4 z_1z_2}\int dq_1d q_2 q_1q_2
J_{\alpha_1}(q_1z_1)J_{\alpha_1}(q_1z_2)
J_{\alpha_2}(q_2z_1)J_{\alpha_2}(q_2z_2)
\frac{K(p;q_1,q_2)}{p^2}\,.
\ee
Details of the calculation are given in Sec.\,\ref{se:MSresult}. An abridged version is provided in App.\,\ref{app:ll}.

We now set $\alpha_{1,2}=1/2$. 
Performing the mass integrals using the kinematic threshold approximation gives
\be {\rm Im}\Pi_{\lambda\lambda}(z_1,z_2)\approx \lambda^2\frac{
\left(
z_2 \cos\left(\frac{pz_2}{2}\right)
\sin\left(\frac{pz_1}{2}\right)-
z_1 \cos\left(\frac{pz_1}{2}\right)
\sin\left(\frac{pz_2}{2}\right)
\right)^2
}{4\pi^3\,k^4\, z_1^2 z_2^2(z_1^2-z_2^2)^2}\,.
\label{eq:ImPi_ll_int}
\ee
Finally, to complete the evaluation of the WKB function $S$ in Eq.\,\eqref{eq:S_final}, we consider ${\rm Im}\Pi$ at same point $z_1=z_2=z$. 
Keeping the leading term in $pz\gg 1$ gives
\be
{\rm Im}\Pi_{\lambda\lambda}(z,z)=\lambda^2 \frac{p^2}{64 \pi^3 (kz)^4}\left(1+O\left(\frac{1}{p z}\right)\right)
\label{eq:ImPi_ll_max}
 \ee
   and
   \be
C_{\lambda\lambda}(z) = \frac{1}{6}+O\left(\frac{1}{p z}\right)  \,.
   \ee
We note ${\rm Im}\Pi_{\lambda\lambda}(z,z)$ has oscillating terms 
but they are subleading in $1/pz$ and thus do not appear in Eq.\,\eqref{eq:ImPi_ll_max}. The same feature is true in the other diagrams we evaluate.

 \subsubsection{$\Phi^3-\Phi (\partial \Phi)^2$ Bubble}
\label{se:lz}

We evaluate the contribution to ${\rm Im}\Pi$ from the bubble induced by one  $\lambda \Phi\Phi_1\Phi_2$ and one $\xi \Phi\partial_M\Phi_1 \partial^M\Phi_2$ vertex. 
We get
      \begin{align}
{\rm Im}\Pi_{\lambda\zeta}(z_1,z_2)\supset & - \lambda \zeta
\frac{z_2^2}{k^2\,z_1z_2} \int dq_1  dq_2 q_1q_2 
\label{eq:ImPi_lchi}
\\ & \quad \quad\quad\quad \frac{1}{16 \pi} 
\bigg(q_1^2+q_2^2 - \frac{p^2}{2} - \partial_{z_2}\partial_{z_2} \bigg) J_{\alpha_1}(q_1z_1)J_\alpha(q_1z_2)
J_{\alpha_2}(q_2z_1)J_\beta(q_2z_2) \nn \\
& +\Big(1\leftrightarrow 2 \Big) \nn
\,.
\end{align}
Details of the calculation are given in App.\,\ref{app:lz}. 
 Taking $z_1=z_2=z$, ${\alpha_{1,2}}=1/2$, and expanding for large $pz$, we obtain
\be
{\rm Im}\Pi_{\lambda\zeta}(z,z)=\lambda\zeta\frac{p^4}{96 \pi^3 (kz)^2}\left(1+O\left(\frac{1}{p z}\right)\right)
 \ee
   and
   \be
C_{\lambda\zeta}(z) = \frac{2}{15}+O\left(\frac{1}{p z}\right)  \,.
   \ee

 \subsubsection{$\Phi (\partial \Phi)^2-\Phi (\partial \Phi)^2$ Bubble}
\label{se:zz}

We evaluate the contribution from two $\xi \Phi\partial_M\Phi_1 \partial^M\Phi_2$ vertices. 
Intermediate details of the derivation are similar to the other contributions but more cumbersome. Details are given in App.\,\ref{app:zz}.    
Proceeding similarly we get 
\be
{\rm Im}\Pi_{\xi \xi}(z,z)=\zeta^2 \frac{113\, p^6}{46080 \pi^3}\left(1+O\left(\frac{1}{p z}\right)\right)
\ee
and
\be
C_{\xi \xi}(z)=\frac{213}{1582}+O\left(\frac{1}{p z}\right)\,.
\ee

\subsection{Opacity from the Dressing}
\label{se:opacity_discussion}
 
We have all the ingredients to describe the behaviour of the dressed propagator in the timelike, $pz_<\ll1$, $pz_<\gg 1$ regime of Poincar\'e position-momentum space induced by the above bubble diagrams.

We plug the various contributions into the WKB formula Eq.\,\eqref{eq:S_final} ---which gives the argument of the exponential WKB ansatz. We integrate from a point $z_0$ to $z$ with $z_0\ll z$ and satisfying $pz_0\gg 1$ such that the asymptotic behaviour is valid.  Using the full functions, the contribution to $S(z)$ from $z_0<1/p$ should be negligible. 
The dependence on $z_0$  will be essentially irrelevant for our purposes, for concreteness 
 we let $z_0\sim 1/p $.

 We obtain
\begin{align}
    S_{\lambda\lambda}(z)& =-\sqrt{\frac{3}{2}}\frac{e^{-3}}{128  \pi^{5/2}}\frac{\lambda^2}{k}\log\left(pz\right)
    \label{eq:S_ll}
\\ 
    & \approx-3\cdot 10^{-5}\,\frac{\lambda^2}{k}\log\left(pz\right) \nn
\\ &\nn    
\\
    S_{\lambda\zeta}(z)& =-\sqrt{\frac{5}{6}}\frac{e^{-15/4}}{128  \pi^{5/2} }\,\lambda\zeta k\, (pz)^2 
\\    
  &   \approx -1\cdot 10^{-5}\,\lambda\zeta k\, (pz)^2 \nn
\\ &\nn    
\\
    S_{\zeta\zeta}(z)& =-\sqrt{\frac{791}{446}}\frac{11e^{-791/223}}{46080 \pi^{5/2} \, }\,\zeta^2 k^3\, (pz)^4  \label{eq:S_zz}
\\    
  &   \approx -5\cdot 10^{-7}\,\zeta^2 k^3\, (pz)^4. \nn
 \end{align}
The $z_0$ dependence is negligible in $S_{\lambda\zeta}$, $S_{\zeta\zeta}$ because of $z_0\ll z$.

Following the WKB analysis of Sec.\,\ref{se:WKB}, we conclude that the dressed propagator in the timelike region behaves as
\be  G_p(z,z')\sim e^{S(z_>)}\,,\ee 
with  $S$ functions  given by Eqs.\,\eqref{eq:S_ll}-\eqref{eq:S_zz}.
Neglecting the effect of $S_{\lambda\lambda}$ which varies very slowly, the dressed propagator is damped for 
\be
S_{\lambda\zeta}\left(pz|_{\rm opacity}\right)\sim -1\,. \label{eq:pz_op}
\ee
The condition Eq.\,\eqref{eq:pz_op} translates as a condition on $pz$ for given couplings and AdS curvature---or vice-versa.

Hence we have found that the bubble diagrams found above dictate the exponential damping of the propagator in the IR region of timelike Poincar\'e position-momentum space.
In the propagator, when $pz_>$ gets larger than the value set by Eq.\,\eqref{eq:pz_op}, the propagator gets exponentially suppressed, making the corresponding IR region effectively opaque to propagation.

\subsubsection{Dimensional Analysis and EFT Censorship}
\label{se:censorship}

We go further by using knowledge from the EFT paradigm, which provides estimate for the couplings.
Dimensional analysis at strong coupling (\textit{i.e.} ``naive dimensional analysis'', here denoted NDA) for an EFT in five dimensions dictates that \cite{Chacko:1999hg,Agashe:2007zd,Ponton:2012bi}
\be
\lambda\sim \sqrt{\ell_5 \Lambda}\,,\,\quad\quad \zeta\sim \sqrt{\frac{\ell_5}{\Lambda^3} }
\label{eq:NDA_lz}
\ee
where $\ell_5=24\pi^3$ is the 5D loop factor. 
Using Eq.\,\eqref{eq:NDA_lz} in $S_{\lambda\zeta}$, $S_{\zeta\zeta}$ we  have 
\be
S_{\lambda\zeta}\approx -1\cdot 10^{-5}\,\ell_5\frac{ k}{\Lambda}\, (pz)^2\,,\quad 
S_{\zeta\zeta} \approx -5\cdot 10^{-7}\,\ell_5\frac{ k^3}{\Lambda^3} (pz)^4\,.
\ee

$\Lambda$ and $k$ can also be related by NDA. It was found in Ref.\,\cite{Costantino:2020msc} that
\be\Lambda \gtrsim \pi k \,.
\label{eq:Lambda_vs_k}
\ee 
This condition follows from  avoiding the strong coupling of  momentum modes in the momentum spectral representation. This relation is also tied to the large $N$ expansion in the dual CFT, see Eq.\,\eqref{eq:N_relation}.   

Taking $\Lambda\sim\pi k$, we find that the damping condition $S_{\lambda\zeta}\sim 1$ is attained for 
\be
pz|_{\rm opacity} \sim 20\,. \label{eq:pz20}
\ee
 For larger $\Lambda$, the value of $pz$ for which the damping occurs gets larger. 

But what of the contribution from the higher dimension operators?
According to the standard EFT paradigm, all contributions should become the same order at the limit of validity of the EFT. Thus by comparing $S_{\lambda\zeta}$ and $S_{\zeta\zeta}$, we robustly determine the validity region of the EFT.  Using $\frac{\zeta}{\lambda}\sim \frac{1}{\Lambda^2}$, it turns out that $S_{\lambda\zeta}\sim S_{\zeta\zeta}$ for
\be
(pz)^2 \sim 10 \frac{\Lambda^2}{k^2}\,, \label{eq:pz_timelike}
\ee
which makes the familiar $z$-dependent cutoff of Poincar\'e position-momentum space appear once again. For $\Lambda\sim\pi k$, the region of EFT breaking starts at $pz=O(10)$ and thus qualitatively matches the region of opacity.  For $\Lambda>\pi k$, the damping occurs before the breaking of the EFT. This is because
\be
pz|_{\rm opacity}\propto \sqrt{\frac{\Lambda}{k}}\,,\quad \quad pz|_{\rm EFT\, breaking}\propto \frac{\Lambda}{k}\,.
\ee
We conclude the damping  induced by interactions prevents the propagator to enter the region of EFT breaking. In short, we can say that opacity \textit{censors} the region of EFT breaking.\,\footnote{Throughout we have used NDA on the renormalizable $\lambda$ coupling.  As an alternative scenario, one could assume that  $\lambda$ is smaller than its strong coupling estimate. In such case the region of EFT breaking cannot be obtained using the $S_{\lambda\zeta}\sim S_{\zeta\zeta}$ criterion. Instead the effect of a next-to-leading higher dimensional operator has to be evaluated and compared to \textit{e.g.} $S_{\lambda\zeta}$ or $S_{\zeta\zeta}$. We expect similar conclusions regarding the exponential damping and EFT breakdown.
}
This is valid in particular for the boundary-to-bulk propagator. 

A first sketch of these features was done in \cite{Fichet:2019hkg} and further insights were given in \cite{Costantino:2020msc}. Censorship of the IR region for timelike momenta was also qualitatively predicted in \cite{ArkaniHamed:2000ds}. Our analysis validates all of these conclusions.

\subsubsection{Aside on the Dual CFT}
Here we mention how the above quantities match to the  dual CFT.  Using dimensional analysis in the holographic action and comparing to  correlators of a gauge theory with adjoint fields, Ref.\,\cite{Costantino:2020msc} finds
\be
\frac{\pi k}{\Lambda}\sim \frac{1}{N^2} \label{eq:N_relation}
\ee
where $N$ is the number of colors. Hence the ubiquitous $\Lambda/k$ ratio  is directly related to $N$.  $N\gg 1$ corresponds to $\Lambda \gg k$, in agreement with Eq.\,\eqref{eq:Lambda_vs_k}. 

\subsubsection{Qualitative Behavior for AdS$_{d+1}$}\label{se:generaldim}

Finally we comment about the behavior of the damping for other dimensions of spacetime. 
To estimate the behavior, we use the kinematic approximation Eq.\,\eqref{eq:ImPi_gen_p_approx},  in which the scaling of the kinematic function is
\be\K(d,p,0,0) \propto p^{d-4}.\ee
In addition, contributions from higher dimensional operators grow with higher powers of $pz$, coming from the extra derivatives in the vertices of the bubble. 
This knowledge is enough to estimate the behavior of ${\rm Im}\Pi$ in any dimension.

In $d=4$, we have obtained that $S_{\lambda\lambda}$,
$S_{\lambda\zeta}$, $S_{\zeta\zeta}$ grow respectively as $\log(pz)$, $(pz)^2$, $(pz)^4$, hence the leading contribution to the exponential damping comes from $S_{\lambda\zeta}$.\,\footnote{
Notice the $ S_{\zeta\zeta}$  contribution is of same order as a mixed contribution from the $\lambda\Phi\Phi_1\Phi_2$ and a higher dimensional operator such as $\Phi \partial_M\partial_N \Phi_1 \partial^M\partial^N\Phi_2$. 
} 
Notice that
  $S_{\lambda\zeta}$ is the leading contribution from the higher dimensional operators of the EFT, since the $\lambda$ coupling is renormalizable in $d=4$.   

In  $d>4$ the contribution from the $\lambda\lambda$ bubble to the $S$ function goes as $S\propto \lambda^2 (pz)^{d-4}$. 
Hence the exponential suppression already happens from the cubic coupling $\lambda$ in these dimensions.  
For $d=3$ the first polynomial  contribution to $S$ is from the $\lambda\zeta$ bubble, which grows  as $\propto pz$. For $d=2$,  it is from the $\zeta\zeta$ bubble which grows as $\propto (pz)^2$.

We conclude that, for any dimension of spacetime, 
 when considering an effective field theory in AdS---implying a tower of higher derivative operators,
the interactions render the IR region of Poincar\'e position-momentum space opaque to propagation. Conversely, the IR region of EFT breakdown, which exists because of the presence of these higher derivative operators, gets censored. These features are summarized in Fig.\,\ref{fig:CartoonBo2B}.

\subsection{Structure of Higher Order Diagrams }

One may wonder what happens to the exponential damping when one takes into account higher order diagrams into account in the self-energy dressing the propagator.  For instance the internal lines of the bubble considered so far are themselves dressed by bubble subdiagrams, and so on. 

A convenient way to  get the answer is to use the momentum spectral representation and introduce a UV and IR brane such that the  momentum modes get discretized---they become the familiar Kaluza-Klein modes. The optical theorem can then be used in the regime of momentum where these modes are narrow. 

Using this approach, it becomes clear that the contribution of higher order diagrams to ${\rm Im}\Pi$ appear when the mass integrals in Eq.\,\eqref{eq:MSresult} go beyond the leading order kinematic threshold, \textit{e.g.} beyond the two-body threshold in case of a bubble.
Instead of being cut at the two-body threshold, the integrand is now non zero for higher masses. These higher order regions of the integrand come with additional loop factor suppression---\textit{i.e.} the additional phase space suppression in the halved diagram. 
The region of the mass integrals where the two-body channel is opened is essentially unaffected by these subleading corrections. 

In summary, higher order corrections give small higher order additions to ${\rm Im}\Pi$, the main contribution to opacity comes from the leading order 1PI diagram. Using results from \cite{Costantino:2020msc} on cascade topologies in the continuum regime where the branes become irrelevant, these conclusions generalize to the  un-truncated Poincar\'e patch.

\subsection{Application to Asymptotically AdS Backgrounds}
\label{se:deformed_ads}

Our study has been focused on the entire Poincar\'e patch $z\in [0,\infty]$ of AdS. In some models, the metric in the IR region can naturally get deformed and become singular. This is instrumental for AdS/QCD,  see \textit{e.g.}
\cite{Karch:2006pv,Gursoy:2007cb,Gursoy:2007er,Gubser:2008ny,Falkowski:2008fz, Batell:2008zm, Batell:2008me,Cabrer:2009we,vonGersdorff:2010ht, Cabrer:2011fb}.
In many instances, models are built which truncate the UV and/or IR regions---leading to UV/IR branes. This well-known setup has applications across physics beyond the Standard Model, for \textit{e.g.} recent holographic dark sector applications see \cite{vonHarling:2012sz,McDonald:2012nc,McDonald:2010fe,McDonald:2010iq, Brax:2019koq, Costantino:2019ixl}. 

How do our findings about AdS opacity apply to such deformed AdS backgrounds? 

In the case of an IR brane, propagation to the brane will be exponentially damped at sufficiently high momentum, such that the propagators become agnostic of the IR region and do not know about the IR brane---neither in the spacelike regime, nor in the timelike regime (see earlier discussion in \cite{Costantino:2020msc,Fichet:2019hkg}). This conclusion readily extends to asymptotically AdS backgrounds with an IR deformation which can be described as an effective IR brane as in \cite{vonGersdorff:2010ht}. 

In the presence of a UV brane, one can check using limits of the propagator from \cite{Costantino:2020msc} that the effects of the UV brane on the propagator become negligible when the endpoints are away from it. Whenever one of the endpoints of the propagator is away from the UV brane, the locality property of $\text{Im}\Pi$ dictates that both endpoints of the relevant dressing diagrams will be away from the brane. The exponential damping is usually relevant deep in the IR, and hence the existence of the UV brane is irrelevant and our conclusions hold.

A wide class of backgrounds is described by a conformally flat metric of the form
\be 
ds^2=e^{-2A(z)}\left(\eta_{\mu\nu}dx^\mu dx^\nu -dz^2 \right)\,. \label{eq:warped_metric}
\ee
In particular, deformed AdS backgrounds typically admit such a description.\,\footnote{This metric appears in certain 5D supergravities, see \textit{e.g.} \cite{Freedman:1999gp}. }
Like in Poincaré position coordinates, this metric has $d$-dimensional Poincaré isometries in the $x^\mu$ coordinates. Hence for an arbitrary warping $A(z)$ we can Fourier-transform to position-momentum space $(p^\mu,z)$ as we did in the exact AdS case. 
This implies that the canonical and spectral representations of the propagator may be used in more general backgrounds. 
The last remaining ingredient is to obtain the solution of the Fourier-transformed free EOM. In any case, the split structure of the spectral functions should automatically hold throughout, as explained in Sec.\,\ref{se:AdS}.

In summary, we conclude that  our results on opacity of the Poincar\'e momentum-space apply to a broad range of  deformed/truncated AdS backgrounds, and that our formalism itself may also be generalized to other metrics of the form Eq.\,\eqref{eq:warped_metric}.

\section{Opacity from Gravity in AdS$_5$}
\label{se:grav}

Even if cubic matter interactions are absent---which could happen for symmetry reasons or if they are vanishing/negligible altogether, there is always a cubic interaction between AdS gravity and matter. This  is true for any matter field since the graviton couples linearly to the matter stress tensor as $h_{MN}T^{MN}$. Hence bulk gravity always  produces a bubble contribution to the self-energy of a matter field, thereby inducing a universal,  1-loop contribution to the ``opacity'' of the IR region (see discussions in  Sec.\,\ref{se:Opacity}).

For this work we consider propagation of a scalar field and we  focus only on the $d=4$ case.  
In $d=4$, the bulk graviton splits into five degrees of freedom, two of helicity-two, two of helicity-one, one of helicity-zero, each with their own coupling to the stress-energy tensor. The trace of the metric is a non-physical, ghosty degree of freedom which also contributes to the loop. 
We  closely follow  the formalism of \cite{Dudas:2012mv}. \footnote{Except that we use the opposite metric signature.}

The pure gravity part of the action is
\be \label{eq:5d_action}
S_{\rm EH}= M^3_* \int d^5X \sqrt{g}  \left({\cal R} - \Lambda_5\right)  \,\ee
with $\Lambda_5=-12k^2$.
The canonically normalized metric fluctuation  around the AdS background $\gamma_{MN}$ is defined by
\be
g_{MN}=\gamma_{MN}+\sqrt{\frac{2}{M^3_*}} h_{MN}\,.
\ee
The expansion of the gravity action Eq.~\eqref{eq:5d_action}  up to quadratic order is well-known (see \textit{e.g.}~\cite{Boos:2002hf,Hinterbichler:2011tt}), giving the action for the 5d graviton $h_{MN}$, 
\begin{align}
 S^h = \int d^5X  \sqrt{\gamma} \,\bigg( 
\frac{1}{2}&\nabla_R h_{MN}\nabla^R h^{MN} - 
\frac{1}{2}\nabla_R h\nabla^R h+\nabla_M h^{MN}\nabla_N h
\\  
-  &\nabla_M h^{MN} \nabla^R h_{RN}+ k^2(h^2_{MN}+h^2) \nonumber 
 + \sqrt{\frac{1}{2 M^3_*}}h^{MN} T_{MN} \bigg) \,
\nonumber 
\end{align}
where
\be
T_{MN} = - 2\frac{\delta {\cal L}_{\Phi}}{\delta \gamma^{MN}} + \gamma_{MN}{\cal L}_{\Phi} \,.
\ee
Following \cite{Dudas:2012mv},  all  degrees of freedom of the graviton can be disentangled using field redefinitions and Faddeev-Popov gauge fixing. 
Defining $\hat h_{MN}=(kz)^2 h_{MN}$  then splitting the graviton components as
\be\label{eq:gravcomponents}
 \tilde h_{\mu\nu}=\hat h_{\mu\nu}-\frac{1}{4}\eta_{\mu\nu}h^\rho_\rho \,, \quad 
  B_\mu = \frac{\sqrt{2}}{kz}\hat h_{\mu 5}\,,\quad 
 \chi=\frac{1}{2}\left(\hat h_\mu^\mu-2 \hat h_{55}\right)\,, \quad
  \phi=\frac{\sqrt{3}}{\sqrt{2}(kz)^2}\hat h_{55}\,,
  \ee
and defining the sources 
\be
\tilde T_{\mu\nu}=T_{\mu\nu}-\frac{1}{4}\eta_{\mu\nu}T_\rho^\rho\,,\quad
\tilde T_{55}=T_{55}+\frac{1}{2}T_\rho^\rho\,, 
\ee
the graviton action $S^h=\int dx^M {\cal L}^h$ takes the simple form
\begin{align}
{\cal L}^h= & 
\frac{1}{2}\bigg(
\frac{1}{(kz)^3}(\partial_R \tilde h_{\mu\nu})^2+\frac{1}{kz}(\partial_R  B_{\mu})^2
+ k z (\partial_R \phi)^2 -\frac{1}{(kz)^3} (\partial_R \chi)^2
\bigg)  \label{eq:Lagh}
 \\ &
+\frac{1}{\sqrt{M_*^3}}
\left(
\frac{1}{\sqrt{2}(kz)^3} \tilde h^{\mu\nu}\tilde T_{\mu\nu}
+
\frac{1}{(kz)^2} B^\mu T_{\mu 5}
+\frac{1}{2\sqrt{2}(kz)^3} \chi T_\mu^\mu
+\frac{1}{\sqrt{3}kz} \phi \tilde T_{55}
\right)\,. \nonumber
\end{align}
In Eqs.\,\eqref{eq:gravcomponents}-\eqref{eq:Lagh} all contractions are done with the Minkowski metric.
 As expected the $\chi$ field has ``wrong-sign'' kinetic term and is an unphysical degree of freedom. Its contribution naturally combines with the one of the $\tilde h_{\mu\nu}$ field. 
 
The graviton degrees of freedom in  Eq.\,\eqref{eq:Lagh} are diagonal, hence their contributions to the self-energy are disentangled such that
\be
{\rm Im} \,\Pi_p(z,z')=
{\rm Im} \,\Pi^{h}_p(z,z')+
{\rm Im} \,\Pi^{B}_p(z,z')+
{\rm Im} \,\Pi^{ \phi}_p(z,z')+
{\rm Im} \,\Pi^{\chi}_p(z,z') \,. 
\ee
In this work  we focus on the contribution of the scalar component $\phi$ of the graviton multiplet.   The Feynman  propagator of $\phi$ is given by 
\begin{align}
 G^\phi(z,z') =\frac{\pi}{2k} J_0(pz_<)H^{(1)}_0(pz_>) 
 \end{align}
with $\left\langle  \phi(p,z) \phi(p',z') \right\rangle=G^\phi(z,z') (2\pi)^4 \delta^{(4)}(p^\mu-p'^\mu)$. 

The stress-energy tensor  of the bulk scalar $\Phi$ is
 \be
T_{MN}= - \partial_M \Phi \partial_N \Phi + \frac{1}{2}\gamma_{MN} \left(\partial_R \Phi \partial^R \Phi  -m^2_\phi \Phi^2 
 \right)\,,
\ee
giving the source 
      \be
   \tilde T_{55}=- \frac{3}{2}(\partial_5 \Phi)^2-\frac{1}{2\,(kz)^2}m_\Phi^2 \Phi^2 \,,
   \ee
 which couples to  the $\phi$ component of the graviton multiplet. Using integration by parts, we rewrite the relevant term of the Lagrangian as
\begin{align}
    {\cal L}^h\supset  \Phi\partial_5\left(\frac{\sqrt{3}}{2kz\sqrt{M_*^3}} \phi
    \partial_5 \Phi\right)-\frac{m_\Phi^2}{2(kz)^3\sqrt{3M_*^3}} \phi  \Phi^2.
\end{align}

We perform the graviton loop calculation in App.~\ref{app:grav_loop}. We find 
  \begin{align}
{\rm Im}\Pi^\phi(z_1,z_2) =  \frac{1}{48\pi\,M_*^3} &\left(
\frac{ m_\Phi^2}{2(kz_1)^2} - \frac{3}{2} \partial_{z_1} \partial_{z_1}^{G_\Phi} \right) 
\left(
\frac{ m_\Phi^2}{2(kz_2)^2} -\frac{3}{2} \partial_{z_2} \partial_{z_2}^{G_\Phi} \right) \label{eq:ImPiphi}   
 \\ &
\frac{1}{z_1z_2}\int dq_1dq_2 q_1q_2\left[z_1^2z_2^2J_\alpha(q_1z_1)J_\alpha(q_1z_2)\right]
J_0(q_2z_1)J_0(q_2z_2)\nn
\,.
\end{align}
Here the $\partial_{z}^{G_\Phi}$ derivatives act on the internal $\Phi$ propagator only (\textit{i.e.} only the term in square brackets), whereas the $\partial_{z}$ derivatives act on everything to the right. 

Using the WKB approximation scheme introduced and justified in Sec.\,\ref{se:Opacity}, we then obtain
\be\label{impigravATzz}
{\rm Im}\Pi^\phi(z,z) = \kappa^2\frac{7 \,p^6}{30720 \pi^3\,k^3}\left(1+O\left(\frac{1}{p z}\right)\right)
\ee
\be\label{eq:cgrav}
C^\phi(z)=\frac{24}{49}+O\left(\frac{1}{p z}\right),
\ee
where we have introduced the 5D gravity coupling strength \be \kappa \equiv \sqrt{\frac{k^3}{M_*^3}}=\frac{k}{M_{Pl}}\,.\ee
This contribution is similar to the effect from the $\Phi (\partial \Phi)^2-\Phi (\partial \Phi)^2$ scalar bubble---which is not  surprising since the number of derivatives in the vertices is the same. 
The argument of the exponential of the WKB ansatz given by Eq.\,\eqref{eq:S_final} is found to be 

  \begin{align}
    S^\phi(z)& =-\frac{49 e^{-49/48}}{491520 \sqrt{3} \pi^{5/2} \, }\,\kappa^2\, (pz)^4  \label{eq:S_phi}
\\    
  &   \approx  -1\cdot 10^{-6}\,\kappa^2 \, (pz)^4 \nn
 \end{align}
 
This shows explicitly that bulk gravity itself induces opacity of the IR region in Poincar\'e position-momentum space, \textit{i.e.} $G_p(z,z')\sim e^{S^\phi(z_>)}$. 
The contribution is universal in the sense that it depends only on the strength of the AdS gravity coupling $\kappa$.\,\footnote{
5D dimensional analysis shows that the maximal value of $\kappa$ is $O(1)$ to avoid higher curvature terms. For completeness we also remind that  $\kappa^2 \sim \frac{1} { N^2}$ when matching  the theory to a 4D CFT  \cite{Aharony:1999ti}. }

In five dimensions, the gravity Lagrangian is understood as  an EFT from which the leading term is the Einstein-Hilbert one. Studying higher order effects from the graviton EFT would require one to include the effect of higher dimension interactions such as  ${\cal R}{\cal L}_{\rm matter}$. This is beyond the scope of this work, in which the main focus has been on scalar fields. Since we do not attempt to evaluate higher order contributions from the  EFT, we do not comment on  the interplay between EFT breakdown and IR opacity in this section. 
It would be interesting to check if the superPlanckian region
is censored as a result of opacity from graviton loops, along the lines of Sec.\,\ref{se:censorship}.

\section{Conclusions }\label{se:Conclusion}

In this work, we have investigated how quantum dynamics affects the propagation of a scalar field in Lorentzian AdS$_{d+1}$. 
Our main results are presented in Sec.\,\ref{se:summary}. This section contains a summary, comments on the implications of this work and possible future directions.

In the first part of this work, we studied the simplest scalar bubble in momentum-position space. 
Different objects appear depending on the representation used.
In the canonical representation, loop integrals with non-integer powers are the key ingredient. 
In the momentum spectral representation, a generalized two-body kinematic threshold  emerges.
In the conformal spectral representation, AdS/CFT arises: CFT correlators naturally appear and combine to form a CFT bubble diagram.\,\footnote{We obtained the anomalous dimension via resummation and verified that it matches exactly the one found in \cite{Giombi:2017hpr} by another method.}
By direct computation, we find that the conformal spectral result for ${\rm Im}\Pi$ takes the form of a sum over double-trace propagators, with coefficients that match exactly those from \cite{Dusedau:1985ue,Fitzpatrick:2011hu}.
We have provided explicit equivalence proofs between these representations.

We prove a number of global and asymptotic properties of the imaginary part of the bubble via the various representations. Asymptotics and convenient approximations are best obtained in the momentum spectral representation. In particular this representation shows that ${\rm Im}\Pi$ tends to be peaked at large $pz$. 

To study concretely how loops affect propagation in AdS, we directly solve the dressed equation of motion. To do so, we introduce a WKB-type method adapted to a convolution term $\Pi* \Gf$ and use a saddle-point approximation of ${\rm Im}\Pi$. The most interesting regime is when the propagator endpoints satisfy $pz_<\ll1 $, $ pz_>\gg 1$, which includes the  boundary-to-bulk propagators with $ pz\gg 1$ as a particular case. This regime has no flat-space equivalent---the $pz\ll1 $ region vanishes if one Weyl-transforms to flat space.

We adopt the EFT viewpoint in which Lagrangians contain an infinite series of higher dimensional operators and  evaluate the effect of bubbles from leading and next-to-leading operators in AdS$_5$.
We find that interactions of high-enough order in the bulk of AdS induce an exponential damping of the propagator for timelike momentum at large $pz$. The damping renders the IR region of the Poincaré patch opaque to propagation. 
This exponential damping is similar to the one induced by the Feynman $i\epsilon$ prescription in the free case. We thus conclude that loop corrections are consistent with the $i\epsilon$ prescription.

By definition, an EFT breaks down when higher dimensional operators  give contributions of the same order. In AdS this occurs at sufficiently large $pz$, \textit{i.e.} in the IR region of the Poincar\'e patch.
 Estimating the operator coefficients using standard dimensional analysis, we find that the region of EFT breakdown falls behind the region of exponential suppression, meaning that this region is effectively \textit{censored}. A qualitative generalization to AdS$_{d+1}$ is presented.
 
A first sketch of these features was proposed in \cite{Fichet:2019hkg}. 
Censorship of the IR for timelike momenta was also qualitatively predicted in \cite{ArkaniHamed:2000ds}, though the mechanism by which their predicted behavior occurs is fundamentally different. Our analysis validates all of these conclusions. 

The damping behavior observed  is not unique to AdS, it may occur in any spacetime background whenever the self-energy of the propagating field develops an imaginary part. In flat space  such damping is associated to decaying particles via the S-matrix optical theorem, but  such interpretation is not required in general---as exemplified by our AdS findings. The key aspect is rather that $\Pi$ develop an imaginary part in some region.

Going beyond scalar loops, we have evaluated a contribution from the leading graviton loop in AdS$_5$. Opacity of the IR region again occurs, and depends only on the strength of AdS gravity controlled by $k/M_{\rm Pl}$. 
We do not comment on EFT breaking in this context since it would require to evaluate the effect of graviton-matter vertices from higher dimensional operators.

\subsection{Possible Implications and Future Directions}

There are some immediate extensions to our study. It would be interesting to derive the approximate behavior of the dressed propagator as obtained in Sec.\,\ref{se:Opacity} from the conformal spectral representation as given in Sec.\,\ref{se:AdSCFT}. Other directions include the study of propagating fields with non-zero spin or to investigate the effect of a larger variety of self-energies including those from fermion or graviton loops. 

The formalism developed here for AdS may be useful for more general backgrounds---namely those admitting 
a warped, conformally-flat metric of the form
described in Eq.\,\eqref{eq:warped_metric}.
A direct consequence of working in the Poincaré momentum space $(p^\mu, z)$ is that the propagators have a split structure in $pz, pz'$, both in the  spectral representations and in the canonical one.
Moreover, whenever the solutions to the free EOM in momentum-position space are analytical, we can expect 
the various representations (canonical, conformal and momentum spectral) of the propagator to have analytical expressions.

This study has been focused on AdS$_{d+1}$. We have essentially not discussed aspects of the dual CFT, apart from using some basic elements which naturally arise in the calculations. 
The bubble topology is the simplest loop diagram, and as such may be seen as a laboratory to study unitarity in AdS. AdS/CFT unitarity methods have recently been developed \cite{Meltzer:2019nbs, Meltzer:2020qbr}. Our detailed analysis of 
$\rm {Im} \Pi$ in its various forms could be useful in these types of studies.\,\footnote{ We have found that  ${\rm Im}(
\mathrm{\mathbf{Cut}} \Pi) = \mathrm{\mathbf{Cut}}( {\rm Im} \Pi ) =   {\rm Im} \Pi  $, where the  AdS $\mathbf{Cut}$ operation acts on poles in the conformal spectral representation. We have also found that upon cutting the bubble in the momentum spectral representation, we obtain the imaginary part in the canonical representation.}

On a more phenomenological note, our results on the exponential damping of the propagator may have implications for extensions of the Standard Model of particles. 
This is for example relevant for extra-dimensional and holographic dark sector models. Consider the so-called ``slice of AdS''---in which there are UV and IR branes that truncate the bulk. Our results indicate that particles in the UV cannot propagate to the IR brane if they have sufficiently high absolute  4-momentum---either timelike or spacelike. This gives rise to the concept of an \textit{emergent} sector which can only interact at sufficiently small values of absolute 4-momentum, as discussed in \cite{Costantino:2020msc}.

Our work may have implications for AdS/QCD, since timelike bulk processes are related to timelike fragmentation functions in the gauge theory. One may note that opacity of AdS in the spacelike regime played a role in the AdS/CFT deep-inelastic study of \cite{Polchinski:2002jw}. Hence  timelike opacity should also be important in the study of the cross-symmetric process, \textit{e.g.}  $e
^+e^-$ annihilation into states of the strongly-coupled  gauge theory.

\section*{Acknowledgments}

We thank David Meltzer, Allic Sivaramakrishnan,  Vasco Goncalves, Mariano Quiros, Flip Tanedo and Kaustubh  Agashe for useful discussions and comments. 
The authors acknowledge the support of their respective cats while finishing this project.
A.C.~is supported by the National Science Foundation Graduate Research Fellowship Program under Grant No.~1840991.
S.F.~is supported by the S\~ao Paulo Research Foundation (FAPESP) under grants \#2011/11973, \#2014/21477-2 and \#2018/11721-4.

\appendix
\section{Dressed Equation of Motion}
\label{app:dressed_EOM}

In this appendix we derive the  equation of motion in the presence of  a generic self-energy using the path integral formalism. The approach amounts to requiring invariance of a given path integral quantity under an infinitesimal change in the field variable. This method was used in \cite{Peskin:257493} in the free case. 

We first derive the homogeneous equation of motion. Start from the partition function of the interacting theory
\be
Z=\int {\cal D}\Phi {\cal D}\varphi e^{i S[\Phi,\varphi]} \,,\quad 
S[\Phi,\varphi]=\int d^{d+1}X \sqrt{|\gamma|}\left(
{\cal L}_{0}[\Phi]+{\cal L}[\varphi]+\lambda \Phi {\cal O}(\varphi)
\right)\,.
\ee
In the fundamental action, ${\cal L}_{0}[\Phi]$ is the free Lagrangian for $\Phi$, given by 
\be
{\cal L}_{0}[\Phi]=-\frac{1}{2} \Phi(X) {\cal D } \Phi(X)\,. 
\ee
${\cal L}[\varphi]$ encodes the properties of other fields $\varphi$ and does not need to be specified. The (generally composite) operator ${\cal O}$ is here assumed to depend on the  $\varphi$ fields only. A very similar calculation can be done if $\Phi$ has self-interactions. 
We  recall the definition of the expectation value of an operator $A$ dictated by the partition function,  
\be \langle A\rangle (X)=Z^{-1}
\int {\cal D} \Phi  {\cal D}\varphi \, A(X)\, e^{i \int d^{d+1}X' \, S[\Phi,\varphi]} 
 \,. \label{eq:A_cl}
\ee

Starting from $Z$, we perform an infinitesimal  change of variable of the field $\Phi(X)\equiv \tilde \Phi(X)+\epsilon(X)$. The path integral measure remains unchanged, ${\cal D} \Phi={\cal D} \tilde \Phi$. Expanding the exponential at first order in  $\epsilon$ gives 
\begin{align}
Z&=\int {\cal D} \Phi {\cal D}\varphi e^{i \int d^{d+1}X' \sqrt{|\gamma|} \left(
{\cal L}_{0}[\Phi]+{\cal L}[\varphi]+  \lambda \Phi {\cal O} \right)} \left(1+
i\int d^{d+1}X \epsilon(X) \sqrt{|\gamma|} \left[-{\cal D} \Phi(X) +\lambda {\cal O}(X)
\right] 
\right) \nn \\
&= Z+
i\int d^{d+1}X \sqrt{|\gamma|}\epsilon(X)\int {\cal D} \Phi {\cal D}\varphi e^{i \int d^{d+1}X' \sqrt{|\gamma|} \left(
{\cal L}_{0}[\Phi]+{\cal L}[\varphi]+\lambda \Phi {\cal O} \right)} 
 \left[-{\cal D} \Phi(X) +\lambda {\cal O}(X)
\right] 
\,.
\end{align}
The second term in the last line must vanish for any $\epsilon(X)$, implying the quantum equation of motion  
\be
\int {\cal D} \Phi {\cal D}\varphi e^{i S[\Phi,\varphi]} 
 \left[{\cal D} \Phi(X) - \lambda {\cal O}(X)
\right] =0\,. \label{eq:qEOM_2}
\ee
Using the definition of the expectation value from Eq.\,\eqref{eq:A_cl} and introducing the notation $\langle \Phi \rangle= \Phi_{\rm cl}  $, $\langle {\cal O} \rangle= {\cal O}_{\rm cl}  $ we have 
\be
{\cal D} \Phi_{\rm cl}(X) - \lambda {\cal O}_{\rm cl}(X)=0\,.
\ee

 The second term involves the expectation value of the composite operator $\cal O$
\be {\cal O}_{\rm cl}(X)=Z^{-1}
\int {\cal D} \Phi  {\cal D}\varphi \, {\cal O}(X)\, e^{i \int d^{d+1}X \sqrt{|\gamma|} \left(
{\cal L}_{0}[\Phi]+{\cal L}[\varphi]+\lambda \Phi {\cal O} \right)} 
 \,.
\ee
We now seek to express ${\cal O}_{\rm cl}$ as a  quantity involving
$\Phi_{\rm cl}$.
We can notice the $\Phi$ field acts as a source for $\cal O$.  It is convenient to  introduce a  generating functional of the ${\cal O}$ correlators $E[\Phi_{\rm cl}]$ such that
\be
E[\Phi_{\rm cl}]=i \log Z[\Phi_{\rm cl}]\,,\quad Z[\Phi_{\rm cl}]= \int  {\cal D}\varphi e^{i \int d^{d+1}X \sqrt{|\gamma|} 
\left({\cal L}_{0}[\Phi]+{\cal L}[\varphi]+  \lambda \Phi_{\rm cl} {\cal O} \right)} 
\ee
From $E[\Phi_{\rm cl}]$ we have the standard identities
\be
  \frac{\delta E[\Phi_{\rm cl}]}{\delta \Phi_{\rm cl}(X)}= - \lambda \sqrt{|\gamma|} {\cal O}_{\rm cl}(X)   \,. \label{eq:Ocl}
\ee
\be
  \frac{\delta^2 E}{\delta \Phi_{\rm cl}(X)\delta \Phi_{\rm cl}(X')}= - i \lambda^2 \sqrt{|\gamma|}_X \sqrt{|\gamma|}_{X'} \langle{\cal O}(X){\cal O}(X')\rangle_{\rm conn}    \, 
\ee
where we display the  connected correlator of $\cal O$. The self-energy $i \Pi$ is defined such as it absorbs the metric factors and the $(i\lambda)^2$ vertices, giving 
\be
  \frac{\delta^2 E}{\delta \Phi_{\rm cl}(X)\delta \Phi_{\rm cl}(X')}= -  \Pi (X,X')   \,. 
\ee
Finally we also have 
\be
 - \lambda \sqrt{|\gamma|} {\Phi}_{\rm cl}(X) =  \frac{\delta E[\Phi_{\rm cl}]}{\delta {\cal O}_{\rm cl}(X)}  \,. \label{eq:Phicl}
\ee

Given these identities, we can rewrite $- \lambda \sqrt{|\gamma|} {\cal O}_{\rm cl}(X)$ as follows. Using the chain rule for functional derivatives in Eq.\,\eqref{eq:Ocl} we get
\be
- \lambda \sqrt{|\gamma|} {\cal O}_{\rm cl}(X) = \int d^{d+1}X' \frac{\delta {\cal O}_{\rm cl}(X')}{\delta \Phi(X)}\frac{\delta E[\Phi]}{\delta {\cal O}_{\rm cl}(X')}
= \int d^{d+1}X' \frac{\delta {\cal O}_{\rm cl}(X')}{\delta \Phi(X)} \left(-\lambda \sqrt{|\gamma|_{X'}}\right)\Phi_{\rm cl}(X')
\,.
\ee
Then using again Eq.\,\eqref{eq:Ocl} notice that
\be
-\lambda \sqrt{|\gamma|_{X'}}\frac{\delta {\cal O}_{\rm cl}(X')}{\delta \Phi(X)} =  \frac{\delta^2 E[\Phi]}{\delta \Phi(X)\delta \Phi(X')} =-\Pi(X,X')  \,.
\ee 
Therefore we have shown that 
\be
\lambda \sqrt{|\gamma|} {\cal O}_{\rm cl}(X)= 
\int d^{d+1}X' \Pi(X,X') \Phi_{\rm cl}(X').
\ee
Hence the dressed equation of motion takes the form 
\be
{\cal D}\Phi_{\rm cl}(X) - \frac{1}{\sqrt{|\gamma|}} \Pi * \Phi_{\rm cl}(X) =0 \,.
\ee

In order to compute the dressed equation of motion for the Feynman propagator, one proceeds similarly with an  extra insertion of $\Phi$ in the path integral such that the starting quantity considered amounts to $Z\Phi_{\rm cl}(X)$.
An infinitesimal shift of the field variable gives 
\begin{align}
& Z\Phi_{\rm cl}(X)= \\ & \nn \int {\cal D} \Phi {\cal D}\varphi e^{i S} \bigg(  \Phi(X)+
\int d^{d+1}X' \epsilon(X') \left(i\sqrt{|\gamma|} \left[-{\cal D} \Phi(X') +\lambda {\cal O}
\right]\Phi(X) 
+\delta^{(d+1)}(X-X')\right)
\bigg) \nn\\
&=Z\Phi_{\rm cl}(X)\,+ \nn \\ & \nn \int d^{d+1}X' \epsilon(X')\int {\cal D} \Phi {\cal D}\varphi e^{i S} 
 \left(i\sqrt{|\gamma|} \left[-{\cal D} \Phi(X') +\lambda {\cal O}
\right]\Phi(X) 
+\delta^{(d+1)}(X-X')\right)
\,.
\end{align}
The second term must vanish for any $\epsilon(X)$, hence we have
\be
\int {\cal D} \Phi {\cal D}\varphi e^{i S} 
 \left(i\sqrt{|\gamma|} \left[-{\cal D} \Phi(X') +\lambda {\cal O}
\right]\Phi(X) 
+\delta^{(d+1)}(X-X')\right) =0.
\ee
In terms of expectation values, we find that the previous statement is equivalent to
\be
\sqrt{|\gamma|}_X {\cal D}_X\langle \Phi(X) \Phi(X') \rangle- \lambda \sqrt{|\gamma|}_X \langle \Phi(X) {\cal O}(X') \rangle_{\rm conn} = -i\delta^{(d+1)}(X-X').
\ee
Using similar manipulations as above\,\footnote{One can for instance introduce a generating functional with sources for $\Phi$ and $\cal O$ which are set to zero at the end of the calculation.} we find 
\be
\lambda\sqrt{|\gamma|}_X\langle \Phi(X) {\cal O}(X') \rangle = i\lambda^2 \sqrt{|\gamma|}_X\sqrt{|\gamma|}_{X'}\langle {\cal O}(X){\cal O}(\tilde X)\rangle_{\rm conn} * G(\tilde X, X')= \Pi * G (X,X')\,.
\ee
The dressed equation of motion is thus 
\be
\sqrt{|\gamma|} {\cal D}_X G(X,X')- \Pi *G(X,X')  = -i\delta^{(d+1)}(X-X'),
\ee
which completes the proof of Eq.\,\eqref{eq:DDelta}.

\section{Proof of Conformal Completeness Relation}\label{se:comp_alpha}

We prove Eq.\,\eqref{eq:comp_alpha} in two steps. In App.\,\ref{se:B1}, we show that the integral vanishes for $z\neq z'$. In App.\,\ref{se:B2}, we integrate Eq.\,\eqref{eq:comp_alpha} over $z'$ to determine the normalization of the delta function.

\subsection{$z\neq z'$}\label{se:B1}
Here we evaluate 
\be
\int_{-i \infty}^{i \infty} d\alpha \Omega_p(\alpha)=\int_{i \infty}^{i \infty}  d\alpha \frac{i\alpha \sin(\pi \alpha)}{\pi^2} (k^2zz')^{d/2} K_\alpha(\sqrt{-p^2}z) K_\alpha(\sqrt{-p^2}z')
\ee
for $z> z'$. This choice is unimportant---it simply controls how we close the contour for some of the terms. We use the series representation of the Bessel functions, Eq.\,\eqref{eq:Kseries}. Four terms emerge. These reduce to two terms under a $\alpha\rightarrow-\alpha$ relabeling. We are left with
\begin{align}
&\int_{-i \infty}^{i \infty} d\alpha \Omega_p(\alpha)=\\&\nonumber(k^2zz')^{d/2}\int_{i \infty}^{i \infty}  d\alpha \frac{i\alpha }{2\sin(\pi \alpha)} \sum_{m,n=0}\frac{1}{m!n!}\left(\frac{\sqrt{-p^2}z'}{2}\right)^{2n}\left(\frac{\sqrt{-p^2}z}{2}\right)^{2m}
\times\\&\nonumber
\left[\frac{1}{\Gamma(n-\alpha+1)}\left(\frac{\sqrt{-p^2}z'}{2}\right)^{-\alpha}\right]\left[\frac{1}{\Gamma(m-\alpha+1)}\left(\frac{\sqrt{-p^2}z}{2}\right)^{-\alpha}-\frac{1}{\Gamma(m+\alpha+1)}\left(\frac{\sqrt{-p^2}z}{2}\right)^{\alpha}\right]\,.
\end{align}
We promote the line integral to a contour and close this contour in a large half-circle towards the negative reals. We then apply the residue theorem and sum over the poles of the $1/\sin\left(\pi\alpha\right)$ function. A sequence of poles are cancelled by the roots of $1/\Gamma(m+\alpha+1)$ in the second term. The remaining terms (labeled by $T_1$ and $T_2$) are
\begin{align}
&T_1=(k^2zz')^{d/2}\times\\ \nn &\sum_{m,n=0}\sum_{l=1}^\infty \frac{il(-1)^{l}}{m!n!}\left(\frac{\sqrt{-p^2}z'}{2}\right)^{2n+l}\left(\frac{\sqrt{-p^2}z}{2}\right)^{2m+l}\frac{1}{\Gamma(n+l+1)\Gamma(m+l+1)}
\\ \nn
\\
&T_2=-(k^2zz')^{d/2} \times\\ \nn &\sum_{m=1,n=0}\sum_{l=1}^m \frac{il(-1)^{l}}{m!n!}\left(\frac{\sqrt{-p^2}z'}{2}\right)^{2n}\left(\frac{\sqrt{-p^2}z}{2}\right)^{2m}\frac{1}{\Gamma(n+l+1)\Gamma(m-l+1)}\left(\frac{z'}{z}\right)^{l}\,.
\end{align}
We introduce $c\equiv m+l$ in $T_1$ and convert the sums over $m,l$ into sums over $c,l$. Upon a relabeling, this gives $T_1=-T_2$ and thus
\be
\int_{-i \infty}^{i \infty} d\alpha\, \Omega_p(\alpha)=0
\ee
for $z\neq z'$.

\subsection{Normalization}
\label{se:B2}

To obtain the normalization and thus prove Eq.\,\eqref{eq:comp_alpha}, we evaluate
\begin{align}
&\int_0^\infty dz'\int_{-i \infty}^{i \infty} d\alpha \Omega_p(\alpha)=\\ \nn &\int_0^\infty dz'\int_{-i \infty}^{i \infty}  d\alpha \frac{i\alpha \sin(\pi \alpha)}{\pi^2} (k^2zz')^{d/2} K_\alpha(\sqrt{-p^2}z) K_\alpha(\sqrt{-p^2}z')\,.
\end{align}
We directly perform the integral over $z'$ and use the series representation Eq.\,\eqref{eq:Kseries} to expand the remaining Bessel function. We then let $\alpha\rightarrow-\alpha$ to reduce the two terms to one. We are left with
\begin{align}
&\int_0^\infty dz'\int_{-i \infty}^{i \infty} d\alpha \Omega_p(\alpha)=\frac{i}{2\pi\sqrt{-p^2} }\left(\frac{2zk^2}{\sqrt{-p^2}}\right)^{d/2} \sum_{m=0}\frac{1}{m!}\left(\frac{\sqrt{-p^2}z}{2}\right)^{2m}
\times\\&\nonumber
\int_{-i \infty}^{i \infty}  d\alpha \left[\frac{\alpha}{\Gamma(m-\alpha+1)}\left(\frac{\sqrt{-p^2}z}{2}\right)^{-\alpha}\right]\Gamma \left(\frac{1}{4} (d-2 \alpha +2)\right) \Gamma \left(\frac{1}{4} (d+2 \alpha +2)\right)\,.
\end{align}

We promote the line integral to a contour and close to the negative reals. A straightforward application of the residue theorem yields
\begin{align}
&\int_0^\infty dz'\int_{-i \infty}^{i \infty} d\alpha \Omega_p(\alpha)=\\ \nn &\frac{(kz)^{d+1}}{k}\sum_{m,n=0} \frac{(-1)^n}{m!n!}\left[\frac{(2n+\frac{d}{2}+1)\Gamma \left(n+\frac{d}{2}+1\right) }{\Gamma(m+(2n+\frac{d}{2}+1)+1)}\left(\frac{\sqrt{-p^2}z}{2}\right)^{2n+2m}\right]\,.
\end{align}
By introducing $c\equiv m+n$ and reorganizing the sums, we obtain
\begin{align}
&\int_0^\infty dz'\int_{-i \infty}^{i \infty} d\alpha \Omega_p(\alpha)=\\ \nn &\frac{(kz)^{d+1}}{k}\sum_{c=0}^\infty\left(\frac{\sqrt{-p^2}z}{2}\right)^{2c}\sum_{n=0}^c \frac{(-1)^n(2n+\frac{d}{2}+1) }{(c-n)!n!}\frac{\Gamma \left(n+\frac{d}{2}+1\right) }{\Gamma(c+(n+\frac{d}{2}+1)+1)}\,.
\end{align}
We directly perform the sum over $n$ with
\begin{align}
\sum_{n=0}^c  \frac{(-1)^n}{(c-n)!n!}\frac{\Gamma \left(n+1+\frac{d}{2}\right)(2n+\frac{d}{2}+1) }{\Gamma(c+(n+\frac{d}{2}+1)+1)}=\delta_{c0}\,,
\end{align}
which we obtained through the use of Mathematica \cite{Mathematica}. This gives
\begin{align}
\int_0^\infty dz'\int_{-i \infty}^{i \infty} d\alpha \Omega_p(\alpha)= \nn \frac{(kz)^{d+1}}{k}\,,
\end{align}
which completes the proof of Eq.\,\eqref{eq:comp_alpha}.

\section{Elements of AdS/CFT in Momentum Space }
\label{se:CFT}

 In this Appendix
 we give elements of CFT in $d$-dimensional Minkowski space  and introduce the AdS bulk-to-boundary propagator. 
All these notions are consistently translated into Lorentzian momentum space. 
We recall the  chosen metric has mostly-minus signature $(+,-,\ldots,-)$. The Fourier transform convention is \be {\cal O} (x)=\int \frac{d^d p}{(2\pi)^d}{\cal O} (p) e^{-i p\cdot x} \label{eq:FT} \,.\ee 

\subsection{Bulk Mass and Conformal Dimensions}
\label{se:Delta}

The relation between a  scalar bulk field in AdS with bulk mass $m^2_\Phi=\left(\alpha^2-\frac{d^2}{4}\right)k^2$
and a scalar primary  operator with scaling dimension $\Delta$ from the dual CFT   is given by (see \textit{e.g.} \cite{Witten:1998qj})
\be
\alpha^2 = \frac{M^2}{k^2} + \frac{d^2}{4} \equiv \left(\Delta-\frac{d}{2}\right)^2 \,.
\ee
The relation implies $\Delta=\frac{d}{2}\pm \alpha$. In general we  have $\alpha \in \mathbb{R} $. As done in Sec.\,\ref{se:AdS}, we restrict to $\alpha \in \mathbb{R}_+$ without loss of generality, and define
\be
\Delta_\pm=\frac{d}{2}\pm \alpha\,. \label{eq:Delta_def}
\ee
Unitarity requires $\Delta \geq d/2-1$. Hence for $\alpha>1$ only the dual CFT operator with $\Delta_+$  exists. For $0<\alpha<1$, both $\Delta_-$ and $\Delta_+$ branches of the correspondence exist.  
\footnote{Notice $\alpha$ is analytically continued to the complex plane in the spectral conformal representation, see Sec.\,\ref{se:alpharep}. 
Also, when studying the Euclidian principal series representation,  the conformal dimension  is usually defined as
$\Delta=\frac{d}{2}+i\nu$, $\nu\in \mathbb{R} $. Here we rather use $\alpha \equiv i \nu$, which  is typically used in other momentum space works such as \cite{Meltzer:2020qbr}. 
}

\subsection{CFT Correlators}

We are interested in scalar primary operators noted ${\cal O}_i$. These representations of the conformal group are labelled by the scaling dimension $\Delta_i$.

\paragraph{2-point}

In position space a 2-point correlator is constrained by conformal invariance to the form 
\be
\langle  {\cal O}_i(x_1)  {\cal O}_i(x_2) \rangle = \frac{1}{(-x^2_{12})^{\Delta_i}}\,,
\ee
up to an overall coefficient.

We introduce the reduced correlator
\be
\langle {\cal O}(p_1) {\cal O}(p_2) \rangle = (2\pi)^d \delta^{(d)}(p_1+p_2)
\llangle {\cal O}(p_1) {\cal O}(p_2) \rrangle \,.
\ee
We get 
$
\langle  {\cal O}(x_1)  {\cal O}(x_2) \rangle = 
\int \frac{d^dp}{(2\pi)^d} \llangle {\cal O}(p) {\cal O}(-p)e^{-ip\cdot x_{12}} \rrangle
$
and obtain 
\be
\llangle  {\cal O}(p)  {\cal O}(-p) \rrangle = - i \frac{\pi^{d/2} \Gamma(d/2-\Delta)}{\Gamma(\Delta)}\left(\frac{4}{-p^2}\right)^{d/2-\Delta}
\ee
Notice that for  $d=4$ and $\Delta=d/2-1=1$, one recovers the usual 4d relation
\be
\frac{-1}{4\pi^2 x^2} = \int \frac{d^4p}{(2\pi)^4} \frac{i}{p^2}e^{-i p\cdot x}\,.
\ee

Using $\Delta=d/2+\alpha$, the correlator reads
\be
\llangle  {\cal O}_i(p)  {\cal O}_i(-p) \rrangle =  -i\frac{\pi^{d/2} \Gamma(-\alpha)}{\Gamma(\alpha+d/2)}\left(\frac{-p^2}{4}\right)^{\alpha} \,.
\label{eq:CFT_2pt}
\ee

\paragraph{3-point}

3-point correlators of the CFT are constrained by conformal invariance to the form
\be
\langle  {\cal O}_1(x_1)  {\cal O}_2(x_2) {\cal O}_3(x_3) \rangle = \frac{1}{\sqrt{-x^2_{12}}^{\Delta_1+\Delta_2-\Delta_3}\sqrt{-x^2_{23}}^{\Delta_2+\Delta_3-\Delta_1}\sqrt{-x^2_{13}}^{\Delta_1+\Delta_3-\Delta_2}}\,.
\ee
Using $\Delta_i=d/2+\alpha_i$, we have
\be
 \langle  {\cal O}_1(x_1)  {\cal O}_2(x_2) {\cal O}_3(x_3) \rangle = \frac{1}{\sqrt{-x^2_{12}}^{\alpha_1+\alpha_2-\alpha_3+d/2}\sqrt{-x^2_{23}}^{\alpha_2+\alpha_3-\alpha_1+d/2}\sqrt{-x^2_{13}}^{\alpha_1+\alpha_3-\alpha_2+d/2}}\,.
\ee

In momentum space one introduces the reduced correlator
\be
\langle  {{\cal O }}(p_1) {\cal O}_1(p_2) {\cal O}_2(p_3)   \rangle= (2\pi)^d
\delta^{(d)}({ p}_1+{ p}_2+{ p}_3)  
\llangle  {{\cal O }}(p_1) {\cal O}_1(p_2) {\cal O}_2(p_3)   \rrangle\,.
\ee

By direct calculation using \textit{e.g.} Schwinger parametrization (see \cite{Bzowski:2013sza}), one obtains
\begin{align} \label{eq:3pt_3K}
&i\llangle  {{\cal O }_1}(p_1) {\cal O}_2(p_2) {\cal O}_3(p_3)   \rrangle = 
\\ & \nn  \frac{\pi^d 2^{4-\alpha_1-\alpha_2-\alpha_3}}{
\Gamma(\frac{\alpha_1+\alpha_2+\alpha_3+d/2}{2})
\Gamma(\frac{\alpha_1+\alpha_2-\alpha_3+d/2}{2})
\Gamma(\frac{\alpha_1-\alpha_2+\alpha_3+d/2}{2})
\Gamma(\frac{-\alpha_1+\alpha_2+\alpha_3+d/2}{2})
}\,\times
\\ & \nn  
  \left(-p^2_1\right)^{\alpha_1/2}\left(-p^2_2\right)^{\alpha_2/2}\left(-p^2_3\right)^{\alpha_3/2}\int^\infty_0 dz z^{d/2-1} K_{\alpha_1}\left(\sqrt{-p^2_1}z\right)K_{\alpha_2}\left(\sqrt{-p^2_2}z\right)K_{\alpha_3}\left(\sqrt{-p^2_3}z\right).
\end{align}

\subsection{Shadow Transform}

\label{se:CFT_shadow}

In conformal field theory a scalar operator $\cal O$ is accompanied by a  ``shadow" operator $\tilde {\cal O}$ with dimension 
\be
\tilde \Delta= d-\Delta\,.
\ee
Equivalently, if $\Delta=d/2+\alpha$, then $\tilde \Delta=d/2-\alpha$. 
$\cal O$ and $\tilde {\cal O}$ have a natural conformal-invariant pairing $\int d^dx {\cal O}(x) \tilde {\cal O}(x) $. Such operation can  connect the legs of different CFT correlators to build loop diagrams,  see Sec.\,\ref{se:CFT_bubble}.

In a given $n$-point correlator, an operator ${\cal O}_i$   is shadow-transformed into $\tilde {\cal O}_i$  by convoluting the correlator with the corresponding shadow 2-pt function $\langle \tilde {\cal O}_i(x) \tilde {\cal O}_i(x') \rangle$ \cite{Karateev:2018oml}. Here we give the transformation for a 3-pt correlator. 
The  shadow transform of the position space 3-pt correlator \cite{Karateev:2018oml, Meltzer:2019nbs} in Minkowski metric is given by
\be
i \int d^d x'_{3} \langle \tilde {\cal O}_3(x_3) \tilde {\cal O}_3(x'_3) \rangle \langle  {\cal O}_1(x_1) {\cal O}_2(x_2) {\cal O}_3(x'_3) \rangle  = S^{\Delta_1,\Delta_2}_{\Delta_3}\langle  {\cal O}_1(x_1) {\cal O}_2(x_2) \tilde {\cal O}_3(x_3) \rangle 
\ee
with the coefficient
\be
S^{\Delta_1,\Delta_2}_{\Delta_3} = 
\frac{\pi^{d/2} \Gamma(\alpha_3)}{\Gamma(d/2-\alpha_3)} \frac{\Gamma(\frac{\alpha_3+\alpha_1-\alpha_2+d/2}{2})\Gamma(\frac{\alpha_3-\alpha_1+\alpha_2+d/2}{2})}{
\Gamma(\frac{-\alpha_3+\alpha_1-\alpha_2+d/2}{2})\Gamma(\frac{-\alpha_3-\alpha_1+\alpha_2+d/2}{2}) }\,.
\ee

We introduce the Fourier transform of the operators as in Eq.\,\eqref{eq:FT}. 
The shadow transform in momentum space becomes a product
\begin{align}
& i \int d^d x'_{3} \langle \tilde {\cal O}_3(x_3) \tilde {\cal O}_3(x'_3) \rangle \langle  {\cal O}_1(x_1) {\cal O}_2(x_2) {\cal O}_3(x'_3) \rangle  \\ \nn & = 
i \int\frac{d^dp_1}{(2\pi)^d}\int\frac{d^dp_2}{(2\pi)^d}\int\frac{d^dp_3}{(2\pi)^d}
e^{i(p_1.x_1+p_2.x_2+p_3.x_3)} \llangle \tilde {\cal O}_3(p_3) \tilde {\cal O}_3(-p_3) \rrangle
\llangle  {\cal O}_1(p_1) {\cal O}_2(p_2) {\cal O}_3(p_3) \rrangle \,
\\ \nn & 
= S^{\Delta_1,\Delta_2}_{\Delta_3}
\int\frac{d^dp_1}{(2\pi)^d}\int\frac{d^dp_2}{(2\pi)^d}\int\frac{d^dp_3}{(2\pi)^d} e^{i(p_1.x_1+p_2.x_2+p_3.x_3)}
\llangle{\cal O}_1(p_1) {\cal O}_2(p_2) \tilde {\cal O}_3(p_3) \rrangle .
\end{align}
The relation in momentum space can be read from the two last lines.

Using the explicit expression for the 2-point shadow correlator (given by Eq.\,\eqref{eq:CFT_2pt} with $\alpha\rightarrow -\alpha_3$) the shadow transform in momentum space becomes simply
\begin{align} \label{eq:ST_3pt}
& \llangle{\cal O}_1(p_1) {\cal O}_2(p_2) \tilde {\cal O}_3(p_3) \rrangle 
= \frac{1}{S^{\Delta_1,\Delta_2}_{\Delta_3}}\frac{\pi^{d/2} \Gamma(\alpha_3)}{\Gamma(d/2-\alpha_3)}\left(\frac{-p_3^2}{4}\right)^{-\alpha_3}
\llangle{\cal O}_1(p_1) {\cal O}_2(p_2)  {\cal O}_3(p_3) \rrangle 
\\ \nn & 
\quad \quad \quad \quad = \frac{
\Gamma(\frac{-\alpha_3+\alpha_1-\alpha_2+d/2}{2})\Gamma(\frac{-\alpha_3-\alpha_1+\alpha_2+d/2}{2}) }{\Gamma(\frac{\alpha_3+\alpha_1-\alpha_2+d/2}{2})\Gamma(\frac{\alpha_3-\alpha_1+\alpha_2+d/2}{2})}
\left(\frac{-p_3^2}{4}\right)^{-\alpha_3}\llangle{\cal O}_1(p_1) {\cal O}_2(p_2)  {\cal O}_3(p_3) \rrangle  \,.
\end{align}

We can use 
 this identity Eq.\,\eqref{eq:ST_3pt} on the explicit representation of the momentum-space 3-pt correlator Eq.\,\eqref{eq:3pt_3K}. When doing so,  one  explicitly sees that
the shadow transform  Eq.\,\eqref{eq:ST_3pt}  precisely amounts to flip the sign of $\alpha_3$ in Eq.\,\eqref{eq:3pt_3K}.

\subsection{CFT Bubble}

\label{se:CFT_bubble}

In the calculation of the  dressed AdS propagator in the conformal spectral representation, a CFT bubble diagram appears. It is made of two 3-pt correlators whose legs are connected  via pairing between ${\cal O}_i$ and $\tilde {\cal O}_i$, or equivalently the legs are connected by 2-pt shadow correlators, see Sec.\,\ref{se:CFT_shadow}.

The CFT bubble evaluation is well known \cite{Karateev:2018oml, Meltzer:2019nbs}. Translating to our conventions we have\,\footnote{One uses $\delta(x_E)=i \delta(x)$, $\delta(\nu)=\delta(-i\alpha)$.  The latter will be integrated along the imaginary line, such that the argument of the Dirac delta is real. However one can check that one  can equivalently use   $\delta(-i\alpha)=i\delta(\alpha)$, that we use throughout.  } 
\begin{align}
&\int d^dx_1 d^dx_2 \langle  {{\cal O }_a}(x)  {\cal O}_1(x_1)  {\cal O}_2(x_2)   \rangle
\langle  { \tilde {\cal O}_1(x_1) \tilde {\cal O}_2(x_2) \tilde {\cal O }_b}(x')
\rangle \\ & \nn
    = {\cal B}_{\cal O} \delta_{ab} 2\pi \left(\delta^{(d)}(x-x')\delta(\alpha-\alpha')\,-i
  \left( \frac{\Gamma(\alpha+d/2)\Gamma(d/2-\alpha)}{\pi^d\Gamma(\alpha)\Gamma(-\alpha)} \right)^{1/2}
    \frac{ \delta(\alpha+\alpha')}{\left(-(x-x')^2\right)^{\alpha+d/2}} \right) \\ & \nn
    = {\cal B}_{\cal O} \delta_{ab} 2\pi \left(\delta^{(d)}(x-x')\delta(\alpha-\alpha')\,+ s.t. \right)
\end{align}
where the second term in the parenthesis corresponds to the shadow transform of the first one---the $-i$ results from Minkowski metric. Below we denote this second term 
by \textit{s.t.} The bubble coefficient is given by 
\be
{\cal B}_{\cal O} = \frac{2\pi^{3d/2}}{\Gamma(d/2)}\frac{\Gamma(\alpha)\Gamma(-\alpha)}{\Gamma(\alpha+d/2)\Gamma(d/2-\alpha)} \,.
\label{eq:BO}
\ee

Introducing the Fourier tranform of the operators, one obtains the CFT bubble in momentum space (involving the reduced correlators). It  is given by
\be \label{eq:CFT_bubble}
\int \frac{d^dq}{(2\pi)^d}
\llangle  {{\cal O }}_a(p) {\cal O}_1(q) {\cal O}_2(-p-q)   \rrangle
\llangle  \tilde{{\cal O }_1}(-q) \tilde{\cal O}_2(p+q) \tilde {\cal O}_b(-p)   \rrangle =  2\pi  {\cal B}_{\cal O} \delta_{ab} \delta(\alpha-\alpha')+s.t.
\ee
On the left-hand side we recognize the integration over the internal  $d$-momentum running inside the loop.

\subsection{Boundary-to-Bulk Propagators }

\label{se:Bo2B}
Here  we come back to AdS.  We introduce the boundary-to-bulk propagator ${\cal K}(z, x, x')$,  a Green function  for sources placed on the AdS boundary (see \textit{e.g.} \cite{Witten:1998qj, Freedman:1999gp} ).
 From the bulk-to-bulk AdS propagator $G^{(0)}(X,X')$, ${\cal K}(z, x, x')$ can  be obtained by sending \textit{e.g.} $z'$  to the AdS boundary, while including an appropriate  scaling factor in $z'$ such that the obtained object  is non zero and finite for $z \neq 0$. 
Boundary-to-bulk propagators  are  closely related to the dual CFT operators and can  be directly labelled with the corresponding scaling dimension $\Delta$.

The boundary-to-bulk propagator associated with a CFT operator with dimension $\Delta_+$ ($\Delta_-$) is defined as 
   \be
{\cal K}_{\Delta_\pm}(x,x', z) =    \frac{\Gamma(\Delta_\mp)}{\pi^{d/2}\Gamma(\Delta_\mp-\frac{d}{2})} \left(\frac{z }{z^2-(x-x')^2}\right)^{\Delta_\mp}
   \ee
where $\Delta_-=d-\Delta_+$, according to  definitions in Eq.\,\eqref{eq:Delta_def}).
\footnote{These propagators satisfy ${\cal K}_{\Delta_\pm}(z\rightarrow 0)= i z^{\Delta_{\pm}}\delta^{(d)}(x)$. 
   A bulk field $\phi(z,x)$ sourced by a boundary field (\textit{i.e.} ``boundary data'') $\phi_0(x)$ satisfies $
 \phi(z,x)   = -iz^{-\Delta_{\pm}}\int d^d y {\cal K}_{\Delta_+}(z,x-y) \phi_0(y)\,.
$ }

 Introducing the Fourier transform, we define the  boundary-to-bulk propagator in position-momentum space as $ {\cal K}_{\Delta_+}(p,z)=k^{d/2}\int d^dx\, {\cal K}_{\Delta_+}(x,x',z) e^{i p.(x-x')}$. \footnote{Defined that way, ${\cal K}_{\Delta_+}$ and ${\cal K}_{\Delta_-}$ have opposite mass dimensions, $[{\cal K}_{\Delta_-}]=-[{\cal K}_{\Delta_+}]=\alpha$.  }
 The result is 
\be
 {\cal K}_{\Delta_\pm}(p,z) = i (kz)^{d/2}     \frac{2}{\Gamma(\Delta_\mp-d/2)}    \left(\frac{\sqrt{-p^2}}{2}\right)^{\Delta_\mp - d/2}K_{d/2-\Delta_\mp}(\sqrt{-p^2}z) \,.
\ee   
   We see that the Bessel $K$ function appears. We also see that multiplying ${\cal K}_{\Delta_+}(p,z)$ and ${\cal K}_{\Delta_-}(p,z)$ cancels out the overall factor of $p$, giving  the dimensionless quantity
    \be
 {\cal K}_{\Delta_+}(p,z)  {\cal K}_{\Delta_-}(p,z')  =  4 \frac{\alpha \sin(\pi \alpha)}{\pi}
      (k^2 zz')^{d/2} K_\alpha(\sqrt{-p^2}z) K_\alpha(\sqrt{-p^2}z')\,. 
\ee   
   This product reproduces precisely the form of the conformal spectral function $\Omega_{\alpha}(p;z,z')$ (see Eq.\,\eqref{eq:Omega_alpha}) such that 
   \be
   \Omega_\alpha (p,z,z')= \frac{i}{4\pi}
   {\cal K}_{\Delta_-}(p,z){\cal K}_{\Delta_+}(p,z')\,.
   \label{eq:Omega_KK}
   \ee
 Such feature is expected---it amounts to introduce the ``split'' representation \cite{Leonhardt:2003qu} in position-momentum space. Since in position space the split representation would involve a convolution on the AdS boundary, in position-momentum space the split representation involves a simple product.

For our calculation of the dressed AdS propagator, the well-known correspondence between ${\cal K}$ and  CFT 2-pt correlator is not needed. 
On the other hand, bulk cubic interactions will involve the 
convolution of three different boundary-to-bulk propagagors. We introduce the notation ${\cal K}_{\Delta_\pm}(p,z)={\cal K}_{\alpha}^\pm(p,z)$.
For the product of three  generic ${\cal K}^-$, using Eq.\,\eqref{eq:3pt_3K} we find
\begin{align} \label{eq:3K_full}
   \int \frac{du} {(k u )^{d+1}} & {\cal K}^-_{\alpha_1 }( p_1; u){\cal K}^-_{\alpha_2 }( p_2; u){\cal K}^-_{\alpha_3 }( p_3;  u) 
\\ \nn &  ~~~~
=  -i \frac{2^{3-\alpha_1-\alpha_2-\alpha_3}}{\Gamma(\alpha_1)\Gamma(\alpha_2)\Gamma(\alpha_3)} (-p^2_1)^{\alpha_1/2} 
(-p^2_2)^{\alpha_2/2} (-p^2_3)^{\alpha_3/2}
\\  \nn &   ~~~~~~
\quad \quad \times \int du \,(ku)^{d/2-1} K_{ \alpha_1}\left(\sqrt{-p_1^2} u\right)K_{ \alpha_2}\left(\sqrt{-p_2^2} u\right)K_{\alpha_3}\left( \sqrt{-p_3^2} u\right)
\\ \nn &  ~~~~
 = c_{\alpha_1,\alpha_2,\alpha_3}k^{d/2-1}
\llangle  {{\cal O }}_1(p_1) {\cal O}_2(p_2) {\cal O}_3(p_3)   \rrangle
\,\,.
\end{align}
with the dimensionless coefficient
\be \label{eq:3K_c}
c_{\alpha_1,\alpha_2,\alpha_3}=
\frac{
\Gamma(\frac{\alpha_1+\alpha_2+\alpha_3+d/2}{2})
\Gamma(\frac{\alpha_1-\alpha_2+\alpha_3+d/2}{2})
\Gamma(\frac{\alpha_1+\alpha_2-\alpha_3+d/2}{2})
\Gamma(\frac{-\alpha_1+\alpha_2+\alpha_3+d/2}{2})
}{2 \pi^d \, \Gamma(\alpha_1)\Gamma( \alpha_2)\Gamma( \alpha_3)} 
\ee
Eq.\,\eqref{eq:3K_full}  relates a 3-pt bulk diagram (first line) to a 3-pt CFT correlator (last line), explicitly showing AdS/CFT at work. 

Convolutions involving ${\cal K}^+$ propagators are obtained from Eq.\,\eqref{eq:3K_full} by a shadow transform. As seen in Sec.\,\ref{se:CFT_shadow}, this amounts to flipping the signs of the corresponding $\alpha_i$ in the $c(\alpha_1, \alpha_2, \alpha_3)$ coefficient.
We have  for instance 
 \begin{align} 
   \int \frac{du}{(k u )^{d+1}} & {\cal K}^-_{\alpha_1 }( p_1; u){\cal K}^-_{\alpha_2 }( p_2; u){\cal K}^+_{\alpha_3 }( p_3;  u) 
 = c(\alpha_1,\alpha_2,-\alpha_3) k^{d/2-1}
\llangle  {{\cal O }}_1(p_1) {\cal O}_2(p_2) \tilde {\cal O}_3(p_3)   \rrangle
\,\,.
\end{align}

\section{Momentum Spectral Integrals  at Large $pz$}
\label{app:large_pz}

We first review aspects of the spectral integral $\int^A_0 dm \Omega_m(z,z')$. We are interested in the $Az\gg1$, $Az'\gg1$ limit at fixed $\alpha$, hence for our purposes the $\alpha$-dependence is negligible and 
we can  set $\alpha=1/2$. The spectral integral is then proportional to
\be
\int^A_0 dm  \sin(mz)\sin(mz')=\frac{1}{2}\frac{\sin(A(z-z'))}{z-z'}   -\frac{1}{2} \frac{\sin(A(z+z'))}{z+z'} \,. \label{eq:Omega1}
\ee

If we let $A\rightarrow \infty$ at fixed $z,z'$, using the definition of the $\rm{sinc}$ function as a nascent Dirac delta, the first term of Eq.\,\eqref{eq:Omega1} becomes proportional to a Dirac delta and the second vanishes---proving Eq.\,\eqref{eq:JJ}. 
We recall this is understood in the distribution sense. The second term oscillates infinitely rapidly for any $z$, $z'$ such that under a $z$ integral, for any appropriate test function  
the  second term averages to zero under the integral. This happens similarly for the first term of Eq.\,\eqref{eq:Omega1} except for $A(z-z')\ll1 $, which asymptotically gives rise to the Dirac delta. 

Instead of letting $A\rightarrow \infty$ in Eq.\,\eqref{eq:Omega1}, we can  introduce  $\bar z = \frac{z+z'}{2}$, $z-z'=\bar z \epsilon$, and let $\bar z \rightarrow \infty$ at fixed $A$ and $\epsilon$. 
This gives the same result as the $A\rightarrow \infty$ limit since 
\be
\frac{\sin(A(z-z'))}{z-z'} = \frac{\sin(A\bar z \epsilon)}{\bar z \epsilon}    =\frac{A}{\bar z}\frac{\sin(A\bar z \epsilon)}{ A \epsilon}  
\xrightarrow[\bar z\rightarrow\infty]{} \frac{\pi A}{\bar z}\delta(A \epsilon  )  =\pi\delta(z-z' )\,.
\ee
Again, this limit is understood in the distribution sense. 
For our purposes, taking the limit of large $\bar z$ at fixed interval $[0,A]$ is more convenient. 

Consider then the spectral integral \be
\int^A_0 dm  \sin(mz)\sin(mz')F(m,A)
\ee
where the $F(m,A)$ function is smooth in $m$ and is independent of $z,z'$. Taking the large $\bar z$ limit at $A$ fixed, for large enough $\bar z$ we have $\partial_m\left(\sin(mz)\sin(mz')\right) \gg \partial_m F(m,A)$ every\-where on the interval and we can approximate the rapidly oscillating function by its average over $[0,A]$,  
\be
\int^A_0 dm  \sin(mz)\sin(mz')F(m,A)\approx\left(\frac{1}{2}\frac{\sin(A(z-z'))}{A(z-z')}   -\frac{1}{2} \frac{\sin(A(z+z'))}{A(z+z')} \right)\int^A_0 dm  F(m,A)
\ee
Taking the large $\bar z$ limit as above gives asymptotically
\be
 \left(\frac{1}{2}\frac{\sin(A(z-z'))}{A(z-z')}    \right)\int^A_0 dm  F(m,A)\xrightarrow[\bar z\rightarrow\infty]{} 
\frac{\pi}{2A}\delta(z-z')  \int^A_0 dm  F(m,A)
\ee
in the distribution sense. 

We can apply the preliminary results above to evaluate the double spectral integral of Eq.\,\eqref{eq:MSresult}
\be
\int^{A_1}_0 dm_1 \int^{A_2}_0 dm_2 \Omega_{m_1}(z,z')\Omega_{m_2}(z,z')F(m_1,m_2,A_i)
\label{eq:OmegaDouble1}
\ee
where $A_1=p$, $A_2=p-m_1$ in the large $pz$ limit.  We keep $p$  finite and take the limit of large $\bar z$. Taking $\alpha=1/2$, the integral is proportional to 
\be
\int^{A_1}_0 dm_1 \int^{A_2}_0 dm_2 \sin (m_1 z)\sin (m_1 z')\sin (m_2 z)\sin (m_2 z') F(m_1,m_2,A_i)
\label{eq:OmegaDouble2}
\ee
$F$ is assumed to be smooth in $m_1$ and $m_2$ such that the limit for rapidly oscillating function introduced above applies. 
We first approximate the $m_2$ integral, giving
\be
\int^{A_1}_0 dm_1  \sin m_1 z\sin m_1 z' \left(\frac{1}{2}\frac{\sin(A_2(z-z'))}{A_2(z-z')}   -\frac{1}{2} \frac{\sin(A_2(z+z'))}{A_2(z+z')} \right)\int^{A_2}_0 dm_2 F(m_1,m_2,A_i)\,.
\ee
Then we use that $\frac{1}{A_2}\int^{A_2}_0 dm_2 F(m_1,m_2,A_i)$ is smooth in $m_1$ with respect to the whole oscillating function such that the latter can be averaged and extracted from the $m_1$ integral.  Combining the sines we have
\begin{align}
=\frac{1}{4A_1}\int^{A_1}_0 dm_1  & \left(\cos(m_1(z-z'))-\cos(m_1(z+z')) \right)  \left(\frac{\sin(A_2(z-z'))}{z-z'}   - \frac{\sin(A_2(z+z'))}{z+z'} \right) \nn \\ & \times \int^{A_1}_0 dm_1\int^{A_2}_0 dm_2\frac{1}{A_2} F(m_1,m_2,A_i) 
\end{align}
The first line contains the average of the oscillating function in $m_1$.
The integrals of the four terms give 
\be
\frac{1}{A_1}\int^{A_1}_0 dm_1  \cos(m_1(z-z')) \frac{\sin(A_2(z-z'))}{z-z'}  =
\frac{1}{2 (z-z')}\sin{p(z-z')} \label{eq:1st_term_osc}
\ee
\be
\frac{1}{A_1}\int^{A_1}_0 dm_1  \cos(m_1(z-z')) \frac{\sin(A_2(z+z'))}{z+z'}  =
\frac{1}{2p zz'}\sin{p z}\sin{p z'}
\ee
\be
\frac{1}{A_1}\int^{A_1}_0 dm_1  \cos(m_1(z+z')) \frac{\sin(A_2(z-z'))}{z-z'}  =
\frac{1}{2p zz'}\sin{p z}\sin{p z'}
\ee
\be
\frac{1}{A_1}\int^{A_1}_0 dm_1  \cos(m_1(z+z')) \frac{\sin(A_2(z+z'))}{z+z'}  =
\frac{1}{2 (z+z')}\sin{p(z+z')}
\ee
In the large $\bar z$ limit the first term Eq\,.\eqref{eq:1st_term_osc}  dominates, the other become negligible. 
Again, this term gives rise a nascent Dirac delta in the large $\bar z$ limit, 
\be
\frac{1}{2 (z-z')}\sin{p(z-z')}= \frac{1}{2 \bar z \epsilon }\sin{p\bar z \epsilon}= \frac{p}{2 \bar z  }\frac{\sin{p\bar z \epsilon}}{p \epsilon} \xrightarrow[\bar z\rightarrow\infty]{} \frac{p}{ \bar z  } \frac{\pi}{2} \delta(p \epsilon)
=   \frac{\pi}{2} \delta( z-z') \,,
\ee
 understood in the distribution sense. 

We have thus obtained that the double spectral integral Eq.\,\eqref{eq:OmegaDouble2}  tends to
\be
\frac{\pi}{8} \delta( z-z')\int^{p}_0 dm_1\int^{p-m_1}_0 dm_2\frac{1}{A_2} F(m_1,m_2,A_i)\,
\ee
at large $\bar z$. This delta function, as mentioned throughout this section, is best thought of as being under an integral. When interested in the value of $\text{Im}\Pi$ itself, this delta may naively seem like a divergence. One should remember that this is a nascent delta and is hence replaced as $z-z'$ becomes arbitrarily small.

In our bubble calculation, $F(m_1,m_2)=\K(d,p, m_1,m_2)$. Even though the remaining double integral may be difficult, we see that all the dependence in $z$ is factored out. Hence the scaling in $p$ is obtained by simple dimensional analysis, giving $p^{d-3}$. Putting together all the factors leads to the asymptotic scaling given in Eq.\,\eqref{eq:ImPiLargep}.

\section{Kinematic Approximation: Numerical Checks}\label{app:kin_approx}

\begin{figure}[th]
\centering
	\includegraphics[width=0.45\linewidth,trim={0cm 0cm 0cm 0cm},clip]{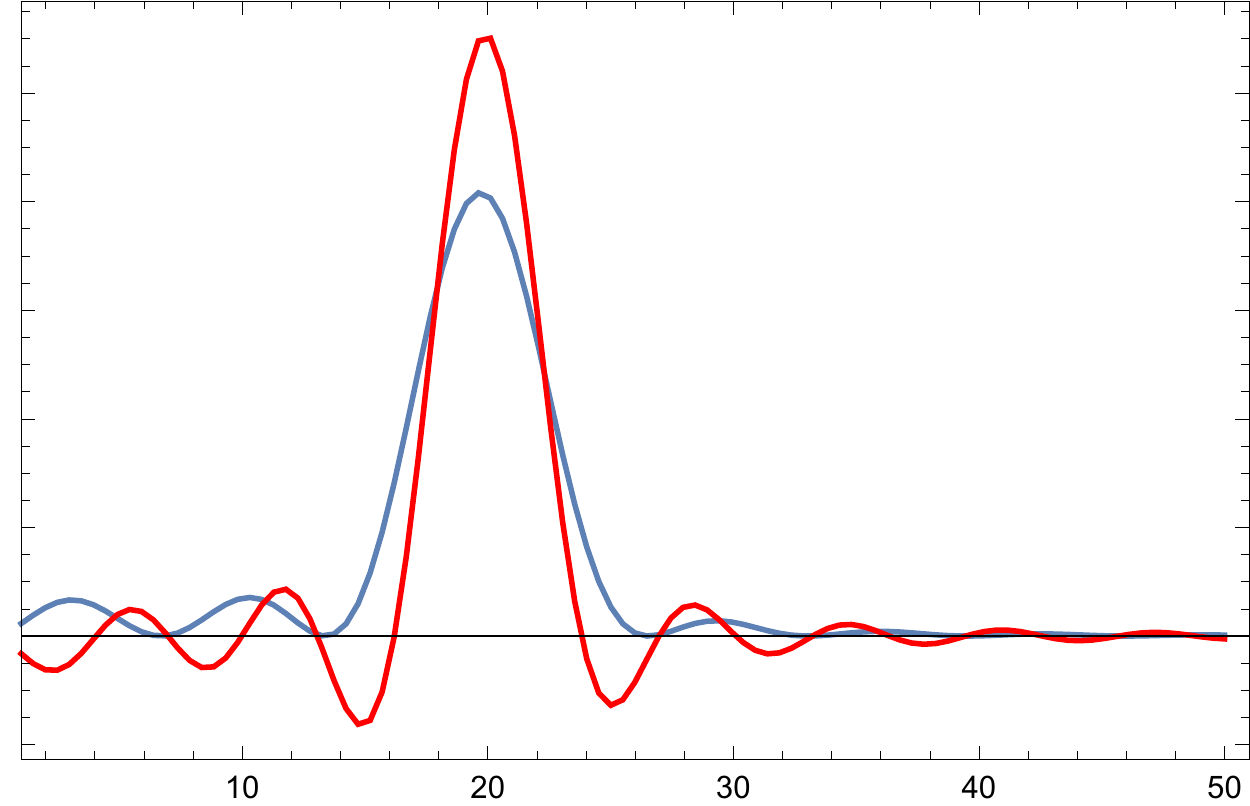}
		\includegraphics[width=0.45\linewidth,trim={0cm 0cm 0cm 0cm},clip]{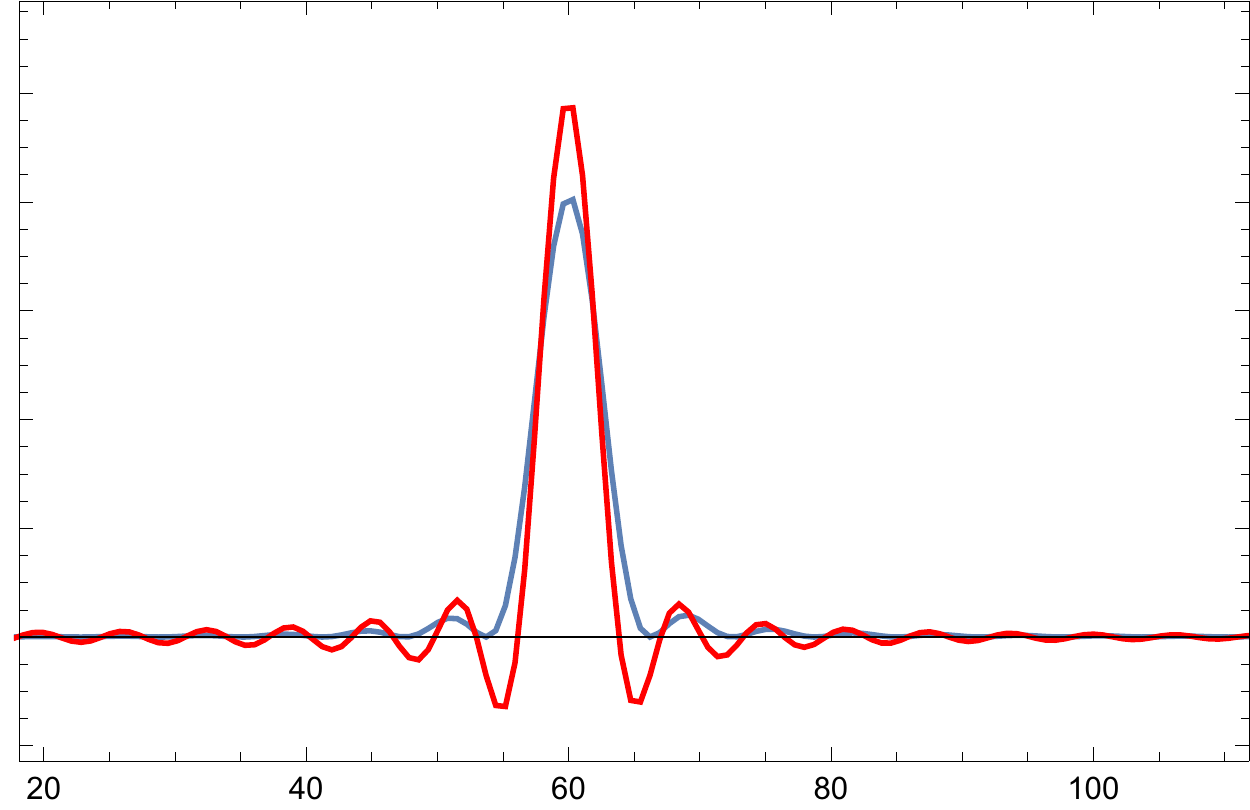}
\caption{
The exact (red) and approximate (blue) profiles of $(zz')^{\frac{d-1}{2}} {\rm Im} \Pi(z,z')$ as a function of $pz$, taking $pz'=20$ (left) and $pz'=60$ (right), and taking   $d=4$.
\label{fig:Kin_approxes}}
\end{figure}

Fig.\,\ref{fig:Kin_approxes} shows examples of ${\rm Im} \Pi(z,z')$ in the kinematic approximation introduced in Sec.\,\ref{se:ImPi_approxes}. 
We can see that a side effect of the kinematic approximation is to somewhat smoothen the oscillations. It also renders ${\rm Im}\Pi$ positive for any $z\neq z'$. These details have a mild impact on the subsequent results. Upon performing the $z$ integrals, the results match with $O(10)$\% accuracy.

\section{Bubble Diagrams} \label{app:bubble} 

In these calculations, we consider bubbles formed of $\lambda \Phi\Phi_1\Phi_2$ and $\xi \Phi\partial_M\Phi_1 \partial^M\Phi_2$ vertices, hence the fields and propagators are not treated as identical.
We use standard manipulations (Feynman parameterization, dimensional regularization) to evaluate the loop. We define
 $\Delta=-x(1-x)p^2+x m_1^2+ (1-x)m^2_2$.
 
 When taking the imaginary part of $\Pi$, we use the kinematic approximation of Sec.\,\ref{se:ImPi_approxes} to obtain
\be
{\rm Im}\int_{x_-}^{x_+}  \log \Delta =   -\pi
\ee
\be
{\rm Im}\int_{x_-}^{x_+} x \log \Delta ={\rm Im}\int_{x_-}^{x_+} (1-x) \log \Delta = -\frac{  \pi }{2}
\ee
\be
{\rm Im}\int_{x_-}^{x_+} x(1-x) \log \Delta = -\frac{ \pi}{6}
\ee
\be
{\rm Im}\int_{x_-}^{x_+} x^2(1-x) \log \Delta ={\rm Im}\int_{x_-}^{x_+} x(1-x)^2 \log \Delta = -\frac{  \pi }{12}
\ee
\be
{\rm Im}\int_{x_-}^{x_+} x^2(1-x)^2 \log \Delta = -\frac{ \pi}{30}\,.
\ee

\subsection{$\Phi^3-\Phi^3 $}
\label{app:ll}

The amplitude of the bubble induced by two $\lambda \Phi\Phi_1\Phi_2$ vertices is
   \be
i\Pi_{\lambda\lambda}(z_1,z_2) =  \lambda^2 
\frac{1}{k^4 z_1 z_2} \int dq_1 dq_2   q_1q_2 \int\frac{d^4q}{(2\pi)^4}
\frac{J_{\alpha_1}(q_1z_1)J_{\alpha_1}(q_1z_2)
J_{\alpha_2}(q_2z_1)J_{\alpha_2}(q_2z_2)}{(q^2-q^2_1)((q+p)^2-q^2_2)}
\,.
\ee
We introduce a Feynman parameter $x$ and use dimensional regularization to evaluate the loop. We have
   \begin{align}
 &i\Pi_{\lambda\lambda}(z_1,z_2) = \\ & \lambda^2 
\frac{1}{k^4z_1z_2} \int dq_1 dq_2   q_1q_2 \int^1_0 dx  \frac{-i}{16 \pi^2} \log\left(\frac{\Delta}{\Lambda^2}\right)
J_{\alpha_1}(q_1z_1)J_{\alpha_1}(q_1z_2)
J_{\alpha_2}(q_2z_1)J_{\alpha_2}(q_2z_2)+\ldots \nn
\end{align}
$\Lambda$ is the regularization scale and the ellipses represent the divergent part of the loop diagram. Both the divergence and the $\Lambda$-dependence vanish upon taking the imaginary part of $\Pi$. 
Taking the imaginary part gives Eq.\,\eqref{eq:ImPi_ll}.

\subsection{$\Phi^3-\Phi (\partial \Phi)^2$}
\label{app:lz}

The amplitude of the bubble induced by one  $\lambda \Phi\Phi_1\Phi_2$ and one $\xi \Phi\partial_M\Phi_1 \partial^M\Phi_2$ vertex is given by
   \begin{align}
& i\Pi_{\lambda\zeta}(z_1,z_2) = \\
&- \lambda \zeta
\frac{(k z_2)^2}{ k^4 z_1  z_2} \int dq_1 dq_2   q_1q_2 \int\frac{d^4q}{(2\pi)^4}
\bigg(q.(p+q)- \partial^{(1)}_{z_2}\partial^{(2)}_{z_2}\bigg)\frac{J_{\alpha_1}(q_1z_1)J_{\alpha_1}(q_1z_2)
J_{\alpha_2}(q_2z_1)J_{\alpha_2}(q_2z_2)}{(q^2-q^2_1)((q+p)^2-q^2_2)} \nn
\,
\end{align}
plus its $(1\leftrightarrow 2)$ counterpart.  The $\partial_z^{(i)}$ derivative acts only on the $G_{q_i}(z_1,z_2)$ propagator, \textit{i.e.} only on the 
$J_{\alpha_i}(q_i z)$ Bessel function in the above expression and subsequent ones. 
We introduce a Feynman parameter $x$ and perform the loop integral via dimensional regularization. We obtain
   \begin{align}
 i\Pi_{\lambda\zeta}&(z_1,z_2) =  - \lambda \zeta
\frac{z_2^2}{k^2 z_1z_2}   \int dq_1  dq_2 q_1q_2 \int^1_0 dx \\  & \frac{-i}{16 \pi^2} \log\left(\frac{\Delta}{\Lambda^2}\right)
\bigg(2 \Delta -x(1-x)p^2- \partial^{(1)}_{z_2}\partial^{(2)}_{z_2}\bigg) J_{\alpha_1}(q_1z_1)J_{\alpha_1}(q_1z_2)
J_{\alpha_2}(q_2z_1)J_{\alpha_2}(q_2z_2)+\ldots
 \nn
\end{align}
$\Lambda$ is the regularization scale and the ellipses represent the divergent part of the loop diagram. Both  vanish upon taking the imaginary part of $\Pi$.
Taking the imaginary part and including the $(1\leftrightarrow 2)$ contribution gives Eq.\,\eqref{eq:ImPi_lchi}.

\subsection{$\Phi (\partial \Phi)^2-\Phi (\partial \Phi)^2$}

\label{app:zz}

The amplitude of the bubble induced by two $\xi \Phi\partial_M\Phi_1 \partial^M\Phi_2$ vertices is
   \begin{align}
i\Pi_{\zeta\zeta}&(z_1,z_2) =  \zeta^2 
\frac{(k z_1)^2(k z_2)^2}{k^4 z_1z_2}    \int dq_1  dq_2 q_1q_2 \int\frac{d^4q}{(2\pi)^4} \\ &
\bigg(q.(p+q)- \partial^{(1)}_{z_1}\partial^{(2)}_{z_1}\bigg)\bigg(q.(p+q)- \partial^{(1)}_{z_2}\partial^{(2)}_{z_2}\bigg)
\frac{J_{\alpha_1}(q_1z_1)J_{\alpha_1}(q_1z_2)
J_{\alpha_2}(q_2z_1)J_{\alpha_2}(q_2z_2)}{(q^2-q^2_1)((q+p)^2-q^2_2)} \nn
\,.
\end{align}
We introduce a Feynman parameter $x$ and perform the loop integral via dimensional regularization to obtain
 \begin{align}
i\Pi_{\zeta\zeta}(z_1,z_2) =  \zeta^2
\frac{z_1^2z_2^2}{z_1z_2}  &  \int dq_1  dq_2 q_1q_2  \int^1_0 dx \frac{-i}{16 \pi^2} 
 \log\left(\frac{\Delta}{\Lambda^2}\right)  \\ & \nn
\bigg(
3\Delta^2+\left(
\frac{1}{2}-6 x(1-x)\right)p^2 \Delta + x^2(1-x)^2p^4  \\ & \nn
\quad\quad-(2 \Delta -x(1-x)p^2) \left(\partial^{(1)}_{z_1}\partial^{(2)}_{z_1}+\partial^{(1)}_{z_2}\partial^{(2)}_{z_2}\right) + \partial^{(1)}_{z_1}\partial^{(2)}_{z_1}\partial^{(1)}_{z_2}\partial^{(2)}_{z_2}\bigg)  \\ &
\quad\quad\quad\quad\quad\quad\quad\quad\quad\quad\quad\quad
J_{\alpha_1}(q_1z_1)J_{\alpha_1}(q_1z_2)
J_{\alpha_2}(q_2z_1)J_{\alpha_2}(q_2z_2)+\ldots \nn
\end{align}
$\Lambda$ is the regularization scale and the ellipses represent the divergent part of the loop diagram.
Evaluating the $x$ integrals and taking the imaginary part of $\Pi$ gives
 \begin{align}
{\rm Im}\Pi_{\zeta\zeta}(z_1,z_2) =  \zeta^2
\frac{z_1^2z_2^2}{z_1z_2} \frac{1}{16 \pi}  &  \int dq_1  dq_2 q_1q_2   
   \\ & \nn
\bigg(
q^4_1+q^4_2+q^2_1q^2_2-\frac{3}{4}p^2(q_1^2+q_2^2)+\frac{p^4}{4}
  \\ & \nn
\quad\quad-\left(q_1^2+q_2^2 - \frac{p^2}{2}\right) \left(\partial^{(1)}_{z_1}\partial^{(2)}_{z_1}+\partial^{(1)}_{z_2}\partial^{(2)}_{z_2}\right) + \partial^{(1)}_{z_1}\partial^{(2)}_{z_1}\partial^{(1)}_{z_2}\partial^{(2)}_{z_2}\bigg)  \\ &
\quad\quad\quad\quad\quad\quad\quad\quad\quad\quad\quad\quad
J_{\alpha_1}(q_1z_1)J_{\alpha_1}(q_1z_2)
J_{\alpha_2}(q_2z_1)J_{\alpha_2}(q_2z_2) \nn
\,.
\end{align}

\subsection{Scalar-Graviton Bubble}
\label{app:grav_loop}
The amplitude induced by the scalar component of the graviton multiplet is 
\begin{align}
i\Pi^\phi(z_1,z_2) =  \frac{-1}{3\,M_*^3} &\left(
\frac{ m_\Phi^2}{2(kz_1)^2} - \frac{3}{2} \partial_{z_1} \partial_{z_1}^{G_\Phi} \right) 
\left(
\frac{ m_\Phi^2}{2(kz_2)^2} -\frac{3}{2} \partial_{z_2} \partial_{z_2}^{G_\Phi} \right)   \\ &\frac{1}{ (k z_1) (k z_2)}
   \int\frac{dq^4}{(2\pi)^4}
G^\Phi_\alpha(q_1,z_1,z_2) G^\phi(q_2,z_2,z_1) \nn
\,.
\end{align}
The $\partial_{z}^{G_\Phi}$ derivatives act only on the internal $\Phi$ propagator (henceforth the term contained in square brackets). Introducing the propagators in their momentum spectral representations, we have
\begin{align}
i\Pi^\phi(z_1,z_2) =  \frac{1}{3\,M_*^3} &\left(
\frac{ m_\Phi^2}{2(kz_1)^2} - \frac{3}{2} \partial_{z_1} \partial_{z_1}^{G_\Phi} \right) 
\left(
\frac{ m_\Phi^2}{2(kz_2)^2} -\frac{3}{2} \partial_{z_2} \partial_{z_2}^{G_\Phi} \right) 
   \\ &\frac{1}{z_1z_2}\int dq_1dq_2 q_1q_2 \int\frac{dq^4}{(2\pi)^4}
\frac{\left[z_1^2z_2^2J_\alpha(q_1z_1)J_\alpha(q_1z_2)\right]
J_0(q_2z_1)J_0(q_2z_2)}{(q^2-q^2_1)((q+p)^2-q^2_2)} \nn
\,.
\end{align}
We introduce a Feynman parameter and perform the loop integral to obtain 
\begin{align}
i\Pi^\phi(z_1,z_2) =  & \frac{-i}{48\pi^2\,M_*^3} 
  \left(
\frac{ m_\Phi^2}{2(kz_1)^2} - \frac{3}{2} \partial_{z_1} \partial_{z_1}^{G_\Phi} \right) 
\left(
\frac{ m_\Phi^2}{2(kz_2)^2} -\frac{3}{2} \partial_{z_2} \partial_{z_2}^{G_\Phi} \right)   \\ &\frac{1}{z_1z_2}\int dq_1dq_2 q_1q_2 \int^1_0 dx 
 \log\left(\frac{\Delta}{\Lambda^2}\right)
\left[z_1^2z_2^2J_\alpha(q_1z_1)J_\alpha(q_1z_2)\right]
J_0(q_2z_1)J_0(q_2z_2)\nn
\,.
\end{align}
Taking the imaginary part gives Eq.\,\eqref{eq:ImPiphi}.

\bibliographystyle{JHEP}
\bibliography{biblio}

\end{document}